%% file: reconfinement.tex
\documentclass{article}
\usepackage[utf8]{inputenc}
\usepackage[english]{babel}
\usepackage[a4paper, margin=3.0cm]{geometry}
\usepackage{amsmath}
\usepackage{amssymb}
\usepackage{xcolor}
\usepackage{braket}
\usepackage{caption}
\usepackage{pdfpages}
\usepackage{comment}

\usepackage{titlesec}
\usepackage{multirow}
\usepackage{adjustbox}
\usepackage{subcaption}
\usepackage{graphicx}
\usepackage{array}
\usepackage{cite}
\usepackage[colorlinks=true,urlcolor=blue,linkcolor=black,citecolor=black]{hyperref}
\usepackage{url}
\usepackage{orcidlink}
\usepackage{booktabs}
\usepackage{float}
\usepackage{tikz}
\usetikzlibrary{arrows.meta}
\usetikzlibrary{patterns}
\usetikzlibrary{calc}

\newcommand{\tmb}[1]{{\mbox{\tiny{#1}}}}
\newcommand{\eq}{\begin{equation}}
\newcommand{\en}{\end{equation}}
\newcommand{\Tr}{\ensuremath{\mathrm{Tr}}}
\newcommand{\SU}{\mathrm{SU}}

\begin{document}

\begin{titlepage}
\begin{center}
{\Large\bf 
    Confining flux tube in the trace deformed (2+1) dimensional $\SU(2)$ gauge theory
}
\end{center}
\vskip1.3cm
\setcounter{footnote}{0}
\renewcommand{\thefootnote}{\fnsymbol{footnote}}
\begin{center}
Claudio Bonati\,\orcidlink{0000-0001-7358-5909}$^{1,}$\footnote{\href{mailto:claudio.bonati@unipi.it}{\texttt{claudio.bonati@unipi.it}}},\quad
Michele~Caselle\,\orcidlink{0000-0001-5488-142X}$^{2,}$\footnote{\href{mailto:michele.caselle@unito.it}{\texttt{michele.caselle@unito.it}}},\quad
Alessio Negro\,\orcidlink{0009-0007-0225-6535}$^{3,4,}$\footnote{\href{mailto:alessio.negro@hiskp.uni-bonn.de}{\texttt{alessio.negro@hiskp.uni-bonn.de}}},\quad
Dario Panfalone\,\orcidlink{0009-0007-6651-7490}$^{2,}$\footnote{\href{mailto:dario.panfalone@unito.it}{\texttt{dario.panfalone@unito.it}}},\quad
and Lorenzo Verzichelli\,\orcidlink{0009-0008-0825-4845}$^{2,}$\footnote{\href{mailto:lorenzo.verzichelli@unito.it}{\texttt{lorenzo.verzichelli@unito.it}}}
\end{center}
\vskip1.5cm

\centerline{$^{1}$ {\sl Department of Physics, University of Pisa and INFN, Pisa}}
\centerline{\sl Largo Pontecorvo 3, I-56127 Pisa, Italy}
\vskip1.0cm

\centerline{$^{2}$ {\sl Department of Physics, University of Torino and INFN, Turin}}
\centerline{\sl Via Pietro Giuria 1, I-10125 Turin, Italy}
\vskip1.0cm

\centerline{$^{3}$ {\sl Helmholtz-Institut f\"ur Strahlen- und Kernphysik, University of Bonn,}}
\centerline{\sl Nussallee 14-16, 53115 Bonn, Germany}
\vskip1.0cm

\centerline{$^{4}$ {\sl Bethe Center for Theoretical Physics, University of Bonn,}}
\centerline{\sl Nussallee 12, 53115 Bonn, Germany}

\vskip1.0cm

\setcounter{footnote}{0}
\renewcommand\thefootnote{\mbox{\arabic{footnote}}}

\begin{abstract}
\noindent
We study the confining flux tube in the reconfined phase of trace deformed $\SU(2)$ Yang-Mills theory in (2+1) dimensions. Using lattice simulations above the standard deconfinement temperature, we analyze Polyakov-loop correlators and extract the ground state energy of the effective string. We show that the usual Nambu-Got\=o effective string description, including its standard higher-order corrections, fails to reproduce the data as the trace deformation is increased. Remarkably, deep in the reconfined regime the results are instead accurately described by the Polchinski-Yang rigid-string solution, corresponding to an effective string dominated by an extrinsic-curvature term. We further investigate the transverse profile of the chromo-electric flux tube and find significant deviations from the standard Yang-Mills behavior, including a substantial modification of the intrinsic width. Finally, we present an exploratory study of the phase diagram, finding evidence for a transition from a continuous to a first order reconfinement line as the deformation parameter increases. These results suggest that the reconfined phase realizes a qualitatively different effective-string regime from ordinary confinement.
\end{abstract}

\end{titlepage}

\tableofcontents

\input{1_introduction}

\input{2_lattsetup.tex}

\input{3_ESTfluxtube}

\input{4_results1}

\input{5_rigidstring.tex}

\input{6_results2}

\input{7_width.tex}

\input{8_conclusion}

\section*{Acknowledgments}

We are delighted to thank 
Claudio Bonanno, 
Aleksey Cherman, 
Margarita Garc\'ia P\'erez, 
Antonio Gonz\'alez-Arroyo, 
Alessandro Mariani, 
Alessandro Nada, 
Johann Ostmeyer, 
Marco Panero, 
Javier Suarez Sucunza, 
Mithat \"Unsal and 
Carsten Urbach 
for useful discussions.

L.\,V. would like to thank Lena Funcke and her group for hospitality in Bonn from May to July 2025.

A.\,N. acknowledges support from the Deutsche Forschungsgemeinschaft (DFG, German Research Foundation) within the CRC~1639 NuMeriQS -- project no.\ 511713970.

C.\,B. acknowledges support from the NPQCD Scientific Initiative of INFN, while M.\,C., D.\,P. and L.\,V. from the SFT Scientific Initiative of the same institution.

The work of M.\,C., D.\,P. and L.\,V. is supported by the Simons Foundation grant 994300 (Simons Collaboration on Confinement and QCD Strings).

The authors gratefully acknowledge the access to the Marvin cluster of the University of Bonn and to the Leonardo cluster at CINECA under the agreement with INFN (project INF25\_sft).

\clearpage

\appendix

\input{appendixA}
\input{appendixB}

\clearpage

\bibliographystyle{JHEP}
\providecommand{\href}[2]{#2}\begingroup\raggedright\endgroup

\end{document}

%% file: 1_introduction.tex
\section{Introduction}
Understanding confinement remains a central unresolved issue in Yang-Mills (YM) theories.
A strategy to address this issue which was proposed long ago is to compactify the theory
on a $\mathbb{R}^3 \times S^1$ manifold, where the compactification radius $N_t$ provides
an adjustable energy scale. The basic expectation is that shrinking this radius can drive
the system into a weak-coupling regime in which perturbative techniques become applicable.
However, in its simplest implementation this reasoning is too naive, because it neglects
the fact that at a critical value $N_{t,c}$ of the compactification radius the theory
experiences a deconfinement transition. Over the last fifty years, numerous proposals have
been explored to remove this transition, or analogous transitions that arise when all
directions are compactified, typically in conjunction with the large-$N$ limit;
see, e.g., Ref.~\cite{Eguchi:1982nm, Bhanot:1982sh, Gross:1982at,
Gonzalez-Arroyo:1982hwr, Gonzalez-Arroyo:1982hyq, Narayanan:2003fc,
Gonzalez-Arroyo:2010omx} for relevant works. More recently, a new proposal has
been put forward to establish an analytic connection between the large-$N_t$
regime, where confinement and other non-perturbative properties are manifest, and
the small-$N_t$ regime, where semiclassical and perturbative methods can be
employed to study the system. Specifically, by adding to the YM action a {\sl
trace deformation} that enforces a vanishing expectation value of the Polyakov
loop, it is possible to stabilize the confined phase even at temperatures above
the conventional deconfinement transition (Ref.~\cite{Unsal:2008ch,
Myers:2007vc}). Recently (Ref.~\cite{Dobozy:2026edc}) a study of the instantons 
in the trace deformaed theory has ben conducted, see also 
Ref.~\cite{Poppitz:2021cxe} for a review on the subject.%
\footnote{The use of adjoint fermions has also been proposed to stabilize center
symmetry (Ref.~\cite{Kovtun:2007py, Myers:2009df}), but numerical simulations
have shown that a spontaneous breaking of center symmetry is still present in
this case, see Ref.~\cite{Cossu:2009sq}.}

The key issue is whether this {\sl reconfined} phase of the trace deformed YM
theory exhibits the same physical features as the usual confining regime.
Several observables seem to be essentially unchanged, including the glueball
spectrum (Ref.~\cite{Athenodorou:2020clr}), the localization/delocalization
transition of the Dirac eigenmodes (Ref.~\cite{Bonati:2020lal}), and the
$\theta$-dependence of the free energy (Ref.~\cite{Bonati:2018rfg, Bonati:2019kmf}).
In addition, phenomena tied to monopole condensation turn out to be largely
equivalent in ordinary YM and in the trace deformed setup
(Ref.~\cite{Bonati:2020lal}). However this is not enough. A clean, unambiguous
answer requires a direct analysis of the confining flux tube in the {\sl
reconfined} phase, and a detailed comparison with its counterpart in the
standard confining regime. This is exactly the goal of the present work. We
tackle this question in the specific setting of the $\SU(2)$ pure gauge theory
in $(2+1)$ dimensions, which represents the simplest lattice gauge theory with a
continuous non-Abelian gauge group and hence provides an ideal testing ground
for long-distance, genuinely non-perturbative aspects of YM dynamics with a
relatively modest numerical cost. A further motivation for this choice is that
the same model has been extensively investigated in the literature, see
Ref.~\cite{Ambjorn:1984me, Teper:1998te, Caselle:2004er, Caselle:2011vk,
Bringoltz:2006zg, Brandt:2010bw,Athenodorou:2016kpd, Brandt:2017yzw,
Brandt:2018fft, Brandt:2021kvt, Bonati:2021vbc,Caristo:2021tbk}, so that the
available results can guide the selection of simulation parameters and
significantly streamline the analysis.

Since the relevant flux-tube observables are conveniently captured by an Effective String
Theory (EST) description, our main goal in this paper will be to study the effective string
model governing the {\sl reconfined} phase and see whether it is the same EST that was recently
shown to reproduce with high precision the flux tube in the ordinary confining regime
(Ref.~\cite{Caristo:2021tbk,Caselle:2024zoh}). With this in mind, we begin by describing our lattice
setup for trace deformed YM in Sec.~\ref{sec:lattsetup}. We then review the EST framework
in Sec.~\ref{sec:ESTfluxtube}, first summarizing its general predictions, then discussing
extensions beyond the Nambu-Got\=o model.
Our numerical results are presented in Sec.~\ref{sec:results}, where we show
that the standard EST description of the confining flux tube fails to reproduce
numerical data when the contribution of the trace deformation is large, while it
is known that it describes very accurately the results of the standard (i.e.,
without trace deformation) $\SU(2)$ model (Ref.~\cite{Caristo:2021tbk, Caselle:2024zoh}).  In
Sec.~\ref{sec:rigidstring} we then propose an alternative description, based on
the so called ``rigid string'' which we then compare with our data in Sec.~\ref{sec:results2}.
In Sec.~\ref{sec:fluxtube} we study the profile and width of the reconfined chromo-electric
flux tube, which display marked differences with respect to that of the pure gauge theory.
We conclude with some final remarks in Sec.~\ref{sec:conclusion}.

The main outcome of our analysis is that in the reconfined phase the flux tube
behaves in a very different way with respect to the ordinary confining regime.
Both its shape and its dependence on the temperature are definitely different
from the ordinary ones. It is interesting to note that this behavior is well
described by a particular (unusual) regime of the rigid string which was studied
years ago by Polchinski and Yang (Ref.~\cite{Polchinski:1992ty}). The argument
that is usually invoked to exclude the presence of a rigidity term in the EST
action is circumvented in this case by the fact that Lorentz invariance is
explicitly broken by the trace deformation. If this picture is correct we may
consider the trace deformed models as the first explicit realization of the
Polchinski-Yang proposal thus opening the way to a detailed study of the many
interesting properties of this model.

Compared to our earlier proceedings~\cite{Bonati:2025hik}, we extend the analysis
in three directions: we explore a wider set of couplings by studying additional values of
$\beta$ and $h$, we perform a dedicated investigation of the flux tube in the {\sl reconfined}
phase, and we provide a systematic study of the corresponding phase diagram.

%% file: 2_lattsetup.tex
\section{Lattice setup}
\label{sec:lattsetup}

We consider the $(2+1)$-dimensional $\SU(2)$ Yang-Mills theory, regularized on a
cubic lattice of spacing $a$ with periodic boundary conditions in the three directions.
The temperature $T$ will be the inverse of the extension of the lattice in the Euclidean
time direction: $1 / T = L_t =  a \, N_t$. In the two spatial directions, the size
of the lattice $L_s = a \, N_s$ will be chosen large enough to neglect finite volume
effects.

We add the trace deformation term to the standard Wilson action in analogy with
Ref.~\cite{Bonati:2018rfg,Bonati:2019kmf,Bonati:2020lal,Athenodorou:2020clr}:
\begin{equation}
    S^{\mathrm{def}} = S_{\rm W} + h \sum_{\vec{x}} |P (\vec{x})| ^2\ ,
\end{equation}
where
\begin{equation}\label{eq:S_W}
    S_{\rm W} = -\frac{\beta}{2} \sum_{x} \sum_{\mu < \nu} \Tr \ U_{\mu\nu} (x).
\end{equation}

In the expressions above, $U_{\mu\nu} (x)$ denotes the plaquette in the $\mu$-$\nu$
plane starting in $x$:
\begin{equation}
    U_{\mu\nu} (x) = U_\mu(x) \, U_\nu(x+\hat{\mu}) \,
                          {U_\mu}^\dagger(x+\hat{\nu}) \, {U_\nu}^\dagger(x)
\end{equation}
and $P(\vec{x})$ the Polyakov loop at spatial position $\vec{x}$:
\begin{equation}
    P(\vec{x}) = \Tr \left [ \prod_{t = 1}^{N_t} U_0(t, \vec{x}) \right ].
\end{equation}

Link variables in space directions ($U_\mu$, $\mu = 1, 2$) are updated with a
standard overrelaxed algorithm, combining heat-bath and microcanonical updates
(Ref.~\cite{Kennedy:1985nu, Creutz:1987xi}), while those in the time direction
($U_0(t,\vec{x})$) are updated with five ``hits'' of a Metropolis algorithm
(Ref.~\cite{Metropolis:1953am}) in each update sweep, similarly to
Ref.~\cite{Bonati:2018rfg}. At each hit, the link is multiplied by a random
$\SU(2)$ matrix $M$ chosen to be close to the identity (with $M$ and
$M^\dagger$ being equiprobable, in order to satisfy detailed balance), with the
maximum deviation of $M$ from the identity matrix fixed in such a way that
$\frac{1}{2}\Tr M \gtrsim 0.75$.

The standard theory, obtained setting $h = 0$, presents a deconfinement phase transition
at a critical temperature $T_c$, which has been accurately determined as a function
of $\beta$ in Ref.~\cite{Edwards:2009qw}. The Polyakov loop is an order
parameter for this phase transition, vanishing for $T < T_c$ (the confined
phase) and acquiring, in the thermodynamic limit, a non-zero expectation value
for $T > T_c$ (the deconfined phase).

It is easy to see that, for $h > 0$, configurations
with non-vanishing value of the Polyakov loop will be suppressed, which allows confinement
to be preserved at temperatures much higher than $T_c$. This regime ($T > T_c$, but $h$ large
enough to ensure $\braket{P} = 0$) is what is usually denoted as the \textit{reconfined} phase
of the model, and will constitute the main focus of the present study. Our analysis suggests that,
for each value of the temperature, the confined and reconfined phases are separated from the
deconfined one by a phase transition  which is continuous for small values of $h$ and becomes a
first order phase transition for $h$ large enough (see the discussion in Appendix~\ref{sec:appendixA}).

We will be particularly interested in comparing the properties of the flux tube in the confined
and reconfined regime. The simplest lattice observable related to the flux
tube is the correlator of two Polyakov loops:
\begin{equation}
    G(R) = \frac{1}{2 \, N_s^{2}} \left\langle\sum_{k = 1, 2} \sum_{\vec{x}}
                                      P\left(\vec{x}\right) \, P\left(\vec{x} + R \, \hat k\right)\right\rangle,
\end{equation}
\noindent
which is related to the free energy of a pair of static color charges separated
by a distance $R$ (Ref.~\cite{Polyakov:1978vu, McLerran:1981pb}).
We computed this observable for different values of $h$ at different
temperatures (changing $N_t$), for two different values of $\beta$ (or,
equivalently, of the lattice spacing). These values were chosen so as to have a
fixed reference value for the critical temperature of the standard $\SU(2)$
model ($N_{t,c}=15$ and $N_{t,c}=20$ respectively) obtained by extrapolating
the values reported in~\cite{Edwards:2009qw}. These values of $N_{t,c}$ are
larger than those explored in previous studies
\cite{Caristo:2021tbk,Athenodorou:2016kpd} to allow a wider exploration of the
reconfined phase of the model. These values are summarized in
Tab.~\ref{tab:info_simulations}. In the same table we also report as reference
the critical value of $N_{t,c}$ (corresponding to the critical temperature in
the ordinary undeformed $\SU(2)$ model) for both values of $\beta$.  As $h$
increases we expect the deconfinement transition to occur at lower values of
$N_t$ and for this reason we systematically studied smaller values of $N_t$. We
verified a posteriori that all the values reported in the table were within the
confined (for $h=0$) or reconfined (for $h\neq 0$) phase of the model.

\begin{table}[ht]
    \centering
    \begin{tabular}{|c|c|c|c|c|}
        \hline
        $\beta$ & $N_{t,c}$ & $h$   & $N_t$     & $N_s$         \\\hline
        \multirow{8}{*}{23.3805} & \multirow{8}{*}{15} &
                              0.000 & $[16,21]$ & \multirow{10}{*}{96} \\\cline{3-4}
                &           & 0.001 & $[15,21]$ &               \\\cline{3-4}
                &           & 0.002 & $[13,21]$ &               \\\cline{3-4}
                &           & 0.003 & $[12,21]$ &               \\\cline{3-4}
                &           & 0.004 & $[10,21]$ &               \\\cline{3-4}
                &           & 0.005 & $[ 8,14]$ &               \\\cline{3-4}
                &           & 0.006 & $[ 7,14]$ &               \\\cline{3-4}
                &           & 0.007 & $[ 7,14]$ &               \\\cline{1-4}
        \multirow{2}{*}{27.4745} & \multirow{2}{*}{20} &
                              0.004 & $[11,19]$ &               \\\cline{3-4}
                &           & 0.005 & $[ 9,17]$ &               \\\hline
    \end{tabular}
    \caption{Summary of our simulation setup. We explored all the integer values of $N_t$ in the indicated
             range (extrema included). For $\beta=23.3805$ and all the combinations of $h$ and $N_t$ we measured all the values of the correlator $G(R)$ in the range $0\leq R\leq 23$, while for $\beta=27.4745$ we measured $G(R)$ in the range $0\leq R\leq 47 $.}
    \label{tab:info_simulations}
\end{table}

%% file: 3_ESTfluxtube.tex
\section{The effective string description of the flux tube}
\label{sec:ESTfluxtube}

In this section, we recall only those Effective String Theory (EST) results that
will be needed in the following sections of the paper. For a more complete
treatment of the subject, see, e.g.,
Ref.~\cite{Aharony:2013ipa,Brandt:2016xsp,Caselle:2021eir}.

Within the EST picture, the flux tube connecting a quark-antiquark pair is
modelled as a thin, fluctuating string, and the Polyakov-loop correlator is
directly related to the free energy of the corresponding string configuration.
This description is expected to hold only for separations between the sources
that are larger than a critical separation $R_c$, and therefore provides an
effective ``low energy'' characterization of confinement. Nevertheless, within
its domain of applicability it reproduces the Polyakov-loop correlator
remarkably well, with an almost perfect agreement between EST predictions and
lattice gauge theory data (see the
reviews~\cite{Aharony:2013ipa,Brandt:2016xsp,Caselle:2021eir}).

The EST description further simplifies in the regime $R \gg N_t$, i.e.\ in the high temperature limit (still below the deconfinement transition, and hence within the confining phase). In this case the {\sl boundary terms} (associated with the quark self-energy) can be neglected~\cite{Caselle:2021eir}, and the Polyakov-loop correlator takes a universal form~\cite{Luscher:2004ib} which, in $(2+1)$ dimensions, reads
\begin{equation}\label{eq:G_R}
    G(R) = \sum_{n=0}^\infty |v_n(N_t)|^2 \, \frac{E_n}{\pi}\, K_{0}(E_n R)\,,
\end{equation}
where $K_0$ is the order zero modified Bessel function of the second kind, $E_n$ is the energy of the $n$-th excited string state, and $v_n(N_t)$ is its amplitude (which also encodes the multiplicity of the $n$-th level). At large $R$ the series is dominated by the ground state, so that the correlator can be approximated as
\begin{equation}\label{eq:G_R_K0}
    G(R) = A(N_t)\; K_{0} \big(E_0(N_t)\, R\big)\,,
\end{equation}
which allows one to determine the ground state energy $E_0(N_t)$ and the pre-factor $A(N_t)$ through a fit.

Eq.~\eqref{eq:G_R_K0} is universal and therefore holds for any EST. To identify which specific effective string theory accounts for our lattice data, one must instead study the $N_t$ dependence of the ground state energy $E_0(N_t)$ and of the amplitude $A(N_t)$. In what follows we will adopt this strategy to determine the EST governing the reconfined phase of the trace deformed model.

In ordinary Yang-Mills, the simplest possible EST which satisfies the constraint
imposed by Lorentz invariance is given by the well
known Nambu-Got\=o action~\cite{Nambu:1974zg,Goto:1971ce} which in three dimensions is defined as
\begin{align}
\label{eq:NGaction}
S_\tmb{NG} = \sigma \int_\Sigma d^2\xi \sqrt{g},
\end{align}
where $g\equiv \det g_{\alpha\beta}$ and
\begin{align}
\label{eq:NGaction2}
g_{\alpha\beta}=\partial_\alpha X_\mu~\partial_\beta X^\mu
\end{align}
is the metric induced on the world-sheet surface $\Sigma$ by the mapping
$X_\mu(\xi)$, where $\xi\equiv(\xi^0,\xi^1)$ are the world-sheet coordinates,
$X_\mu$ (with $\mu=0,1,2$) are the string coordinates in the three dimensional
target space and $\sigma$ is the string tension. This action is explicitly
reparametrization invariant. In the EST approach this invariance is fixed using
the so-called ``physical gauge'', which identifies the first two degrees of
freedom of the string as the world-sheet coordinates $\xi^0=X^0$,
$\xi^1=X^1$. In this gauge the only remaining degree of freedom is the
transverse displacement $X^2$, which we shall denote in the following simply as
$X(\xi^0,\xi^1)$, which is assumed to be a single-valued function of
$(\xi^0,\xi^1)$. As it is well known this gauge fixing is anomalous, but it can
be shown that the anomaly is irrelevant in the
large $R$ limit~\cite{Aharony:2013ipa, Brandt:2016xsp} in which the
Nambu-Got\=o action can be used as an effective description of the interquark
potential.
In the physical gauge the Nambu-Got\=o action can be rewritten as:
\begin{align}
\label{eq:NGaction3}
S_\tmb{NG}[X] = \sigma \int_0^L d\xi^0 \int_0^R d\xi^1 \sqrt{1+(\partial_{\xi^0}X)^2+(\partial_{\xi^1}X)^2},
\end{align}
and despite its apparent complexity, it can be integrated exactly
(Ref.~\cite{Arvis:1983fp}), leading to the following
expression for $E_0(N_t)$:
\begin{equation}\label{eq:E0_NG}
    E_0(N_t)=\sigma N_t \sqrt{1-\frac{\pi}{3\sigma N_t^2}}\ .
\end{equation}
This model has only one free parameter, i.e., the string tension $\sigma$.
Once this parameter is fixed, for example by a large distance fit at zero temperature, the theory becomes highly predictive.

It is however well known that the NG string cannot be the end of the story,
since it gives the same answer for all LGTs independently from their gauge
group; moreover, it fails to predict the correct critical exponent of the
deconfinement phase transition.

The NG action should be considered only as the first term of a large distance
(or equivalently low energy) expansion of the real EST. The most general form
of this expansion is of this type:
\begin{equation}\label{eq:S_BNG}
    S=\sigma RN_t+\frac{\sigma}{2}\int \mathrm{d}^2\xi \left[\partial_\alpha X_i \partial_\alpha X^i+c_2(\partial_\alpha X_i \partial_\alpha X_i)^2+c_3(\partial_\alpha X_i \partial_\beta X^i)^2+\dots\right],
\end{equation}
however it can be shown (Ref.~\cite{Aharony:2013ipa,Dubovsky:2012sh}) that
the $c_i$ coefficients are strongly constrained by the Lorentz invariance of
the string in the target space, and coincide with those of the NG action (in
three dimensional models) up to terms proportional to $1/N_t^7$. This result is
known as the ``low energy universality'' of the EST
(Ref.~\cite{Aharony:2013ipa,Dubovsky:2012sh}) and explains why the NG action,
even if it is not the exact EST, approximates so well the Polyakov loop
correlator data.

Taking into account these constraints, the EST prediction for the ground state energy at the first non-trivial order --- which in the following we refer to as the {\sl Beyond Nambu-Got\=o} (BNG) expression --- has the following form
\begin{equation}\label{eq:E0_BNG_Nt7}
    E_0(N_t)=\sigma N_t-\frac{\pi}{6N_t}-\frac{\pi^2}{72\sigma N_t^3}-\frac{\pi^3}{432\sigma^2N_t^5}-\frac{5\pi^4}{10368\sigma^3N_t^7}+\frac{k_4}{\sigma^3N_t^7}\ ,
\end{equation}
where the first terms are simply the expansion of Eq.~\eqref{eq:E0_NG} to the
$1/N_t^7$ order and $k_4$ is a new non-universal parameter, which depends on
the gauge group and, in principle, on any other relevant feature of the model
(for instance, on $h$ in the trace deformed model).

\begin{figure}[h]
    \begin{center}
        \includegraphics[width=0.70\textwidth]{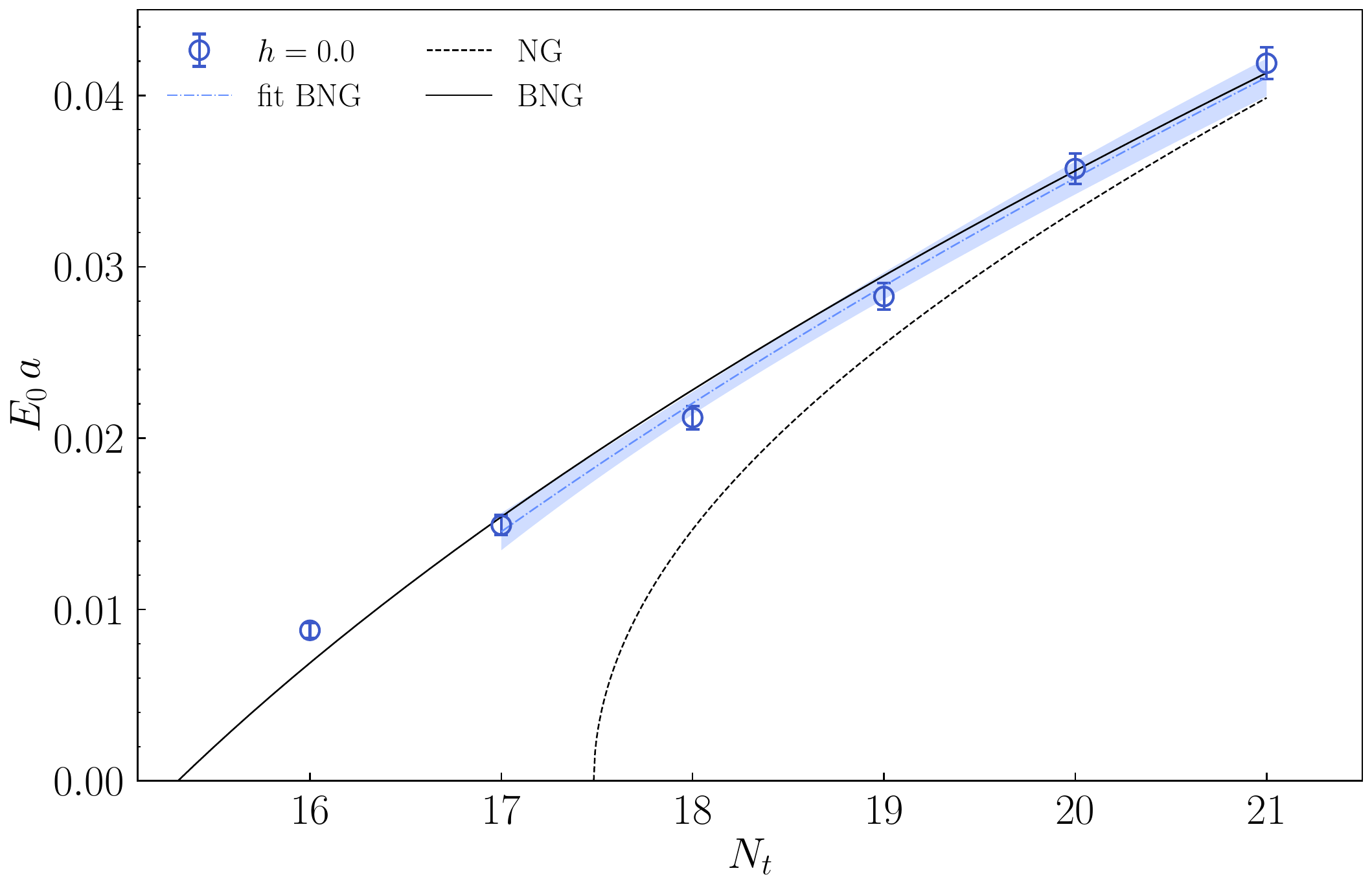}
    \end{center}
    \caption{Ground state at $h = 0$ and their fit according to Eq.~\eqref{eq:E0_BNG_Nt7} (dash-dotted blue line, with confidence band). 
             The black solid line is not a fit to the data, but is obtained assuming the corrections to NG numerically determined in \cite{Caselle:2024zoh}.
             The formula reported in Eq.~\eqref{eq:E0_BNG_Nt7} is able to fit the small $N_t$ (high temperature)
             behavior of $E_0(N_t)$ remarkably well.}
    \label{fig:E0_T_BNG}
\end{figure}

The agreement of this expression with the results of simulations of the
ordinary $\SU(2)$ model is impressive. We report an example of this agreement in
Fig.~\ref{fig:E0_T_BNG} (from Ref.~\cite{Caristo:2021tbk}, to which we refer
for further details). Here blue open circles are the results of simulations
in the ordinary $\SU(2)$ model, while the black dashed line is the Nambu-Got\=o
prediction, which, in agreement with the low energy universality is able to fit the data only for large values of $N_t$. The deviations are then perfectly fitted by Eq.~\eqref{eq:E0_BNG_Nt7} which allows one to fix the value of the new parameter
$k_4$ with good precision\footnote{The analysis can be pursued to even higher
orders and also the next to leading coefficient $k_5$ can be extracted from the
simulations, (see \cite{Caselle:2024zoh} for further details). In the present
case the precision of our data is not enough to evaluate this higher order
term.} (black continuous line).

Remarkably enough the $k_4$ parameter is not completely free. It can be shown
using a bootstrap analysis (see Ref.~\cite{EliasMiro:2019kyf,EliasMiro:2021nul})
that $k_4$ must satisfy the following bound:
\begin{equation}
    k_4 \le \frac{32 \, \pi^6}{225 \cdot 3 \cdot 2^8} = 0.178 \dots\ .
\end{equation}
For the ordinary $\SU(2)$ model in (2+1) dimensions one finds $k_4\sim 0.04$
which, as expected, satisfies this bound (Ref.~\cite{Caselle:2024zoh}). This
constraint will play a major role in the following.

%% file: 4_results1.tex
\section{Analysis of the Polyakov loop correlator in the reconfined phase}\label{sec:results}

\subsection{The phase diagram of the model} \label{subsec:phase}

\begin{figure}
    \centering
    \includegraphics[width=0.7\textwidth]{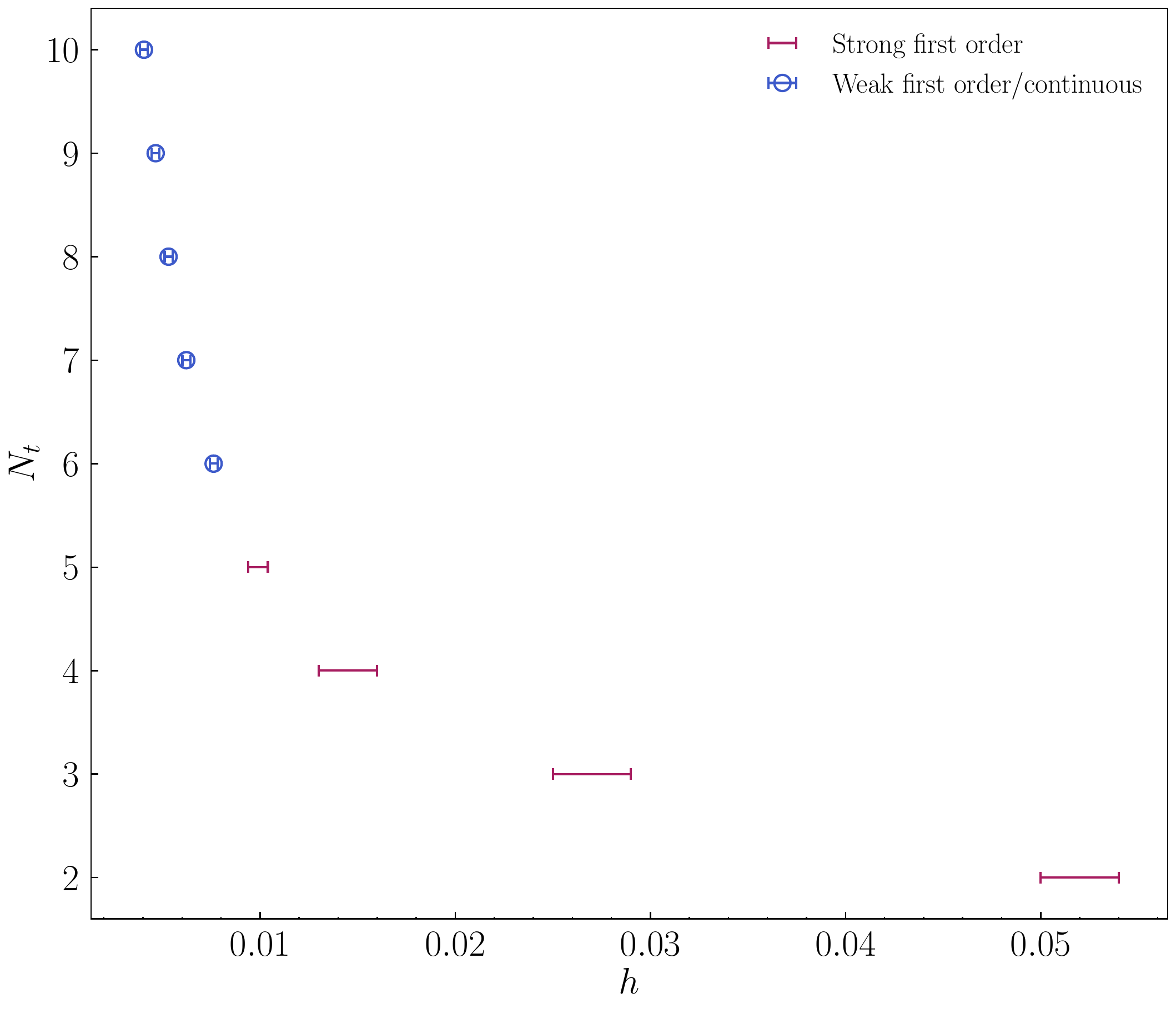}
    \caption{Sketch of the phase diagram, obtained with the data in Tab.~\ref{tab:hcrit}.
             The hysteresis region (when observed) is marked by the large red bars.}
    \label{fig:hcrit}
\end{figure}

As a preliminary step we performed a set of simulations to identify the
reconfined region in the $(h,N_t)$ plane for $\beta=23.3805$. For each value of
$N_t$ in the range $N_t\in[2,10]$ we evaluated the Polyakov loop expectation
value as a function of $h$ looking for the value $h_c(N_t)$ at which $\langle P
\rangle$ vanishes (note that the spatial volume is large enough for tunneling
events between different center sectors to be strongly suppressed when center
symmetry is spontaneously broken). At small $N_t$, strong hysteresis makes the
value of $h$ at which the center symmetry appears to be restored depend on the
initial configuration (ordered or disordered). For larger values of $N_t$
($N_t \ge 6$), it is possible to identify a peak in the Polyakov loop
susceptibility and extract a pseudo-critical value of $h$ from it. From these
data we obtained approximate (since an infinite volume extrapolation is not
performed) estimates of the values $N_{t,c}(h)$ which correspond to the inverse
of the deconfinement temperature as a function of
$h:\hskip 0.3cm T_c(h)=1/N_{t,c}(h)$. Results are reported in
Tab.~\ref{tab:hcrit} and shown in Fig.~\ref{fig:hcrit}.

\begin{table}
    \centering
    \begin{tabular}{|c|c|}
                              \hline
        $N_t$ & $h_{pc}$   \\ \hline
         2 & 0.0500-0.0540 \\ \hline
         3 & 0.0250-0.0290 \\ \hline
         4 & 0.0130-0.0160 \\ \hline
         5 & 0.0094-0.0104 \\ \hline
         6 & 0.00762(20)   \\ \hline
         7 & 0.00622(20)   \\ \hline
         8 & 0.00531(20)   \\ \hline
         9 & 0.00465(20)   \\ \hline
        10 & 0.00405(20)   \\ \hline
    \end{tabular}
    \caption{Pseudocritical values of $h$ for $N_s = 96$ and $N_t \in [2, 10]$, for $\beta = 23.3805$.
             For $N_t \ge 6$  the phase transition is continuous or weakly discontinuous
             (we will show some evidence of a first order phase transition at
             $N_t = 6$, in the Appendix~\ref{sec:appendixA}),
             while for $N_t \le 5$ hysteresis is observed. In the former case we
             report the position of the peak of the Polyakov loop susceptibility,
             together with an uncertainty associated with its determination.
             In the case of strong first order transition, instead, we indicate the
             extrema of the region where hysteresis is observed in MC histories of
             $\mathcal{O}(10^6)$ updates.}
    \label{tab:hcrit}
\end{table}

Then, for the values of $h$ reported in Tab.~\ref{tab:info_simulations} we selected
a set of values of $N_t$ in the reconfined region, just above $N_{t,c}(h)$  (see
Tab.~\ref{tab:info_simulations}) and for each combination of $h, N_t$
we evaluated the Polyakov loop correlators for a wide range of interquark distances:
$0\leq R \leq 23$. We then performed a similar analysis at $\beta=27.4745$ for the
values of $h$ and $N_t$ reported in Tab.~\ref{tab:info_simulations}.
We added this second value of $\beta$ to test the scaling properties of our results.
For this second value of $\beta$  we studied a larger range of values of
$R$: $0\leq R \leq 47$.

\subsection{Extracting the ground state energy of the string}
We extracted the ground state energy of the string by fitting the $R$ dependence of the correlator with the EST
prediction of Eq.~\eqref{eq:G_R_K0}, modified so as to take into account the
periodic boundary conditions:
\begin{equation}\label{eq:fS}
    G(R)=A\left(K_0(E_0 R)+K_0(E_0(N_s-R))\right)\ .
\end{equation}

This formula neglects the contribution given by string excited states. For this
reason, it is valid only for values of $R$ large enough to suppress the
contamination due to excited states.
Thus, when fitting our data, we only included correlators with $R\!>\!R_{\min}$,
and verified a posteriori that the result does not depend on the choice of
$R_{\min}$ within error bars. For the data obtained using $\beta=23.3805$ we
found that $R_{\min} = 15$ is sufficient for all values of $N_t$, whereas for
$\beta=27.4745$, we need to use the larger value $R_{\min}=20$.

Data extracted from single MCMC streams exhibit strong cross-correlations among
different separations $R$. We account for this at the first stage of the analysis with a
blocked bootstrap procedure; the details and the associated caveats are discussed in
Appendix~\ref{sec:appendixB}. The fit results are listed in
Tabs.~\ref{tab:E0_beta23_h000}--\ref{tab:E0_beta27_h005}, and the parameter of primary
interest is the ground state energy $E_0$. As discussed in Sec.~\ref{sec:ESTfluxtube}, its
dependence on the temporal extent of the lattice will be our probe to discriminate between
different ESTs.
\subsection{Fitting \texorpdfstring{$E_0(N_t)$}{E0(Nt)} with Eq.~\eqref{eq:E0_BNG_Nt7}}

As a first step we tried to fit our data with Eq.~\eqref{eq:E0_BNG_Nt7}.
Fitting our data at $h = 0$ and $\beta = 23.3805$ for $N_t \ge 17$ we found,
as expected, a good agreement between the model and numerical results (with a
reduced $\chi^2$ of order unity). In particular, we estimate
$k_4 = 0.039(18)$, which nicely agrees with the value $k_4= 0.050(8)$ that was
found in~\cite{Caristo:2021tbk,Caselle:2024zoh}. However, when repeating the
same analysis with $h > 0$ a very different phenomenology was observed.

For $h$ in the range $[0.001, 0.004]$ it was still possible to perform fits with
acceptable $\chi^2$ values, however the value of $k_4$ was
rapidly growing with $h$. In Tab.~\ref{tab:fitsbng} we present the results of the fits at
each value of $h$, including all the values of $N_t \ge {N_t}^{\min}$, where
${N_t}^{\min}$ has been chosen to obtain an acceptable $\chi^2$ value, i.e.,
smaller than twice the number of degrees of freedom in the fit. As mentioned in
Sec.~\ref{sec:ESTfluxtube}, positive values of $k_4$ are constrained by the
bootstrap analysis of \cite{EliasMiro:2019kyf,EliasMiro:2021nul} to satisfy
$k_4\lesssim 0.178$.
The value we obtained for $h = 0.004$ is already multiple standard deviations beyond the bound.
Furthermore, for still larger values of $h$ it was impossible to model
numerical data using Eq.~\eqref{eq:E0_BNG_Nt7}, even allowing for very large
values of $k_4$. This suggests that, at least for $h\geq 0.004$ the flux tube
in the reconfined phase cannot be described by an ordinary EST.

\begin{table}[h]
    \centering
    \begin{tabular}{|c|c|c|c|c|}
    \hline
    $h$ & $[N_t^{\min},N_t^{\max}]$ & $\sigma$ & $k_4$ & $\chi^2/\text{dof}$ \\ \hline
    0.000 & \multirow{4}{*}{$[17,21]$} & 0.003426(34) & 0.039(18) & 1.46 \\ \cline{1-1}\cline{3-5}
    0.001 & & 0.003631(37) & 0.098(21) & 0.35 \\ \cline{1-1}\cline{3-5}
    0.002 & & 0.003843(40) & 0.185(27) & 0.92 \\ \cline{1-1}\cline{3-5}
    0.003 & & 0.003941(42) & 0.360(27) & 0.79 \\ \hline
    0.004 & $[17,20]$ & 0.004115(60) & 0.527(34) & 0.14 \\ \hline
    \end{tabular}
    \caption{Parameters of the fit of Eq.~\eqref{eq:E0_BNG_Nt7} to data generated at $\beta=23.3805$.}
    \label{tab:fitsbng}
\end{table}

\subsection{Testing the role of the bulk degrees of freedom of the model}
\label{subsec:u1compare}
To better understand the origin of this behavior it might be useful to compare
our situation with another model for which strong deviations from the expected
EST behavior are observed~\cite{Caselle:2014eka}: the 3d $U(1)$ model, in which
these corrections are understood as due to the interaction of the flux tube
degrees of freedom with the bulk degrees of freedom of the
theory~\cite{Aharony:2024ctf}.

In fact, an underlying assumption in the EST discussed in the previous section
is the possibility of neglecting the interactions of the flux tube with the bulk
degrees of freedom of the model (in particular with the lightest glueball).
This assumption is expected to be correct in the case of ordinary non-Abelian
LGTs due to the large value of the ratio $m/\sqrt{\sigma}$, where $m$ is the
mass of the lightest bulk degree of freedom. This is confirmed not only by the
good agreement between EST predictions and simulations, but also by the fact
that (at least for $SU(N)$ models in three dimensions) the lattice data suggest
a very smooth dependence of EST parameters as a function of $N$, with a large
$N$ limit almost reached already for $N=3$ \cite{Teper:1998te}, and it is known that in the large
$N$ limit the glueballs decouple from the flux tube degrees of freedom
\cite{Teper:2009uf, Aharony:2009gg}. However there are situations in which neglecting the
interaction with the bulk degrees of freedom is not justified. This is the case
for instance of the 3d $U(1)$ gauge model, in which large deviations from the
expected EST behavior are observed.
In this model, thanks to the exact solution
of~\cite{Polyakov:1976fu,Gopfert:1981er}, the ratio $m/\sqrt{\sigma}$ can be
evaluated exactly and goes to zero as the continuum limit is approached (i.e.
as $\beta$ is increased). When $m\sim \sqrt{\sigma}$ the effect of the
interactions of the flux tube with the bulk degrees of freedom becomes
important and strongly affects the value of the ground state energy
$E_0(N_t)$~\cite{Caselle:2014eka}. In this case, thanks to the exact knowledge
of the confining mechanism, the effect of this interaction can be evaluated
perturbatively~\cite{Aharony:2024ctf} and is in good agreement with the results
of numerical simulations~\cite{Caselle:2016mqu}. From the perturbative
calculation of ~\cite{Aharony:2024ctf} we see that the main effect on the
ground state energy is a change of coefficient of the so-called ``L\"uscher
term'' (i.e. the $1/N_t$ correction), whose value decreases as the ratio
$m/\sqrt{\sigma}$ decreases and reaches zero for $m=0$.

To see if a similar mechanism is at work in our case  we tried to fit the data with a function of this type:
\begin{equation}\label{eq:E0_L}
    E_0(N_t)=\sigma_\mathrm{eff} N_t-c\frac{\pi}{6N_t}
\end{equation}
with $\sigma$ and $c$ as free parameters. Results of this analysis are reported in Tab.~\ref{tab:newfit}.
The fit works reasonably well for large values of $N_t$, and indeed the value
of the L\"uscher term decreases as $h$ increases, however it becomes {\sl
negative} for $h>0.004$, and becomes more and more negative as $h$ increases.
An explanation for this behavior comes from the fact that the L\"uscher term is likely not the end of the story. If the true form $E_0(N_t)$ were to also contain an apparently harmless positive constant term, this would be reabsorbed into the L\"uscher coefficient, making it negative\footnote{In hindsight, the effective string model we are going to introduce in Sec.~\ref{sec:rigidstring} reproduces, when we expand $E_0(N_t)$ around $N_t=\infty$, a correction to the L\"uscher term that matches our data.}. However, in our case the confining mechanism is not known and we cannot exclude that a
calculation similar to the one discussed in~\cite{Aharony:2024ctf} could lead
to negative values of $c$, but it seems rather unlikely and for this reason we
tried to explore also other possible explanations.

\begin{figure}
    \begin{center}
        \includegraphics[width=0.70\textwidth]{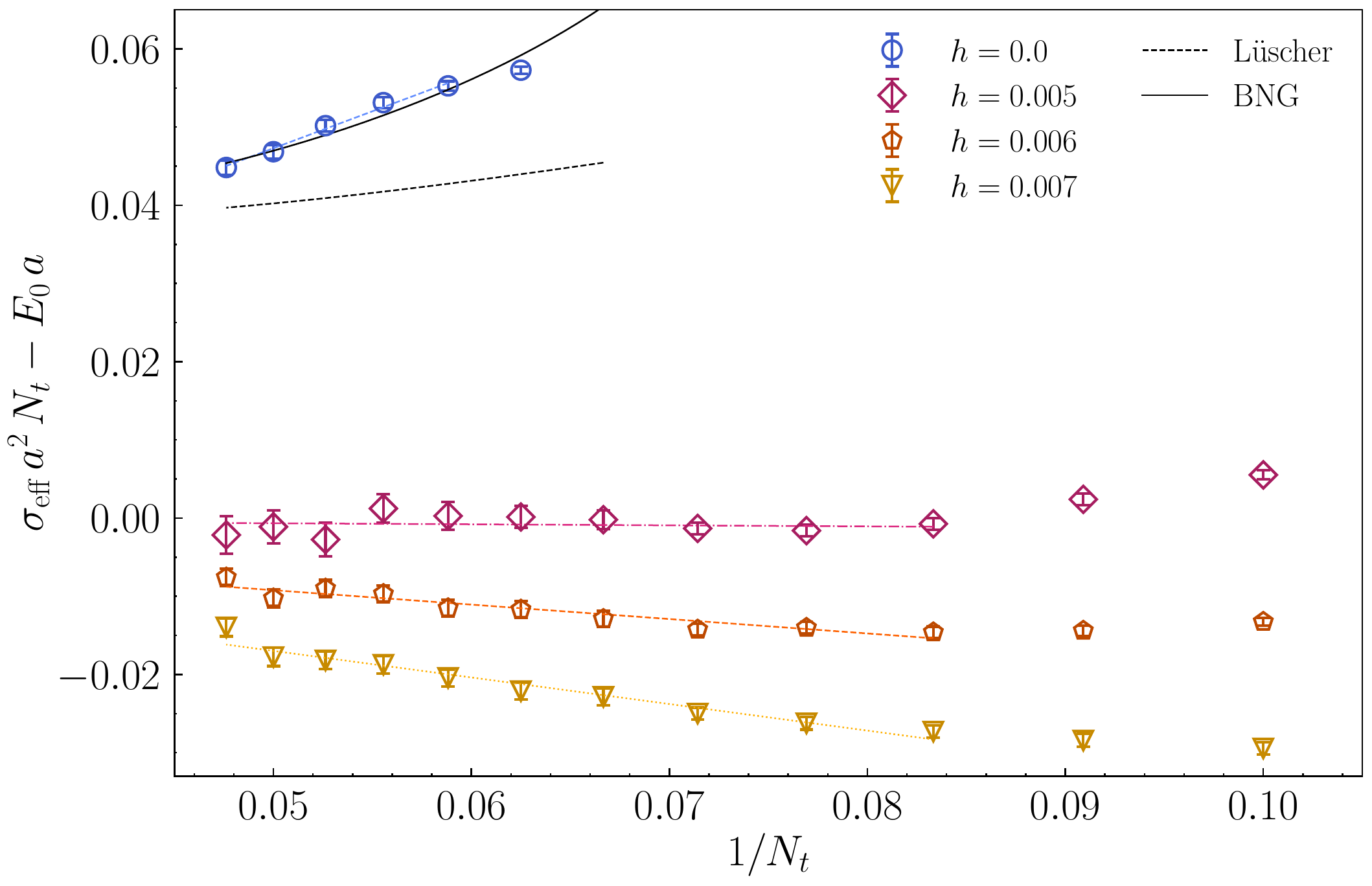}
    \end{center}
    \caption{We plot our numerical determinations of $E_0$ for $h \ge 0.005$
    (and $h = 0$ for comparison) highlighting the fitted term proportional to $1 /
    N_t$ in Eq.~\eqref{eq:E0_L}. With this aim, we subtract the ground state from
    the term $\sigma_\mathrm{eff} N_t$, so that the fit model becomes a line with
    slope $c$. Notice the unusual negative sign of the $c$
    correction for large values of $h$. For $h = 0$, we see how the fit according
    to Eq.~\eqref{eq:E0_L} (blue dashed line) mimics the known result of NG plus
    the first two corrections (black solid line), although higher corrections than
    $1 / N_t$ are dominant at the examined temperatures. Their importance is
    evident from the distance between the data and the black dashed line that only
    takes into account the correct L\"uscher term (with $c = 1$) and the known
    string tension. It is evident from the figure that, for the truncated formula
    (Eq.~\eqref{eq:E0_L}) to provide a good fit, both parameters ($\sigma_\mathrm{eff}$ 
    and $c$) must deviate from their true values ($\sigma \ne \sigma_\mathrm{eff}$ 
    as it is clear comparing Tab.~\ref{tab:newfit} with Tab.~\ref{tab:fitsbng}).}
    \label{fig:E0_T_LUSCH}
\end{figure}

\begin{table}[h]
    \centering
    \begin{tabular}{|c|c|c|c|c|}
    \hline
    $h$ & $[N_t^{\min},N_t^{\max}]$ & $\sigma_\mathrm{eff}$ & $c$ & $\chi^2/\text{dof}$ \\ \hline
    0.000 &  $[17,21]$ & 0.00413(12) & 1.807(78) & 0.57 \\ \hline
    0.001 &  $[15,21]$ & 0.003864(68) & 1.408(37) & 0.80 \\  \hline
    0.002 &  $[13,21]$ & 0.003718(44) & 1.076(18) & 0.83 \\  \hline
    0.003 &  $[12,21]$ & 0.003617(42) & 0.800(17) & 0.67 \\ \hline
    0.004 &  $[12,21]$ & 0.003401(50) & 0.437(22) & 0.75 \\ \hline
    0.005 &  $[12,21]$ & 0.002948(86) & -0.263(33) & 0.58 \\ \hline
    0.006 &  $[12,21]$ & 0.002784(53) & -0.354(24) & 0.81 \\ \hline
    0.007 &  $[12,21]$ & 0.002629(55) & -0.651(25) & 0.99 \\ \hline
    \end{tabular}
    \caption{Results of the fit of the data generated at $\beta=23.3805$  with Eq.~\eqref{eq:E0_L}.}
    \label{tab:newfit}
\end{table}
Before discussing these alternative explanations, let us add a few comments on
the results of these fits. First, it is interesting to observe that for $h=0$
with this fitting function we find completely wrong values for $\sigma$ and
$c$. This is due to the fact that we are neglecting higher order corrections in
the EST which are reabsorbed by a larger value of $c$ and by an increase in
$\sigma$ (see Fig.~\ref{fig:E0_T_LUSCH} for a detailed discussion of this
shift). The effect is particularly large in our case because we are near the
continuum limit, with a very small string tension, a regime in which EST
corrections are particularly important. This is a clear warning: when trying to
perform precision tests for very small values of the lattice spacing it is
mandatory to properly take into account EST corrections otherwise the
determination of the physically relevant parameters, like the string tension,
can be biased by a large amount (even if apparently the $\chi^2$ values of the
fits are good!).

\begin{figure}[h]
    \begin{center}
        \includegraphics[width=0.70\textwidth]{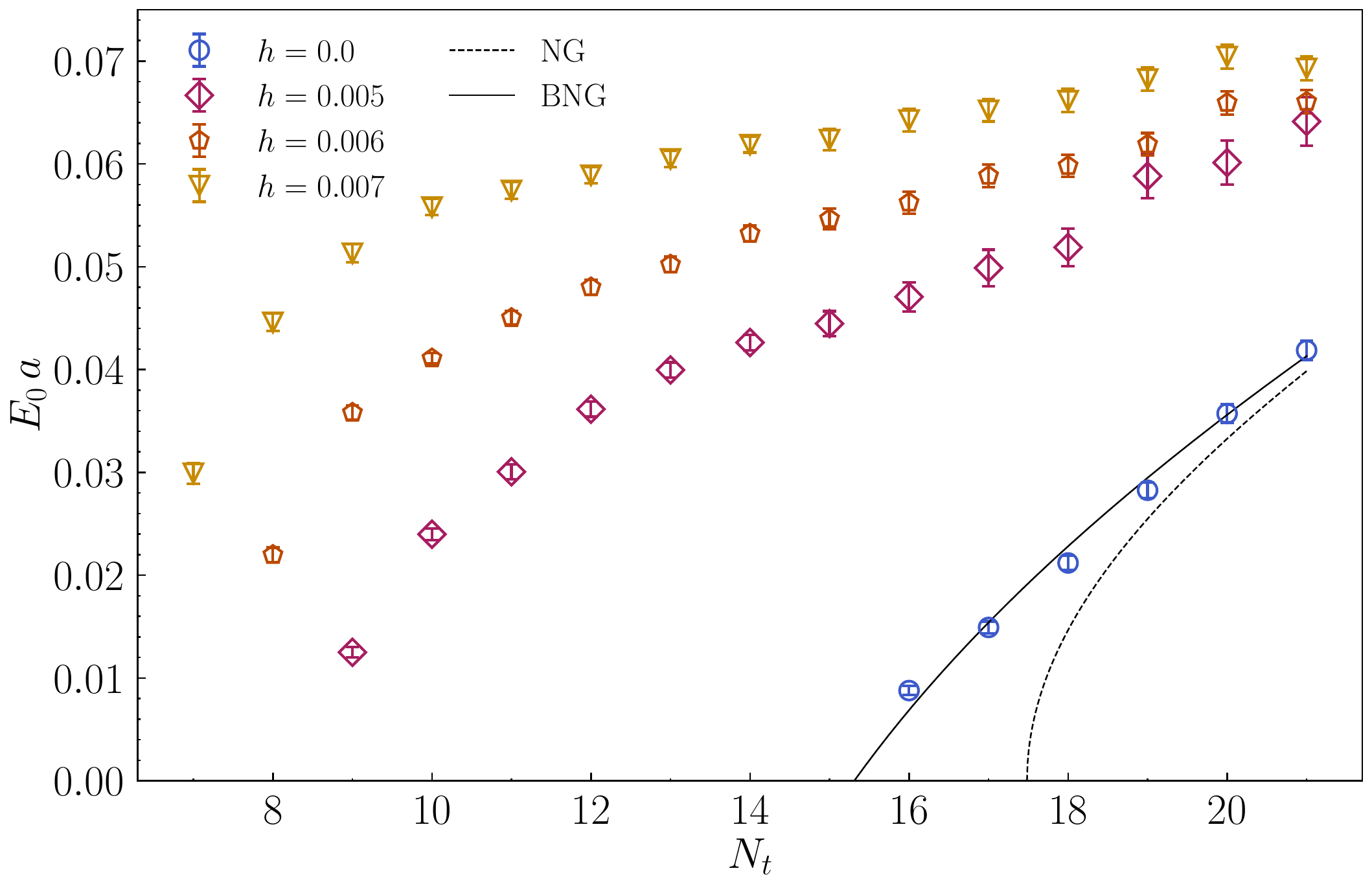}
    \end{center}
    \caption{Ground state at large $h$ and $h = 0$ for comparison. The black solid line is not a fit to the $h = 0$ data, 
             but is obtained assuming the corrections to NG numerically determined in \cite{Caselle:2024zoh}.}
    \label{fig:E0_T_BNG_TD}
\end{figure}

Furthermore, as $h$ increases the minimum value of $N_t$ that can be used in the
fit obtaining reasonable $\chi^2$ values (denoted by $N_t^{\min}$) decreases,
consistently with the fact we have confinement for larger values of the
temperature, but for $h\geq 0.003$ it stabilizes at $N_t^{\min}=12$. For smaller
values of $N_t$ we still have confinement (compare with Tab.~\ref{tab:hcrit})
but, clearly, with a completely different EST behavior. This is clearly visible
from the data reported in Fig.~\ref{fig:E0_T_BNG_TD} where we compare the $h=0$
data with those obtained at $h>0.004$.

Finally, in the 3d $U(1)$ model a strong decrease of the ratio
$m/\sqrt{\sigma}$ is observed while $\beta$ increases~\cite{Caselle:2014eka},
as predicted by the exact solution of the
model~\cite{Polyakov:1976fu,Gopfert:1981er}. In the present case we see instead
an opposite behavior: while the lowest glueball mass is essentially unchanged as
$h$ increases \cite{Athenodorou:2020clr}, we see from Tab.~\ref{tab:newfit}
that the string tension {\sl decreases} as $h$ increases, thus increasing the
value of the ratio $m/\sqrt{\sigma}$. This makes it particularly difficult to
interpret the deviations from the EST behavior as due to the interaction
between bulk and flux tube degrees of freedom, and further supports the attempt
to find an alternative model.

%% file: 5_rigidstring.tex
\section{The rigid string}
\label{sec:rigidstring}

An interesting candidate to describe the behavior of the flux tube for $h>0$ is
the so called ``rigid string'' which is obtained by adding to the Nambu-Got\=o
action a term proportional to (the square of) the extrinsic curvature, which has
the effect of increasing the stiffness of the configuration. The rigid string
has a long history: it was originally introduced to describe the physics of
fluid membranes~\cite{Peliti:1985eo,Helfrich:1985eo,Forster:1986ot} and was
later proposed by Polyakov and Kleinert as a way to stabilize the Nambu-Got\=o
action~\cite{Polyakov:1986cs, Kleinert:1986bk}. With the notations of
Eqs.~\eqref{eq:NGaction} and \eqref{eq:NGaction2} its action can be written as

\begin{equation}
    S_{R}= \int_\Sigma d^2\xi \, \sqrt{g} \, \left[\sigma  + \gamma_2 \mathcal{K}^2 +  \dots \right] \ ,
    \label{rigid}
\end{equation}

where $\mathcal{K}=\Delta(g)X$ is the extrinsic curvature, $X$ is the transverse
displacement in the physical gauge, see Sec.~\ref{sec:ESTfluxtube}, and
\begin{equation}
    \Delta(g)=\frac{1}{\sqrt{g}} \, \partial_a \, \left[\sqrt{g} \, g^{ab}\partial_b \right] \ .
\end{equation}

Evaluating this expression in the physical gauge for the $(2+1)$ dimensional
case in which we are interested, and keeping only leading-order terms, one finds:

\begin{equation}
S_{R}= \int_\Sigma d^2\xi \; \left[ \sigma \, \partial X \partial X + \gamma_2 \, \partial^2 X \partial^2 X + \dots \right] \ .
    \label{rigid2}
\end{equation}

The standard approach to study the rigid string model was to treat the {$\mathcal{K}^2$}
term as a perturbation of the Gaussian (Nambu-Got\=o) one, see for
instance~\cite{Braaten:1986bz} and~\cite{German:1989vk}. Using this approach an
estimate for the corrections to the Nambu-Got\=o prediction for $E_0(N_t)$  was
obtained soon after this action was proposed~\cite{Braaten:1986bz,German:1989vk}.
However, when high precision simulations for the interquark potential became
available, it was soon realized that $E_0(N_t)$, at least in non-Abelian LGTs in
(2+1) and in (3+1) dimensions, was in substantial agreement with the prediction
of the simple Nambu-Got\=o action, thus suggesting an almost negligible value of
the rigidity correction.

In recent years it has been understood that this rigidity term is actually
absent from any Poincar\'e invariant EST. This is known as ``low energy
universality'' \cite{Aharony:2013ipa,Dubovsky:2012sh} and is ultimately due to
the fact that the {$\mathcal{K}^2$} term is proportional to the equation of motion
of the NG string. Thus, in agreement with the low energy universality, the first
allowed perturbative correction to the NG action must involve the {\sl fourth
power} of the extrinsic curvature and is described by the following action
\begin{equation}
S_{BNG}= \int_\Sigma d^2\xi \sqrt{g}\,\left[\sigma +  \gamma_3 \mathcal{K}^4 + \dots \right] \, ,
\label{SBNG}
\end{equation}
It is exactly this term which is responsible for the tiny deviations with
respect to the Nambu-Got\=o predictions in ordinary confining gauge theories
that we mentioned in Sec.~\ref{sec:ESTfluxtube} and that are represented in
Fig.~\ref{fig:E0_T_BNG_TD} by the blue dashed line. The coefficient $\gamma_3$
in Eq.~\eqref{SBNG} is related to the $k_4$ parameter of
Eq.~\eqref{eq:E0_BNG_Nt7} by the following expression:
\begin{equation}
    \gamma_3 = - \frac{225}{32 \pi^6} k_4
\end{equation}

\subsection{The Polchinski-Yang solution}

However this is not the end of the story. An alternative framework for
analyzing $S_R$ was proposed by Polchinski and Yang in 1992
\cite{Polchinski:1992ty}. This approach treats the four-derivative term as the dominant
one and the two-derivative (NG) term as a small perturbation. This requires for
consistency that $\gamma_2\gg N_t^2\sigma$ and $N_t^2\sigma \ll 1$, where
$\gamma_2$ is the coefficient multiplying the square of the extrinsic curvature
in Eq.~\eqref{rigid}. This regime corresponds to a completely different vacuum,
which is ``unphysical'' for ordinary non-Abelian Yang-Mills theories. Indeed,
for these theories (the inverse of) the critical temperature is
$N_{t,c}\sim 1/\sqrt{\sigma}$ and the constraint $N_t^2\sigma\ll1$ implies that
the model is in the deconfined phase, where we do not expect the presence of a
confining flux tube and thus of an effective string description\footnote{Despite
this fact, this regime was studied in great detail for completely different
reasons. The goal was to show that in this particular regime the (unphysical)
high temperature behavior of the model was the same as that of QCD in the large
$N$ limit \cite{Polchinski:1992ty}.}.

Although the solution proposed by Polchinski-Yang (PY) is unphysical in
ordinary YM models, the regime $N_t^2\sigma \ll 1$ is exactly the one we are
interested in to describe the deep reconfined phase of the trace deformed
models: for large enough values of $h$, the critical temperature is shifted to
higher values and eventually the condition {$N_t^2\sigma \ll 1$} is fulfilled
even in the confining regime.

Following \cite{Polchinski:1992ty} we have (for generic values of the
transverse dimensions) (see also \cite{Ambjorn:2014rwa} for a slightly
different formulation):
\begin{equation}
    \label{eq:e0_PY}
    E_0 = w \, \lambda \ ,
\end{equation}
where
\begin{equation}
    \label{eq:e0_PY1}
    w=\sqrt{N_t^2-\frac{b(d-2)N_t}{2\sqrt{\lambda}}}\ ,
\end{equation}
and
\begin{equation}
    \label{eq:e0_PY2}
    \sqrt{\lambda}=\frac{3b}{8}\frac{(d-2)}{N_t }+\sqrt{\frac{9b^2}{64}\frac{(d-2)^2}{N_t^2}+{\sigma'}-\frac{\pi(d-2)}{3N_t^2}}
\end{equation}

where $b$ is related to the $\gamma_2$ coefficient which appears in
Eq.~\eqref{rigid} by $b=1/\sqrt{2\gamma_2}$, and we denoted the string tension
which appears in Eq.~\eqref{rigid} by $\sigma'$, because there is no reason for
it to coincide with the string tension $\sigma$ of the ordinary YM theory at
the same coupling $\beta$.

This analytical solution yields a modified critical deconfinement temperature:
\begin{equation}
    \label{eq:ntc_PY}
    \left( N^{(PY)}_{t,c} \right)^2 \sigma'=\frac{\pi(d-2)}{3}-\frac{b(d-2)^2 }{8}\ ,
\end{equation}
which, as expected, is larger than the canonical Nambu-Got\=o prediction of $N_{t,c}^2\sigma=\pi(d-2)/3$.

Let us note, as a side remark, that the PY solution shows deviations from NG
which would be forbidden by the ``low energy universality''
\cite{Aharony:2013ipa}. In particular, it violates the conditions imposed by
the Lorentz invariance of the theory, on the coefficients multiplying the terms
proportional to $1 / {N_t}^k$ in the expansion of the ground state energy, for
$k \le 5$. To our knowledge, in the context of the trace deformed theory, this
does not lead to a contradiction since the model we are simulating explicitly
breaks the Lorentz invariance. The constraints arise from the non linear
realization of the symmetry and in particular (see also
\cite{Aharony:2010db,Dubovsky:2012sh,Billo:2012da}), from the ``redundancy'' of
spontaneously broken generators: the result is that, in the Lorentz-invariant
theory, the number of Goldstone bosons ($d - 2$) is smaller than the number of
symmetry generators broken by the worldsheet ($d - 2$ translations and $2(d -
2)$ space-time rotations), leading to non-trivial constraints on the dynamics
of the Goldstone modes. In the trace deformed theory, however, some of these
non-linearly realized rotations are explicitly broken (specifically, in our 
$(2+1)$-dimensional case, the rotation that mixes Euclidean time with the 
direction orthogonal to the worldsheet)
by a term (the trace deformation) which appears to be relevant in the continuum
limit, so that the assumptions underlying the low energy universality no longer
hold. 
In principle it is possible to write down all the new allowed terms in the 
effective string theory, for example those with derivatives only in one of the two
worldsheet directions, and try to fit our results at all the values of $h$.

%% file: 6_results2.tex
\section{Fitting the data with the Polchinski-Yang solution}\label{sec:results2}

We fitted our data with Eq.~\eqref{eq:e0_PY}. Results of the fits are reported
in Tab.~\ref{tab:newfitPY} and Tab.~\ref{tab:newfitPY_beta27} for
$\beta=23.3805$ and $\beta=27.4745$ respectively. It is worth noting that fits
performed by using Eq.~\eqref{eq:e0_PY} depend on just two parameters
($\sigma'$ and $b$), exactly as the fits to Eq.~\eqref{eq:E0_L} discussed in
Sec.~\ref{sec:results}. Consequently, the improved agreement between the data
and the model cannot be ascribed to an increase in the number of fitting
parameters.

\begin{table}[h]
    \centering
    \begin{tabular}{|c|c|c|c|c|c|c|}
    \hline
    $h$ & $[N_t^{\min},N_t^{\max}]$ & $\sigma'$ & $b$ & $\gamma_2$ & $N_{t, c}^{(PY)}$ &$\chi^2/\text{dof}$ \\ \hline
    0.005 &  $[10,18]$ & 0.000728(18) & 2.8117(30) & 0.06325(13)  & 9.00(6) & 0.78 \\ \hline
    0.006 &  $[9,19]$  & 0.000871(14) & 2.8298(22) & 0.062440(98) & 7.29(7) & 0.79 \\ \hline
    0.007 &  $[8,21]$  & 0.000986(12) & 2.8516(20) & 0.061487(86) & 5.58(10) & 1.02 \\ \hline
    \end{tabular}
    \caption{Results of the fit of the data at $\beta = 23.3805$  with Eq.~\eqref{eq:e0_PY}. We also report the value for (the inverse of) the critical temperature obtained from Eq.~\eqref{eq:ntc_PY} (note that it is not integer in units of the lattice spacing). }
    \label{tab:newfitPY}
\end{table}

\begin{table}[h]
    \centering
    \begin{tabular}{|c|c|c|c|c|c|c|}
    \hline
    $h$ & $[N_t^{\min},N_t^{\max}]$ & $\sigma'$ & $b$ & $\gamma_2$ & $N_{t, c}^{(PY)}$ &$\chi^2/\text{dof}$ \\ \hline
    0.004 & [11, 19] & 0.000532(12) & 2.8199(26) & 0.06287(12) & 10.00(8) & 0.69 \\  \hline
    0.005 & [ 9, 17] & 0.000604(25) & 2.8556(44) & 0.06131(19) &  6.79(25) & 1.36 \\  \hline
    \end{tabular}
    \caption{Same as Tab.~\ref{tab:newfitPY}, with data at $\beta = 27.4745$.}
    \label{tab:newfitPY_beta27}
\end{table}

\begin{figure}[htb!]
    \centering
    \begin{center}
        \includegraphics[width=0.70\textwidth]{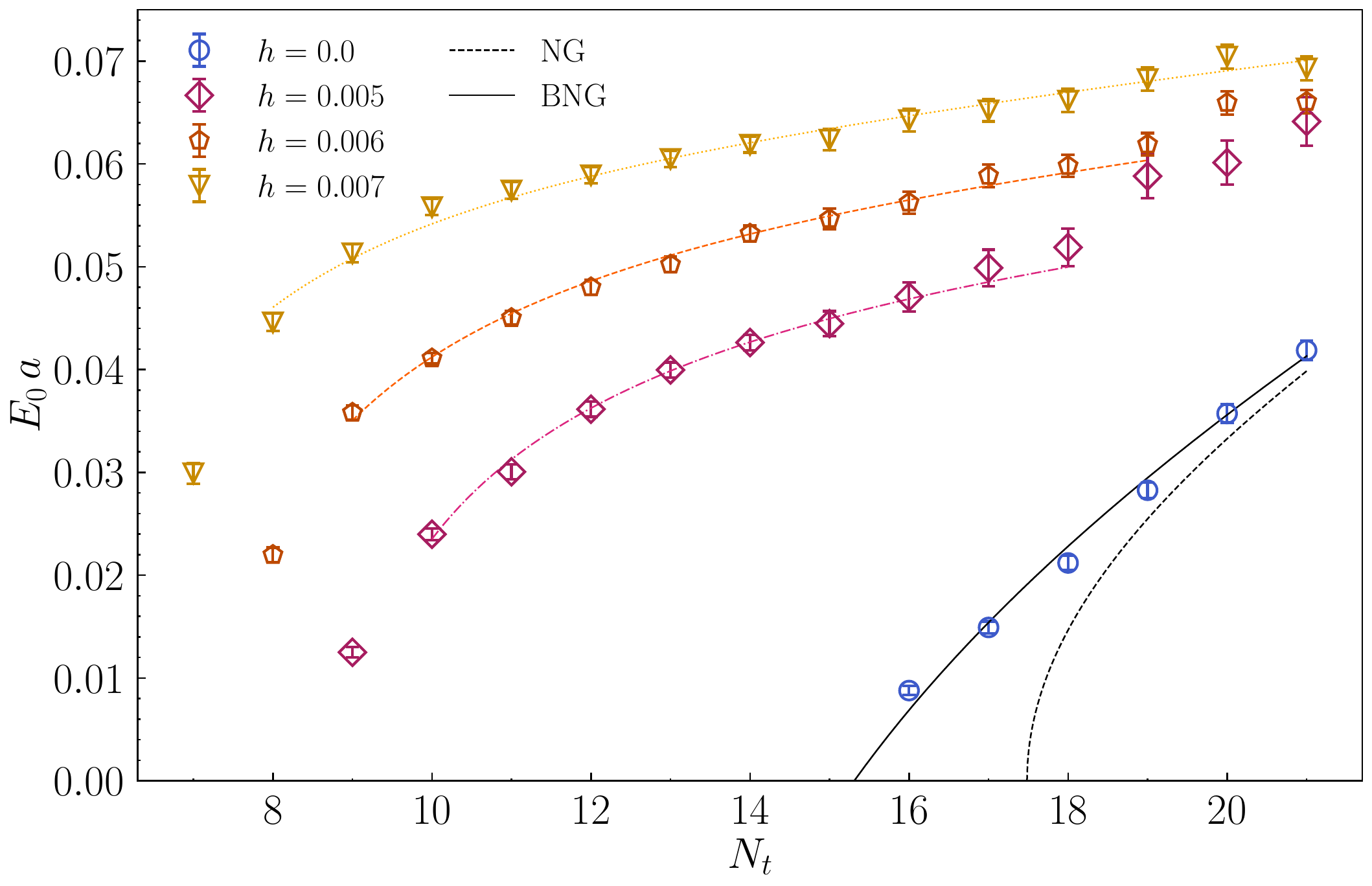}
    \end{center}
    \caption{Polchinski-Yang solution describing the reconfined data ($h \ge 0.005$), to be confronted with the BNG expression describing data coming from pure YM ($h = 0$).}
    \label{fig:PYvsNG1}
\end{figure}

Let us briefly comment on these results, considering for now only
$\beta=23.3805$. The first remarkable feature is that, for all the values
$h\geq 0.005$ we could fit our data practically in the whole range of $N_t$
values available, always finding good $\chi^2$ values, as illustrated in
Fig.~\ref{fig:PYvsNG1}. In contrast, for $h<0.005$, Eq.~\eqref{eq:e0_PY} yields
acceptable fits only for the largest values $N_t$. For this reason, in the
following we shall concentrate only on the values $h\geq 0.005$. Note that this
behavior is exactly the opposite of that found in Sec.~\ref{sec:results} when
discussing fits to the model in Eq.~\eqref{eq:E0_L}.

The second interesting feature is that the predicted (inverse of the) critical
temperature $N_{t, c}^{(PY)}$ extracted from these fits using Eq.~\eqref{eq:ntc_PY} 
differs from the deconfinement temperature (let us remind, $N_{t, c} = 15$ 
at $\beta = 23.3805$ and $N_{t, c} = 20$ at $\beta = 27.4745$) of the undeformed 
theory by a quantity that increases with $h$.

\begin{figure}[htb!]
    \centering
    \includegraphics[width=0.70\textwidth]{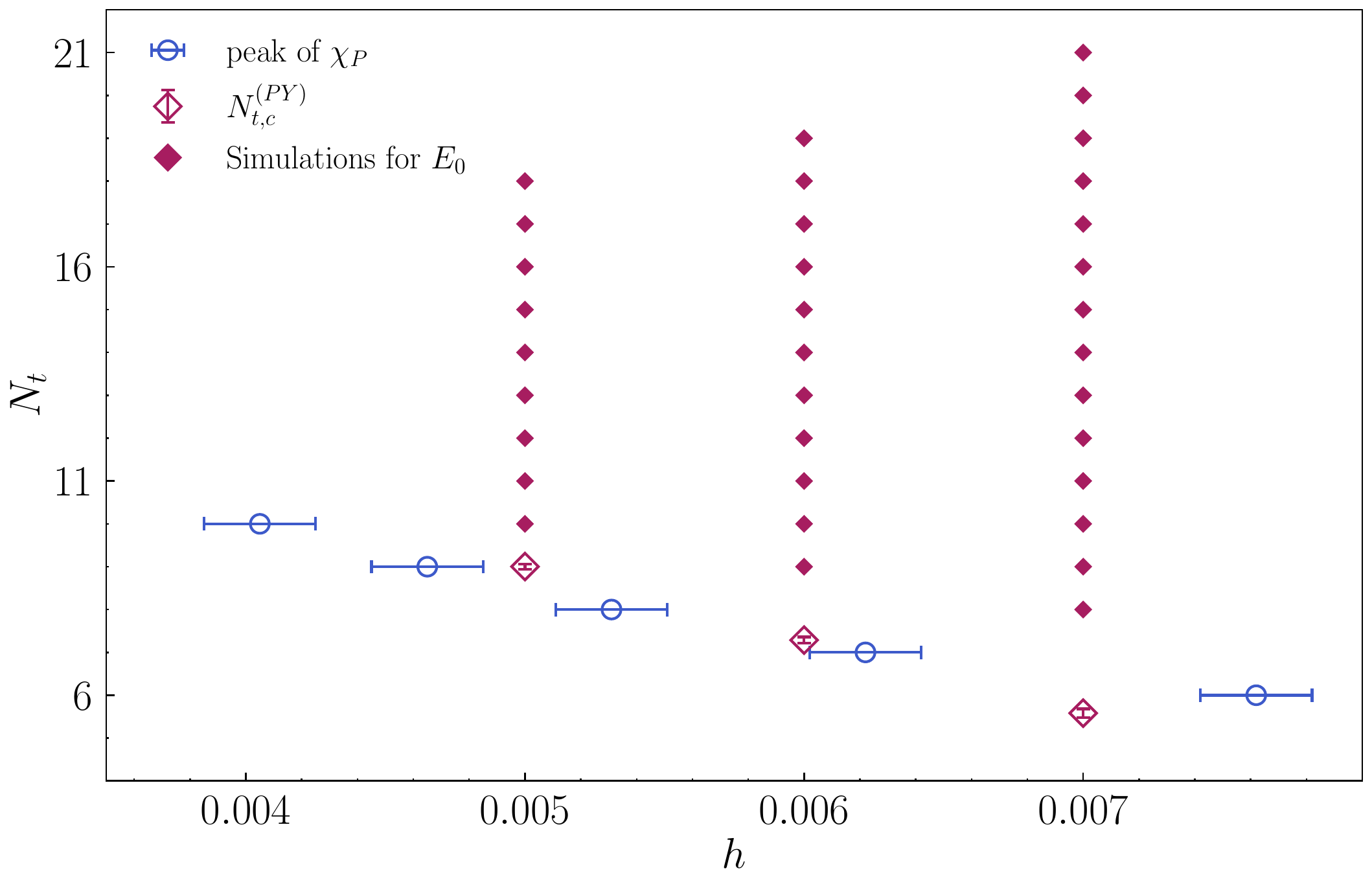}
\caption{Comparison between the estimates of the critical values of $h$ and $N_t$ obtained by two different methods.
             All the data shown refer to simulations at $\beta = 23.3805$, with $N_s = 96$.
             Blue circles represent the peak of the Polyakov loop susceptibility obtained varying $h$ with fixed lattice size.
             Open red diamonds represent the extrapolated critical value $N_{t,c}$ at which the ground state energy $E_0$ vanishes,
             according to the PY formula in Eq.~\eqref{eq:e0_PY}, for each of the three largest values of $h$.
             The red filled diamonds, instead, are the value of $h$ and $N_t$ at which the values of $E_0$ included in our fits were measured.}
    \label{fig:compare_ntc}
\end{figure}

Moreover, the (inverse of the) critical temperature $N_{t, c}^{(PY)}$ extracted
from the fits is in good agreement with the critical temperatures that we found
studying the phase diagram of the model (see Sec.~\ref{subsec:phase}).
The correspondence is visible in Fig.~\ref{fig:compare_ntc},
where we plotted the value of $h$ at which we observe a peak in the susceptibility
of the Polyakov loop at fixed size of the lattice and the critical inverse temperature
predicted from the PY formula, given the fitted parameters.
This represents a strong consistency check of the whole procedure, since the two
sets of values were obtained with completely different methods.

Additionally, from the fits we see a well-defined dependence of $\gamma_2$ 
and $\sigma'$ as a function of $h$. While $\gamma_2$ decreases, $\sigma'$ increases.
As a consequence, the range of values of $N_t$ for which the condition $\gamma_2 \gg N_t^2\sigma'$
is fulfilled (and thus the range of validity of Eq.~\eqref{eq:e0_PY}) decreases
with $h$, in agreement with the fact that also $N_{t, c}^{(PY)}$ decreases
with $h$.

Finally, a puzzling feature of our results is that, while the constraint
$\gamma_2 \gg N_t^2\sigma'$ is fulfilled only for values of $N_t$ in the vicinity
of the deconfinement temperature $N_{t, c}^{(PY)}$, the fits seem to show
a very good behavior for much larger values of $N_t$, for which the ratio
$N_t^2\sigma'/\gamma_2$, which is the parameter of the perturbative expansion, is
relatively large. We have no explanation for this good behavior of Eq.~\eqref{eq:e0_PY}
beyond the expected range of validity.

\begin{figure}[htb!]
    \begin{center}
        \includegraphics[width=0.70\textwidth]{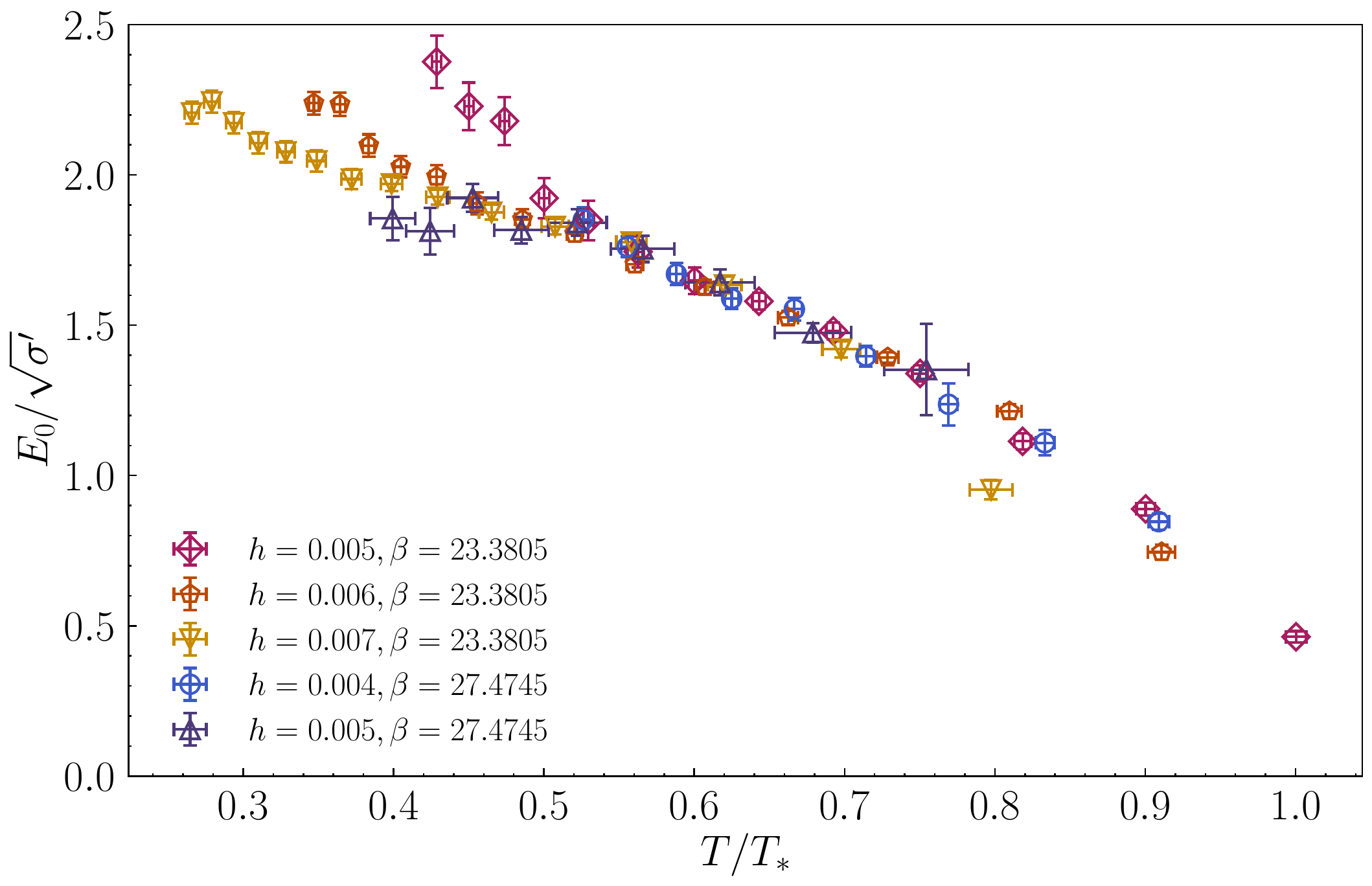}
        \caption{Collapse plot of the data in the reconfined regime (large
                 $h$), with the axes scaled according to the PY natural scales.
                 In particular, we plotted the ground state divided by the
                 square root of the PY string tension $\sigma '$ against the
                 temperature in units of the temperature $T_*$ at which the PY
                 ground state vanishes (the reciprocal of the sixth column in
                 Tabs.~\ref{tab:newfitPY} and \ref{tab:newfitPY_beta27}).}
        \label{fig:PYuniversal}
    \end{center}
\end{figure}

We repeated the same analysis at $\beta=27.4745$; the corresponding fit
parameters are collected in Tab.~\ref{tab:newfitPY_beta27}. Also at this value
of $\beta$ the PY solution describes the data with good $\chi^2$ values, and the
behavior of the fitted $\sigma'$, $\gamma_2$ and $N_{t, c}^{(PY)}$ as a function
of $h$ is analogous to that observed at $\beta=23.3805$. The comparison is
summarized in Fig.~\ref{fig:PYuniversal}, where the data at all the values of
$\beta$ and $h$ we studied are collapsed together, with the axes scaled
according to the natural scales of the PY solution.

To better understand this issue and, more in general,  to gain a better insight
into the behavior of the flux tube, we decided to study its shape and width.

%% file: 7_width.tex
\section{The chromo-electric flux tube profile in the reconfined phase}
\label{sec:fluxtube}

We find another remarkable difference with respect to the standard gauge theory
analyzing the ``profile'' of the flux tube, defined as in
Ref.~\cite{Caselle:2026coc}. This lattice observable can provide particularly
insightful information for the study of effective descriptions of confinement,
although it is also well known to be affected by significant statistical noise.

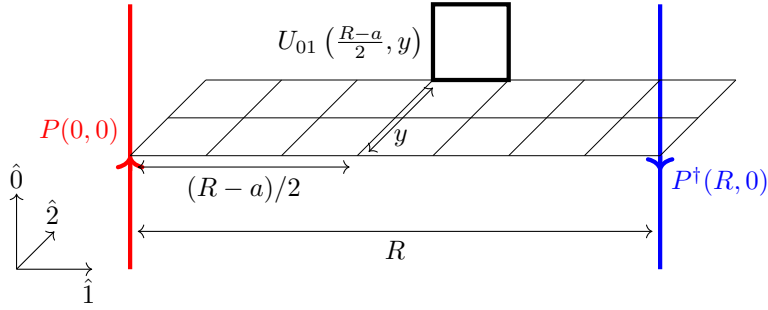
\begin{figure}[h]
\centering
    \input{poly_poly_plaq}
    \caption{Geometry of the three-point correlator $F_{01}$ in Eq.~\eqref{eq:threepoints} ($\hat{0}=\hat{t}$, $\hat{1}=\hat{x}$). The Polyakov loops at separation $R$ are shown in red and blue, while the plaquette operator is indicated in black; thick lines represent the corresponding traced Wilson lines.}
    \label{fig:poly_poly_plaq}
\end{figure}

Following the original proposal of Ref.~\cite{DiGiacomo:1990hc}, we define the
(disconnected) profile of the flux tube as
\begin{equation}
    \rho(R,y) = \frac{F_{01}(R,y)}{G(R)}-\langle U_{01}\rangle\ ,
\end{equation}
where
\begin{equation}\label{eq:threepoints}
    F_{01}(R, y) = \frac{1}{{N_s}^2} \, \left\langle  \sum_{\vec{x}}
        P \! \left(\vec{x}\right) \,
        \tfrac{1}{2} U_{01} \! \left(\vec{x} + \vec{l}\, \right) \,
        P^\dagger \! \left(\vec{x} + R \hat 1\right)
    \right\rangle\ ,
\end{equation}
and $\vec{l} = \frac{1}{2}(R - a) \hat1 + y \hat2$, for values of $R$ that are
odd in units of the lattice spacing $a$, see Fig.~\ref{fig:poly_poly_plaq} for
a graphical illustration.

As discussed in Ref.~\cite{Caselle:2026coc}, at high temperatures (close to the deconfinement
phase transition, yet still in the confining phase), the profile $\rho$ for Yang-Mills
$\SU(2)$ can be derived from the spin-spin-energy correlator of the two dimensional
Ising model, in agreement with the Svetitsky-Yaffe (SY) mapping~\cite{Svetitsky:1982gs}.
That correlator is modelled as
\begin{equation}
    \label{eq:rho_SY}
    \rho(R,y) = A^{\mathrm{(SY)}} \, \frac{2 \pi R}{4 l^2} \,
           \frac{\exp(-l / \lambda)}{K_0(R / (2 \, \lambda) )}\ ,
\end{equation}
where $l=|\vec{l}|$ is the distance between each of the Polyakov loops and the
plaquette, and $\lambda$ is a length parameter to be fitted.
According to the SY mapping, when the distance $R$ is large enough, the scale $\lambda$
should be independent from $R$, and equal to $1 / (2 E_0)$.

\begin{figure}[htb!]
\centering
\includegraphics[width=0.70\textwidth]{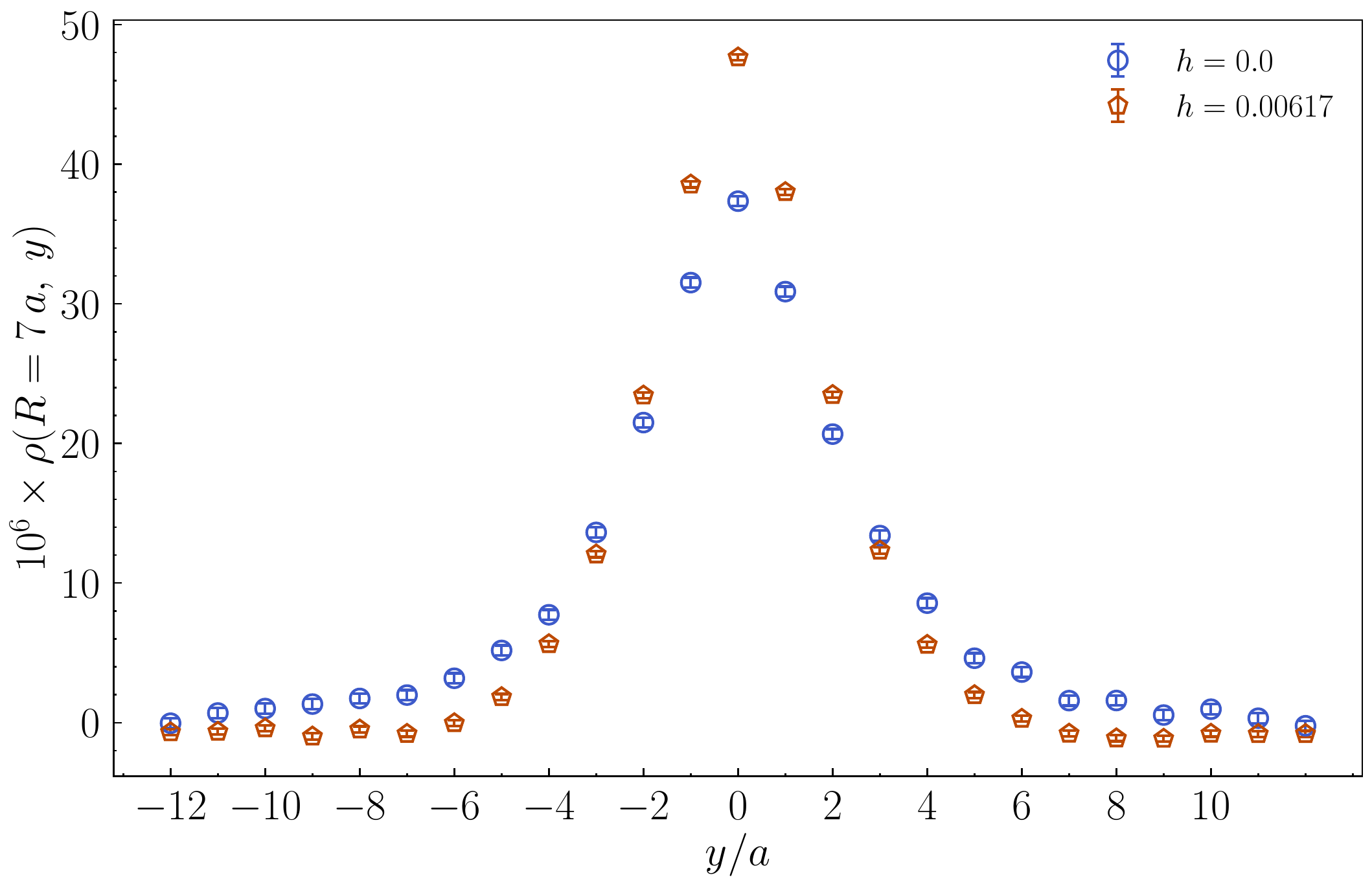}
\caption{The flux tube in the standard theory and with $h = 0.00617$. The values
         of $N_t$ (respectively 23 and 13) were chosen such that $E_0$ is roughly
         the same (we measured 0.0519(14) and 0.0528(3)). At large $h$ it is evident
         the squeezing of the flux tube, as indicated by the fitted values of $\lambda$,
         5.7(1.2) and 1.47(18).}
\label{fig:squeeze_fluxtube}
\end{figure}

We compute the profile of the flux tube for different combinations of $N_t$ and
$h$, so that the ground state $E_0$ of the string is approximately constant (see
Tab.~\ref{tab:width_params} for some details of the simulations). The first 
qualitative observation we can make is that in the reconfined regime the flux tube
appears ``squeezed'', when compared with the corresponding observable at $h = 0$, 
as we can see in Fig.~\ref{fig:squeeze_fluxtube}.

\begin{figure}[htb!]
    \begin{center}
        \includegraphics[width=0.70\textwidth]{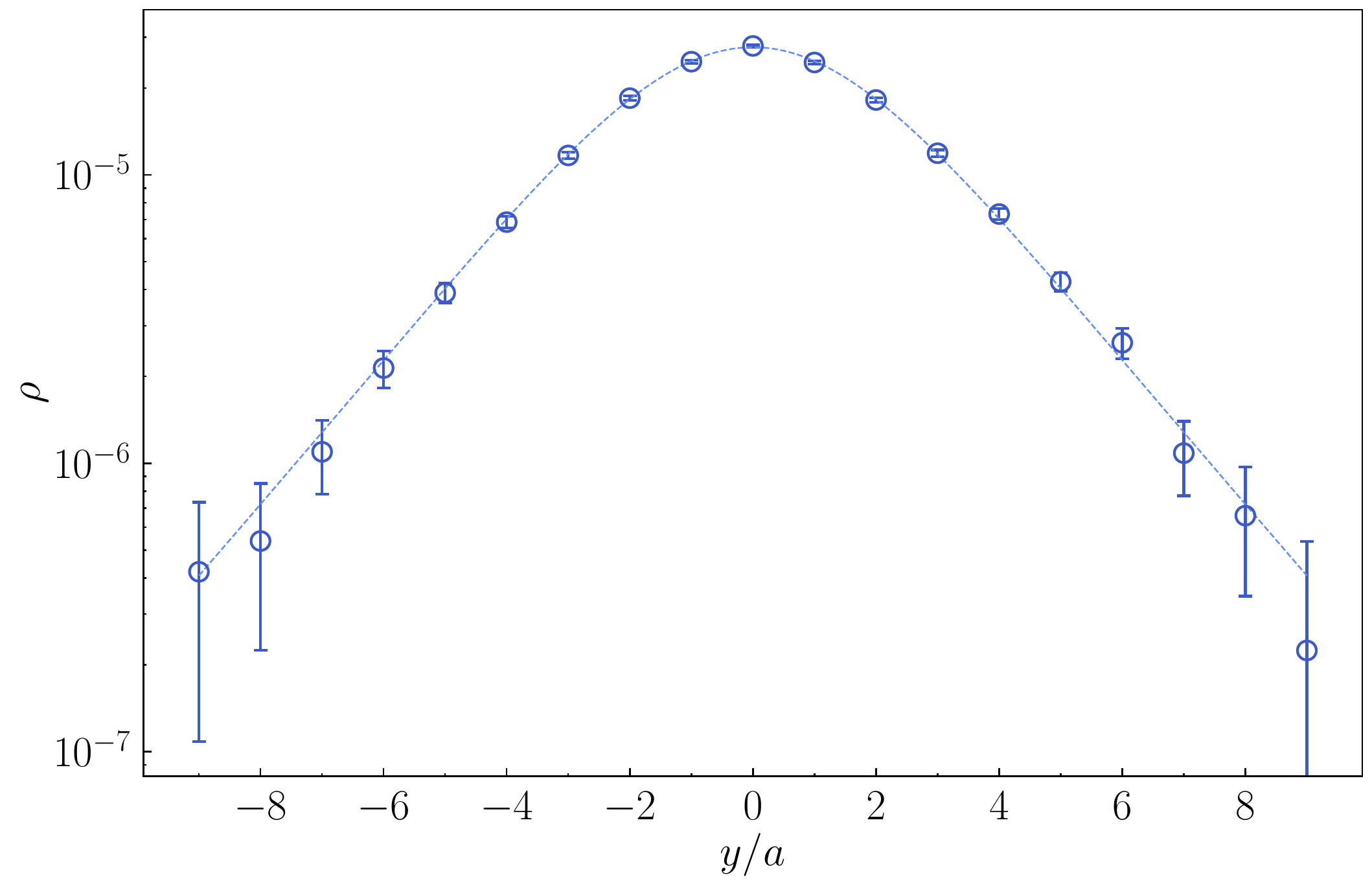}
    \end{center}
    \caption{Shape of the flux tube $\rho (R,y)$ for  $ N_{t} = 16$, $ h = 0.00548 $ and a distance between Polyakov loops $R / a = 9$. The continuous line represents the best fit curve of Eq.~\eqref{eq:rho_fitmodel}.}
    \label{fig:tube}
\end{figure}

At $h > 0$ we have no reason to assume the SY mapping holds, however we tried fitting
our data with a similar model:
\begin{equation}
    \label{eq:rho_fitmodel}
    \rho(R, y) = \tilde{A} \, \frac{\exp\left( -l / \lambda \right)}{l^p}
\end{equation}
where we leave $p$ as a free parameter since we do not know a priori if the same description
\textit{\`a la} Svetitsky-Yaffe holds in the reconfined regime as well as at $h = 0$.
Thus we fit with three free parameters ($\lambda$, $p$ and $\tilde{A}$) and repeat the fit
assuming $\lambda = 1 / (2 E_0)$, without any assumption on the dependence of $\tilde{A}$ on $R$.
An example of fit is shown in Fig.~\ref{fig:tube}.

\begin{table}
\centering
\begin{tabular}{|c|c|c|c|c|c|c|}
\hline
$h$ & $N_t$ & $N_s$ & samples & $R_{\text{min}}$ & $R_{\text{max}}$ & $E_0$ \\
\hline
0.00637 & 12 & \multirow{11}{*}{120} & \multirow{4}{*}{40000} & \multirow{11}{*}{7} & \multirow{11}{*}{23} & 0.05431(83) \\ \cline{1-2}\cline{7-7}
0.00617 & 13 &  &  &  &  & 0.05275(30) \\ \cline{1-2}\cline{7-7}
0.00596 & 14 &  &  &  &  & 0.05260(79) \\ \cline{1-2}\cline{7-7}
0.00572 & 15 &  &  &  &  & 0.05263(84) \\ \cline{1-2}\cline{4-4}\cline{7-7}
0.00548 & 16 &  & \multirow{4}{*}{30000} &  &  & 0.05286(94) \\ \cline{1-2}\cline{7-7}
0.00496 & 17 &  &  &  &  & 0.05222(94) \\ \cline{1-2}\cline{7-7}
0.00466 & 18 &  &  &  &  & 0.05356(94) \\ \cline{1-2}\cline{7-7}
0.00405 & 19 &  &  &  &  & 0.05297(94) \\ \cline{1-2}\cline{4-4}\cline{7-7}
0.003   & 20 &  & \multirow{3}{*}{40000} &  &  & 0.0528(14)  \\ \cline{1-2}\cline{7-7}
0.002   & 21 &  &  &  &  & 0.0522(12)  \\ \cline{1-2}\cline{7-7}
0.000   & 23 &  &  &  &  & 0.0519(14)  \\
\hline
\end{tabular}
\caption{Simulation parameters for the determination of  $\rho (R,y)$ in Eq.~\eqref{eq:rho_SY} for $\beta=23.3805$.
         At $h = 0$, for this value of $\beta$, $N_t = 23$ corresponds to a temperature
         $T \simeq 0.66 T_c$, close to the lowest temperature at which the Svetitsky-Yaffe
         model for the flux tube was tested in Ref.~\cite{Caselle:2026coc}.}
\label{tab:width_params}
\end{table}

\begin{figure}[htb!]
    \centering
    \includegraphics[width=0.7\textwidth]{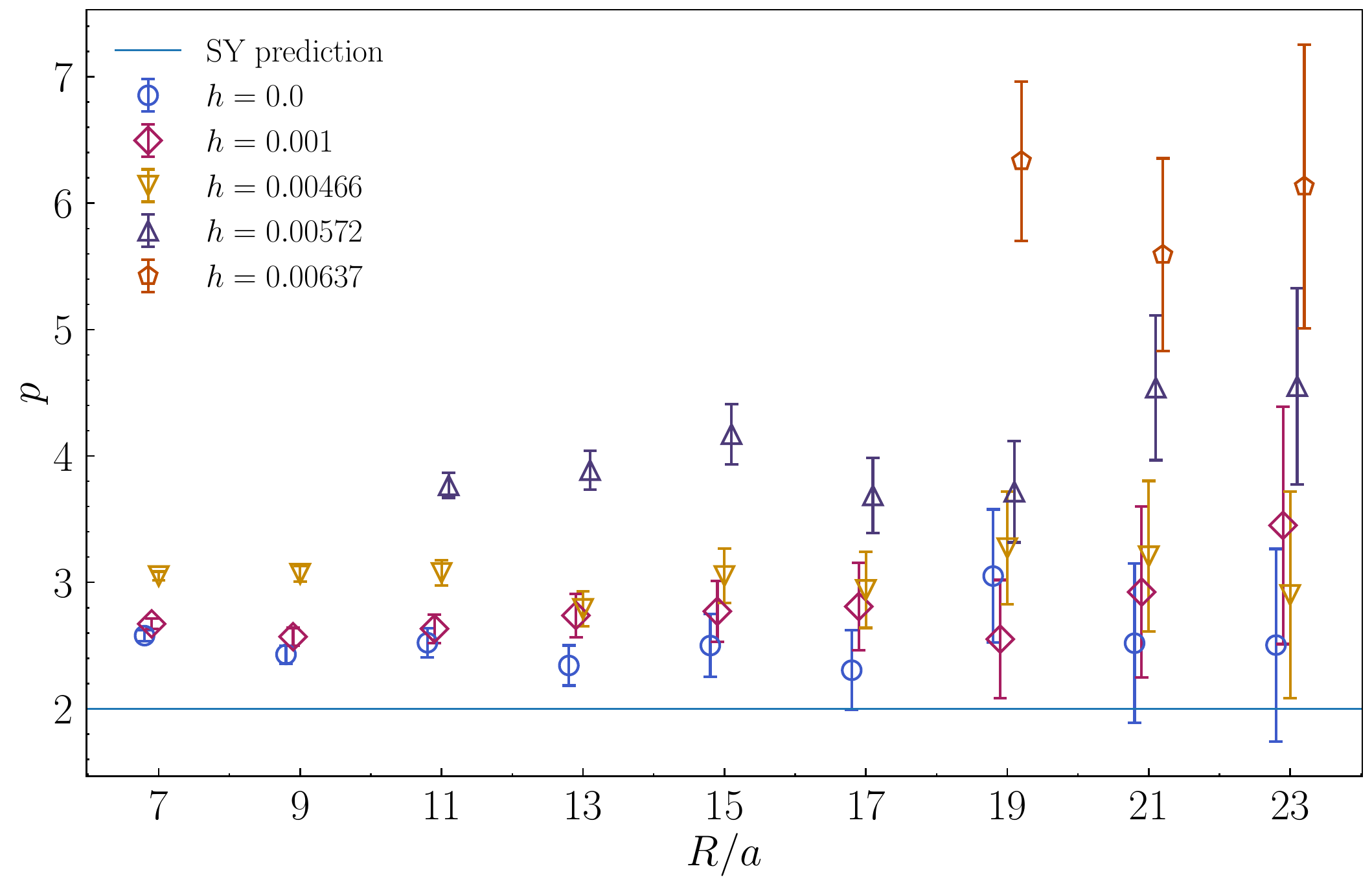}
    \caption{Best fit value of the power $p$ in Eq.~\eqref{eq:rho_fitmodel} when
             $\lambda$ is fixed to $1 / (2 E_0)$ for some values of $h$.
             We only show results of acceptable fits.
             The solid blue line is $p = 2$, as expected for $h = 0$ at large $R$
             (see Ref.~\cite{Caselle:2026coc}).
    }
    \label{fig:power_profile}
\end{figure}

As we can see from
Tabs.~\ref{tab:fitres_profile_nt23}~to~\ref{tab:fitres_profile_nt12}, the model
(even fixing $\lambda$ from the two point function of the Polyakov loops) fits
the data accurately only for small values of $h$. As soon as $h$ grows above
$0.005$, that is, once we are firmly inside the reconfined regime, the model
reproduces the data only at large separations $R$ between the Polyakov loops,
where the statistical precision is poorest, and only when very large powers $p$
are allowed. The need for such large powers lacks at the moment a clear
physical motivation. No improvement is obtained leaving $\lambda$ as a free
parameter in Eq.~\eqref{eq:rho_fitmodel}. The fitted values of the power $p$
for those fit that can be considered of ``acceptable'' quality ($\chi^2 < 2
\times \text{number of degrees of freedom}$) are shown in
Fig.~\ref{fig:power_profile}.

\begin{figure}[htb!]
\centering
\includegraphics[width=0.7\textwidth]{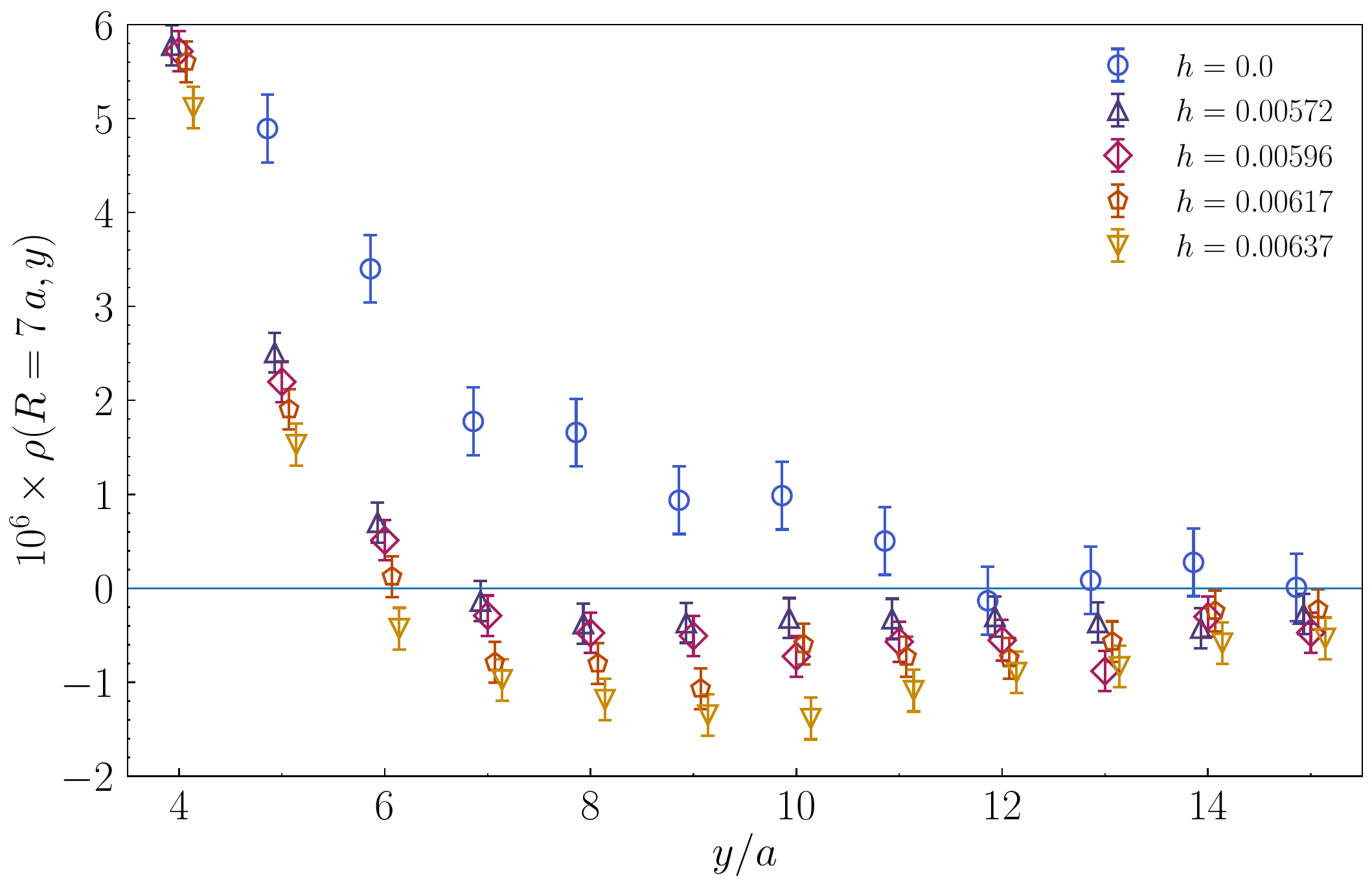}
\caption{Comparison between the profiles at $R = 7 \, a$ for different values of $h$ (and $N_t$, see
         Tab.~\ref{tab:width_params}). In particular, we chose the values of $h$ for which we observe
         $\rho < 0$ and added $h = 0$ for comparison. We averaged the values for positive and negative
         $y$ and zoomed on the region where we find the ``dip'' into the $\rho < 0$ region. For clarity,
         we slightly shifted the value of $y / a$ by a different value for each $h$.}
\label{fig:deep_fluxtube}
\end{figure}

A more careful analysis of the data at large $h$ reveals an intriguing fact for
which we have at present no theoretical explanation, but which is clearly the
reason for the bad quality of our fit. With the statistical precision of our
data we can see that there is a region, corresponding to $7 \le y / a \le 14$,
where $\rho(R,y)$ is negative (see Fig.~\ref{fig:deep_fluxtube}). This peculiar
behavior seems to be present for all the values of $R$ explored, or (to be
conservative) at least
for those where our data are precise enough (roughly $7 \le R / a \le 13$).

\clearpage

%% file: poly_poly_plaq.tex
\begin{tikzpicture}[scale=1., transform shape]
    \draw [ -> ] (-1, 0) -- (-1, 1) node [pos=0.95, anchor=south] {$\hat 0$};
    \draw [ -> ] (-1, 0) -- (0, 0) node [pos=0.95, anchor=north] {$\hat 1$};
    \draw [ -> ] (-1, 0) -- (-0.5, 0.5) node [pos=0.95, anchor=south] {$\hat 2$};

    \draw [red, ultra thick, ->] (0.5, 0) -- (0.5, 1.5) node [anchor=south east] {$P(0, 0)$};
    \draw [red, ultra thick] (0.5, 1.3) -- (0.5, 3.5);

    \draw [blue, ultra thick] (7.5, 0) -- (7.5, 1.5) node [anchor=north west] {$P^\dagger(R, 0)$};
    \draw [blue, ultra thick, ->] (7.5, 3.5) -- (7.5, 1.3);

    \draw [ultra thin] (0.5, 1.5) -- (7.5, 1.5);
    \draw [ultra thin] (1.0, 2.0) -- (8.0, 2.0);
    \draw [ultra thin] (1.5, 2.5) -- (8.5, 2.5);

    \draw [ultra thin] (0.5, 1.5) -- (1.5, 2.5);
    \draw [ultra thin] (1.5, 1.5) -- (2.5, 2.5);
    \draw [ultra thin] (2.5, 1.5) -- (3.5, 2.5);
    \draw [ultra thin] (3.5, 1.5) -- (4.5, 2.5);
    \draw [ultra thin] (4.5, 1.5) -- (5.5, 2.5);
    \draw [ultra thin] (5.5, 1.5) -- (6.5, 2.5);
    \draw [ultra thin] (6.5, 1.5) -- (7.5, 2.5);
    \draw [ultra thin] (7.5, 1.5) -- (8.5, 2.5);

    \draw [ultra thick] (4.5, 2.5) -- (5.5, 2.5) -- (5.5, 3.5) -- (4.5, 3.5) -- cycle;
    \node at (4.5, 3) [anchor=east] {$U_{01} \left( \frac{R - a}{2}, y \right)$};

    \draw [ <-> ] (0.6, 0.5) -- (7.4, 0.5) node [pos=0.5, anchor=north] {$R$};
    \draw [ <-> ] (0.6, 1.35) -- (3.4, 1.35) node [pos=0.5, anchor=north] {$(R - a) / 2$};
    \draw [ <-> ] (3.65, 1.55) -- (4.5, 2.4) node [pos=0.5, anchor=north] {$y$};

\end{tikzpicture}

%% file: 8_conclusion.tex
\section{Conclusion}\label{sec:conclusion}

In this work we addressed a fundamental question about trace deformed
Yang-Mills theories, i.e., whether the reconfined phase of these models
(realized at temperatures above the standard deconfinement transition for
positive values of the trace deformation parameter) shares the same long
distance physics as ordinary confinement.

Focusing on the simplest setting, the $(2+1)$ dimensional trace deformed
$\SU(2)$ model, we studied the confining flux tube through Polyakov-loop
correlators, and compared the resulting ground state energy $E_0(N_t)$ with
effective-string theory expectations. Our data show that, as the deformation is
increased, the behavior of the confining strings in the reconfined phase
significantly departs from the Nambu-Got\=o description, even including the
standard higher-derivative corrections that successfully account for the pure
Yang-Mills case.

The main outcome is that, in the reconfined regime and for sufficiently large
deformation parameters, numerical data are instead accurately described by the
Polchinski-Yang solution of the rigid string, i.e.\ an effective string
dominated by an extrinsic-curvature term. This is particularly remarkable
because the corresponding high temperature rigid-string regime is essentially
inaccessible in ordinary Yang-Mills, as it would lie beyond the deconfinement
point, while the trace deformation provides a genuine lattice realization of it
that still shows confinement. Hence the reconfined phase offers a new laboratory
in which the otherwise elusive properties of the rigid string can be probed
quantitatively, and our results provide direct evidence for the peculiar
large-$T$ behavior predicted by the Polchinski-Yang framework.

Independent support for this picture comes from the study of the transverse
structure of the flux tube. While at small deformation the profile is broadly
compatible with the usual near-$T_c$ behavior derived from the Svetitsky-Yaffe
conjecture, increasing $h$ leads to a clear change in the intrinsic width,
signalling a qualitatively different flux-tube dynamics. Together with the
exploratory phase-diagram study reported in Appendix~\ref{sec:appendixA}, these
findings point to a confinement mechanism in the reconfined phase that, while
sharing several global features with ordinary confinement, is naturally captured
by a different effective-string description.

A major consequence of our analysis is thus the indication that the original
program of using the trace deformation to reach a perturbative description of
confinement is most probably unfeasible. However, at the same time, our results
suggest that it could be possible to obtain a perturbative description of
confinement in a model described by the Polchinski-Yang rigid string, a goal
which would certainly deserve attention and further efforts.

An accurate investigation of the $(N_t,h)$ phase diagram of the model could help
in characterizing the transition between the Nambu-Got\=o and the
Polchinski-Yang regimes, but other developments are also clearly possible and
worthwhile. Extending the analysis to excited string states would allow us to
further constrain the rigid string interpretation. Moreover we know that
trace deformed Yang-Mills is only one way to evade deconfinement; it would be
interesting to repeat analogous flux-tube tests in other constructions and
check whether they realize the same (or different) effective-string regime.
Confinement in the semiclassical regime of compact $\mathrm{U}(1)$ gauge theory
is associated with monopole condensation, and a rigid string description is
naturally expected to arise. Clarifying whether the reconfined $\SU(2)$ string
realizes an analogous monopole-driven rigidity mechanism could provide a more
direct analytic underpinning of the effective theory observed here.

%% file: appendixA.tex
\section{Phase diagram of the reconfined theory}\label{sec:appendixA}

\begin{figure}[h!]
    \begin{center}
    \resizebox{0.70\textwidth}{!}{%
    \input{phase_diagram_sketch.tex}%
    }
    \end{center}
    \caption{Sketch of the conjectured phase diagram in the $(h,N_t)$ plane. The critical line is continuous (weak first order/second order) for $h\le h^{\star}$ and discontinuous (first order, dashed) for $h>h^{\star}$, the two branches meeting at the tricritical point $h^{\star}$ (magenta). The two regions are described by a Beyond Nambu-Got\=o (BNG) and a Polchinski-Yang (PY) effective string respectively.}
    \label{fig:phase_diagram_sketch}
\end{figure}
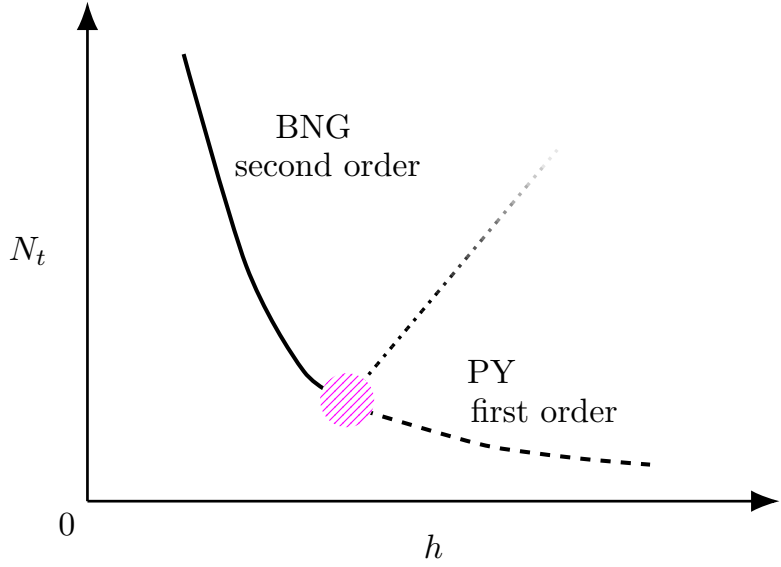

From the discussion in the main text we see how the effective string smoothly
departs from the BNG action known to be accurate at $h=0$: as we increase the
deformation parameter $h$ the NG prediction for the intrinsic width of the flux
tube $\lambda$ ceases to be valid. This suggests that some kind of phase
transition of the model could be driving the change in the string. We thus set
out to study the phase diagram of our reconfined theory. We conjectured that
the critical line could change nature along its extent --- continuous up to a
tricritical point $h^{\star}$ and discontinuous beyond it (a scenario realised,
e.g., in the Blume-Capel model~\cite{Zierenberg:2016zoh,Mozolenko:2024blc}), as
sketched in Fig.~\ref{fig:phase_diagram_sketch}. The $h\le h^{\star}$
continuous branch and the $h>h^{\star}$ discontinuous one would then correspond
to areas described by either a Nambu-Got\=o or a Polchinski-Yang effective
string.

In this appendix we shift the focus from the effective string
description of the Polyakov loop correlators to the properties of the critical line of
the model. Sec.~\ref{sec:appB_thmeth} reviews the lattice techniques we
use to characterize the phase transition; the results obtained at low and high
temperature are presented in Sec.~\ref{sec:appB_lowT} and
Sec.~\ref{sec:appB_highT} respectively, and we conclude in
Sec.~\ref{sec:appB_concl}.

\subsection{Theory and methods}\label{sec:appB_thmeth}

To characterize the transition we investigate the finite-size scaling (FSS) of two
RG invariant quantities, together with direct diagnostics of phase coexistence.
The two RG invariant quantities are the Binder cumulant of the Polyakov loop $P$
\begin{equation}
    U=\frac{\langle |P|^{4} \rangle}{\langle |P|^2 \rangle^2}\ ,
\end{equation}
and the ratio 
\begin{equation}
    R_{\xi }=\xi/N_s\ ,
\end{equation}
where the second-moment correlation length is defined as
\begin{equation}
 \xi^{2} = \frac{1}{4\sin^2(p_{\min}/2)}\frac{\tilde{G}(\textbf{0})-\tilde{G}(\textbf{p})}{\tilde{G}(\textbf{p})}\ ,
\end{equation}
where $p_{\min}=2\pi/N_s$, $\textbf{p}$ is a vector with only one non-vanishing
component equal to $p_{\min}$, $\textbf{0}$ denotes zero momentum, and
$\tilde{G}(\textbf{p})$ is the Fourier transform of the
spatial correlator of Polyakov loops $G(\textbf{x})$. We refer the reader to
Ref.~\cite{condmat0012164,Bonati:2024gh} for more details.

Since at the deconfinement phase
transition the spontaneous symmetry breaking pattern of the trace-deformed
theory is the same of the standard Yang-Mills theory, the Svetitsky-Yaffe
conjecture~\cite{Svetitsky:1982gs} suggests that, when the phase transition is
continuous, the critical exponents should be those of the 2D Ising universality
class: $\nu\!=\!1$, $\omega\!=\!2$ (see, e.g., \cite{condmat0012164}). We thus start at a small value of $h$,
expecting to recover 2D Ising scaling, then increase $h$ to look for
distinctive signs of a discontinuous critical line, such as phase coexistence
(see Sec.~\ref{sec:appB_highT}).

\subsection{Compatibility with the 2D Ising universality class at small $h$}\label{sec:appB_lowT}

\begin{figure}[htbp]
    \begin{center}
        \includegraphics[width=0.70\textwidth]{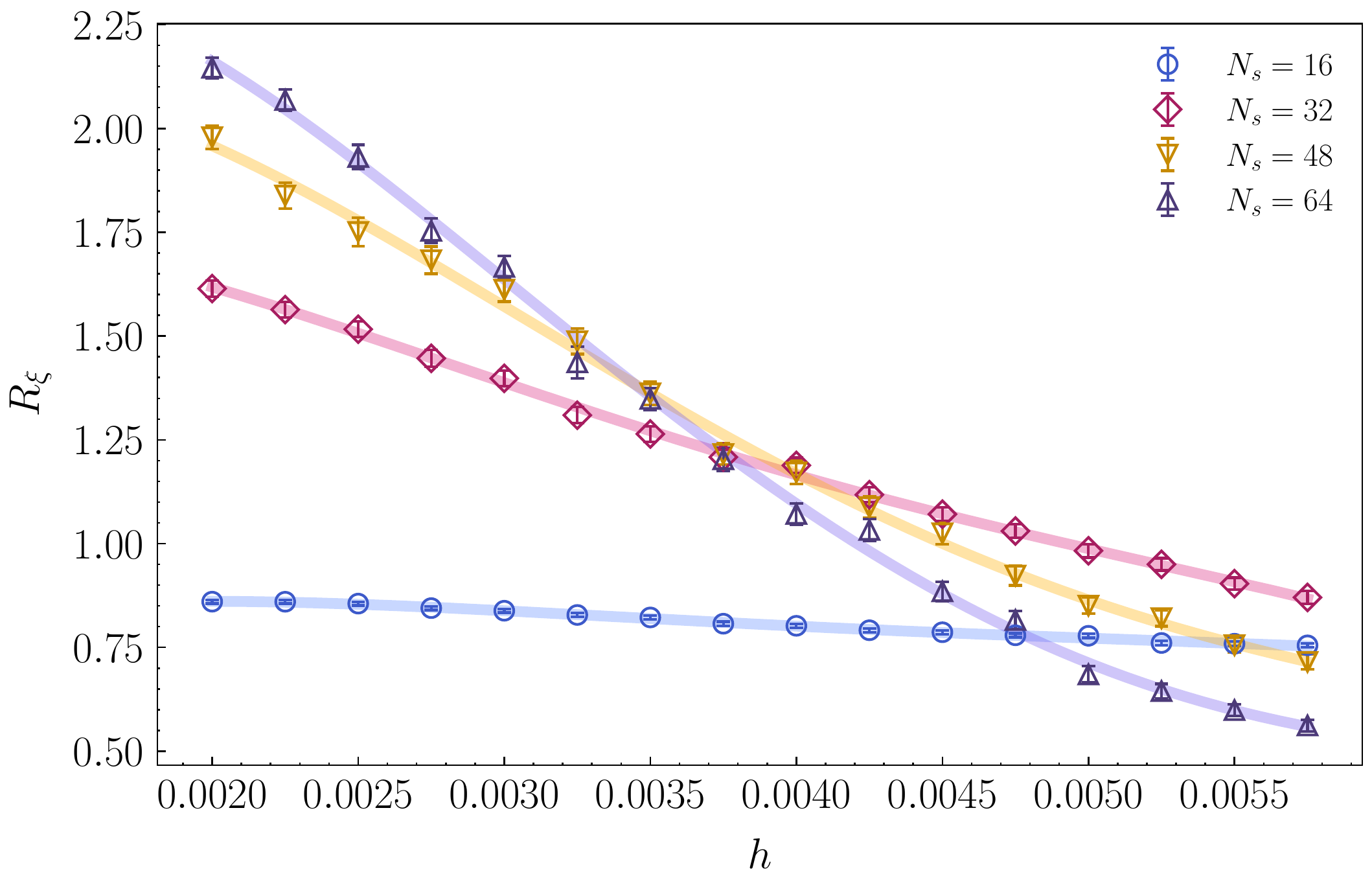}
    \end{center}
    \caption{Means, errors and multi-histogram \cite{Ferrenberg:1988yz} interpolation of $R_{\xi }$ as $h$ is taken across the reconfinement phase transition, at $N_{t}=10$. The plot suggests that $N_s=16$ might be too small, and the scaling corrections might be too large for such a small volume to make any meaningful FSS analysis.}
    \label{fig:Rxi_h}
\end{figure}

\begin{figure}[htbp]
    \begin{center}
        \includegraphics[width=0.70\textwidth]{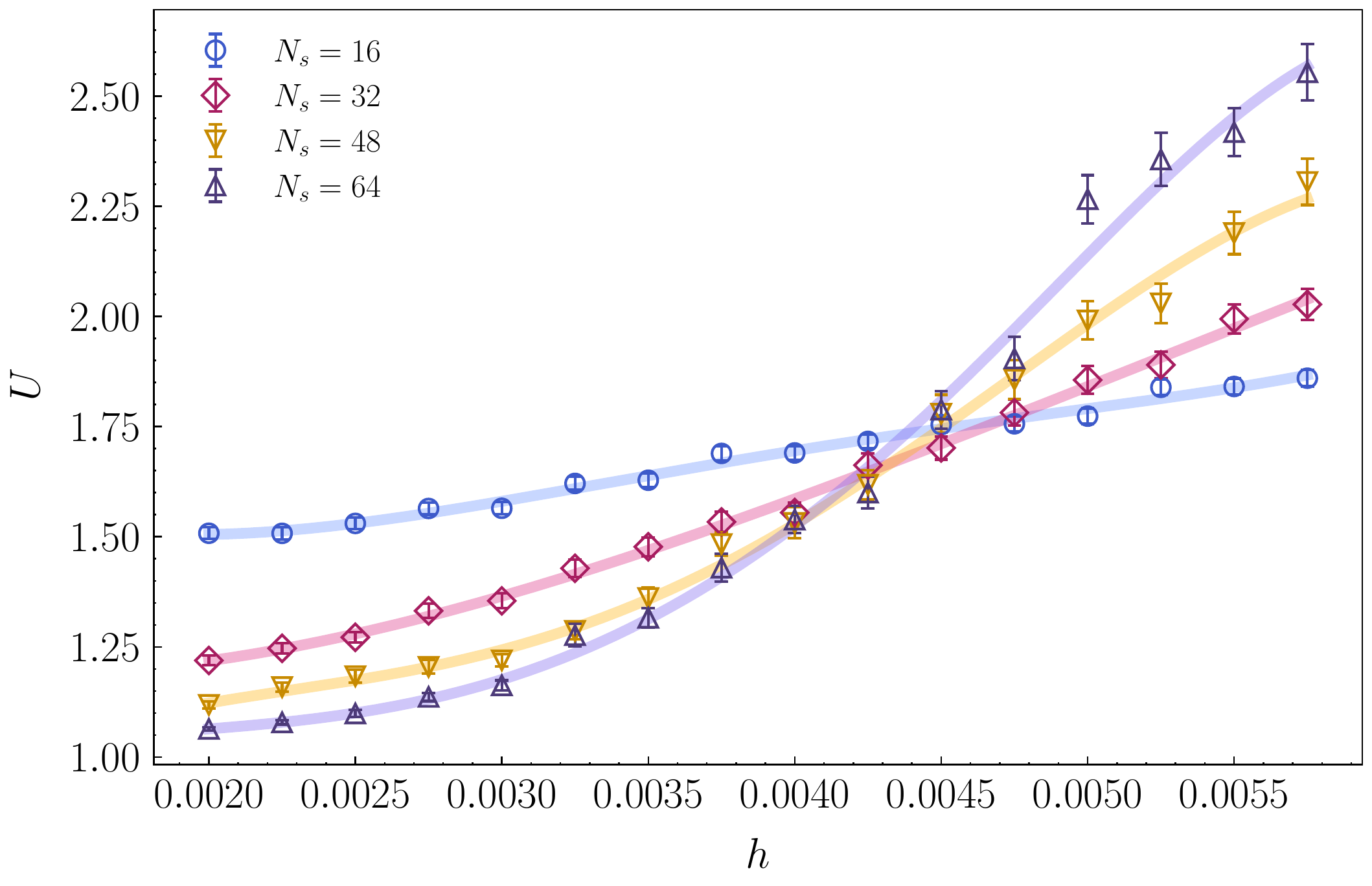}
    \end{center}
    \caption{Means, errors and multi-histogram interpolation of $U$ as $h$ is taken across the reconfinement phase transition, at $N_{t}=10$. As for $R_{\xi }$, the $N_s=16$ volume appears too small to be included in a meaningful FSS analysis.}
    \label{fig:U_h}
\end{figure}

We probed the system at $N_t\!\lesssim\!N_{t,c}$, where $N_{t,c}$ is the
standard 2D $\SU (2)$ confinement/deconfinement phase transition critical
radius. In this subsection we focus on $N_{t}=10$, the case at the lower
temperature among the two we studied in this work.
The higher-temperature case, $N_t=6$, in which the phase-coexistence
phenomenon typical of a discontinuous phase transition has been observed, is
discussed in Sec.~\ref{sec:appB_highT}.
In Figs.~\ref{fig:Rxi_h} and~\ref{fig:U_h} we plot the measurements of
$R_{\xi}$ and $U$ respectively, varying $h$ across the reconfinement
transition, for several values of the lattice spatial volume $N_s$.
It is clear from these data, see in particular Fig.~\ref{fig:Rxi_h}, that large corrections to scaling 
are present for the lattice with $N_s\!=\!16$.

\begin{figure}[h!]
    \begin{center}
        \includegraphics[width=0.70\textwidth]{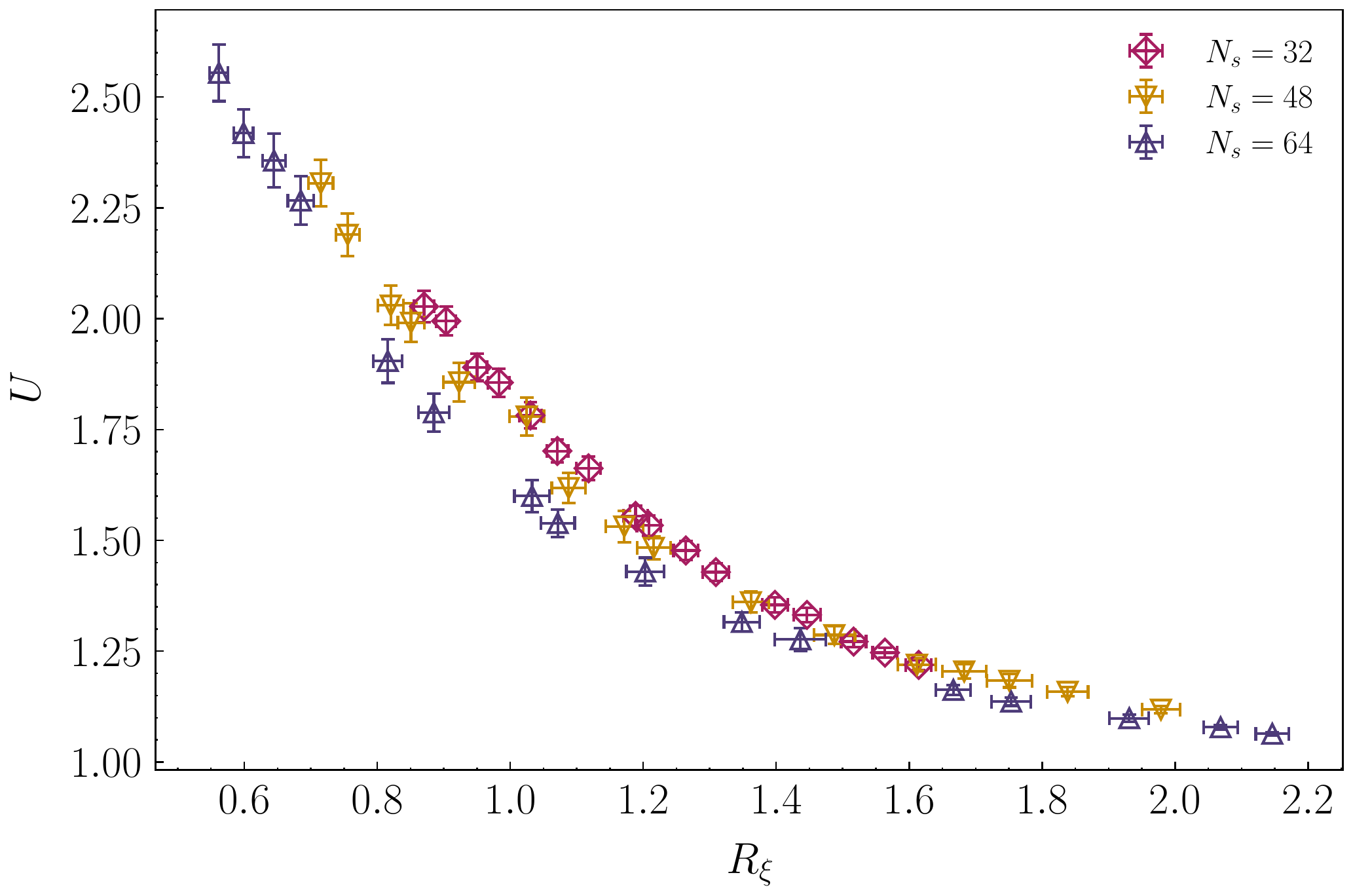}
    \end{center}
    \caption{Universal scaling curve, showing the Binder cumulant as a function of the second-moment correlation length for $N_{t} = 10$.}
    \label{fig:U_xi_lowT}
\end{figure}

We fit the curves with the FSS ansatz
\begin{equation}
    \mathcal{F}(h;h_c)=f((h-h_{c})N_{s}^{1 / \nu })+g((h-h_{c})N_{s}^{1 / \nu })N_{s}^{-\omega }\ ,
\end{equation}
where $\mathcal{F}$ denotes a generic RG
invariant quantity (which in our case can be either $U$ or $R_{\xi}$) and
$f,g$ are polynomials in $x=(h-h_{c})N_{s}^{1/\nu}$ of maximum degrees five and
two respectively. To control systematic errors we either fix both $\nu$ and
$\omega$ to the 2D Ising values, or fix $\omega\!=\!2$ and fit $\nu$; each fit
is performed taking into account or
excluding from the fit data obtained on the smallest available volume. The
results are collected in Tab.~\ref{tab:fitlowT}.
Taking into account the systematics of the
fitting procedure, we quote as final values for the critical reconfinement
parameter and the correlation-length exponent $h_c = 0.0035(5)$ and $\nu =
0.9(1)$.

\begin{table}[!h]
\centering
\begin{tabular}{|c|c|c|c|c|c|}
\hline
$\mathcal{F}$ & $h_c$ & $\nu$ & $\omega$ & $N_s$ & $\chi^{2}/\text{dof}$ \\ \hline
$R_{\xi}$ & $0.00331(4)$  & \multirow{4}{*}{$1$} & \multirow{8}{*}{$2$} & \multirow{2}{*}{$16,32,48,64$} & $1.01$ \\ \cline{1-2}\cline{6-6}
$U$       & $0.00411(6)$  &                      &                      &                                & $1.04$ \\ \cline{1-2}\cline{5-6}
$R_{\xi}$ & $0.00306(13)$ &                      &                      & \multirow{2}{*}{$32,48,64$}    & $0.73$ \\ \cline{1-2}\cline{6-6}
$U$       & $0.00396(11)$ &                      &                      &                                & $1.10$ \\ \cline{1-3}\cline{5-6}
$R_{\xi}$ & $0.00335(4)$  & $0.86(4)$            &                      & \multirow{2}{*}{$16,32,48,64$} & $0.80$ \\ \cline{1-3}\cline{6-6}
$U$       & $0.00407(5)$  & $0.86(3)$            &                      &                                & $1.04$ \\ \cline{1-3}\cline{5-6}
$R_{\xi}$ & $0.00307(16)$ & $0.98(16)$           &                      & \multirow{2}{*}{$32,48,64$}    & $0.75$ \\ \cline{1-3}\cline{6-6}
$U$       & $0.00409(14)$ & $0.90(7)$            &                      &                                & $1.07$ \\ \hline
\end{tabular}
\caption{Fit results at $N_{t}=10$ for the critical reconfinement value $h_{c}$ from the RG-invariant observables $\mathcal{F}=R_{\xi}=\xi/N_s$ and $\mathcal{F}=U$. The first four rows keep both $\nu$ and $\omega$ fixed at the 2D Ising values; the last four keep only $\omega$ fixed.}
\label{tab:fitlowT}
\end{table}

Even though large scaling corrections are present for data with $h>0$, the curves obtained seem to approach, as $N_s\to\infty$, the one corresponding to the $h=0$ case, as shown in Fig.~\ref{fig:U_xi_lowT}.   

\begin{figure}[h!]
    \begin{center}
        \includegraphics[width=0.70\textwidth]{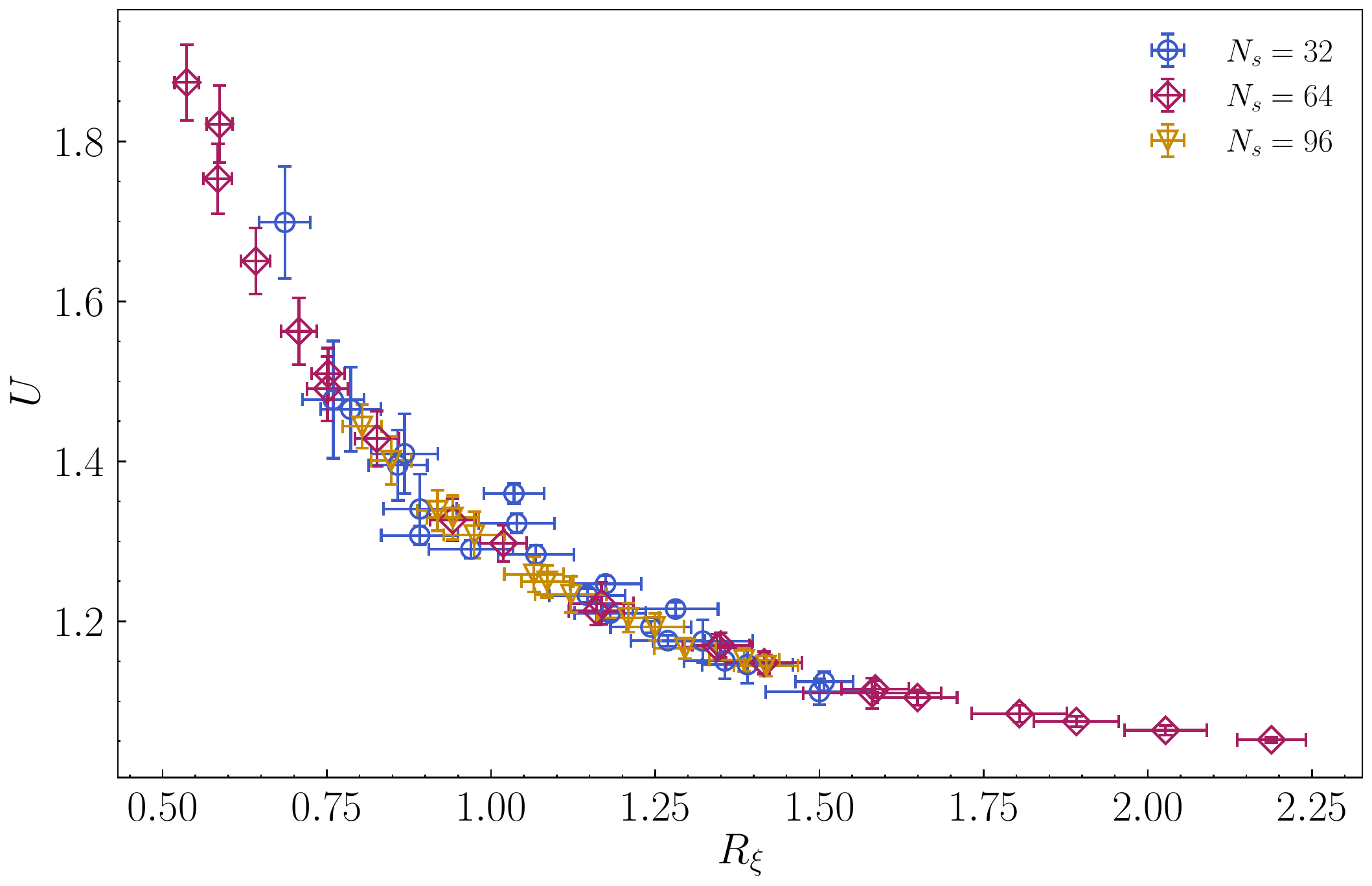}
    \end{center}
    \caption{Universal scaling curve $U(R_{\xi })$ for the non-trace deformed theory at $h=0$ and $N_t=5$, swept across the standard $\SU(2)$ deconfinement transition.}
    \label{fig:U_xi_h0_Nt5}
\end{figure}

We then benchmarked the same $U(R_{\xi })$ analysis against the well-understood
$h=0$ deconfinement transition of pure $\SU(2)$ Yang-Mills, where the
Svetitsky-Yaffe conjecture is firmly established and the transition is known to
belong to the 2D Ising universality class. We worked at $N_t=5$, sweeping
$\beta$ across the critical point at three spatial volumes $N_s\in
\{32,64,96\}$, and computed the same Binder cumulant and second-moment
correlation length. The resulting universal scaling curve is shown in
Fig.~\ref{fig:U_xi_h0_Nt5}, and is to be compared with
Fig.~\ref{fig:U_xi_lowT}: at $h=0$ the curves at different volumes collapse as
expected, exhibiting the same qualitative behavior as the small-$h$ data at
$N_{t}=10$. The close resemblance of the two scaling curves furnishes an
independent cross-check of the universality assignment of the small-$h$ branch
of the reconfinement transition.

\subsection{Phase coexistence and multimodality as $h$ is increased}\label{sec:appB_highT}

After the $N_{t}=10$ analysis of Sec.~\ref{sec:appB_lowT} we moved to higher
temperatures, focusing on $N_{t}=6$, and pushed $h$ to larger values. A
qualitatively different pattern already emerges from finite-size scaling:
repeating in this case the same procedure discussed in
Sec.~\ref{sec:appB_lowT}, we get the results reported in Tab.~\ref{tab:fitlowT_Nt6}

\begin{table}[!h]
\centering
\begin{tabular}{|c|c|c|c|c|c|}
\hline
$\mathcal{F}$ & $h_c$ & $\nu$ & $\omega$ & $N_s$ & $\chi^{2}/\text{dof}$ \\ \hline
$R_{\xi}$ & $0.00769(14)$ & \multirow{2}{*}{$1$} & \multirow{4}{*}{$2$} & \multirow{6}{*}{$32,50,80$} & $5.07$ \\ \cline{1-2}\cline{6-6}
$U$       & $0.00919(8)$  &                      &                      &                              & $4.53$ \\ \cline{1-3}\cline{6-6}
$R_{\xi}$ & $0.00761(5)$  & $0.42(2)$            &                      &                              & $1.45$ \\ \cline{1-3}\cline{6-6}
$U$       & $0.00766(4)$  & $0.53(3)$            &                      &                              & $1.73$ \\ \cline{1-4}\cline{6-6}
$R_{\xi}$ & $0.00764(3)$  & $0.32(2)$            & $6.0(15)$            &                              & $0.85$ \\ \cline{1-4}\cline{6-6}
$U$       & $0.00784(7)$  & $0.53(4)$            & $6.9(18)$            &                              & $1.24$ \\ \hline
\end{tabular}
\caption{Same as Tab.~\ref{tab:fitlowT} but at $N_{t}=6$, with $N_{s}\in\{32,50,80\}$. The last two rows leave both $\nu$ and $\omega$ free.}
\label{tab:fitlowT_Nt6}
\end{table}

Imposing the Ising values $\nu=1$, $\omega=2$ yields $\chi^{2}/\text{dof}\sim
5$; fitting the value of the critical exponent $\nu$
instead returns $\nu\simeq 0.4$--$0.5\approx 1/d = 1/2$, the value expected at
a first order transition. Combining the fits we quote $h_c=0.0081(8)$ and
$\nu=0.45(10)$ as the final $N_{t}=6$ estimates.

This led us to look directly for signals of metastability, such as hot/cold
dependence, plateaus separated by rare tunnelling events, bimodal distributions.

\begin{figure}[h]
    \begin{center}
        \includegraphics[width=0.98\textwidth]{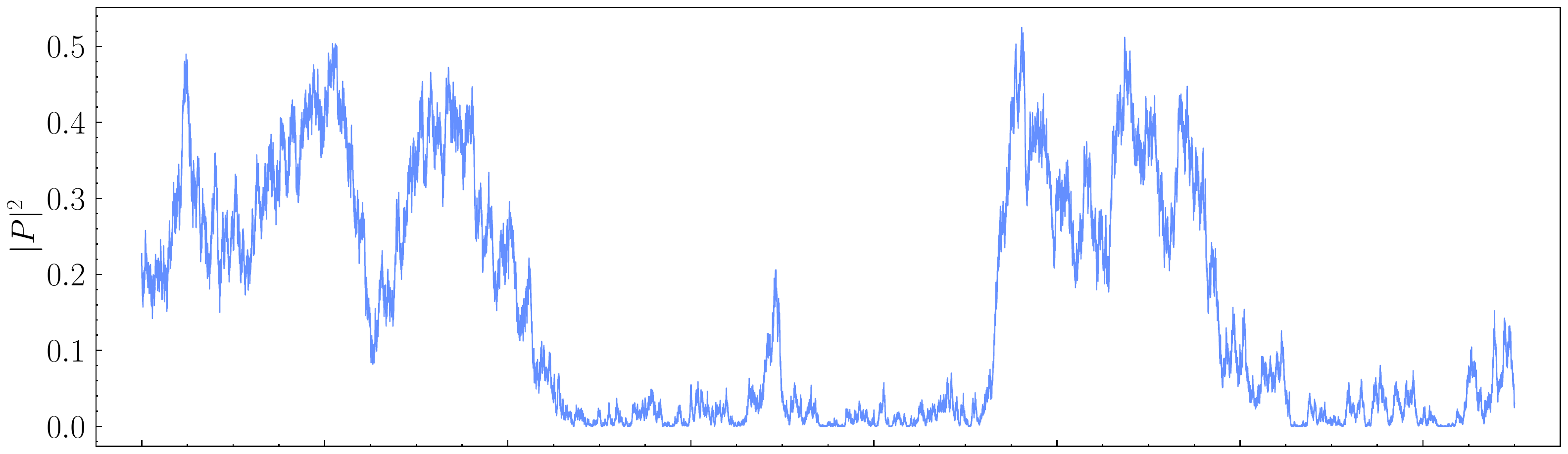}
        \includegraphics[width=0.98\textwidth]{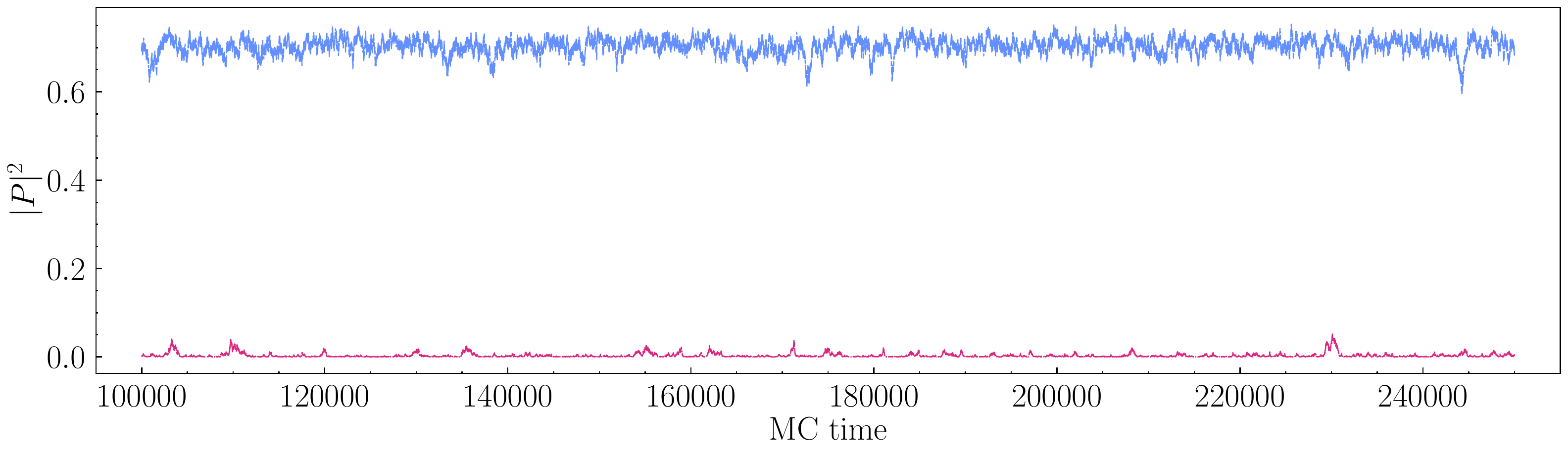}
    \end{center}
    \caption{Metastable phases at two points on the phase diagram.
             Top panel: $\left( N_t=6,h=0.0076 \right)$ with $N_s = 80$.
             Bottom panel: $\left( N_t=3, h=0.0028 \right)$ with $N_s = 96$, with two different initializations.
             In blue (with large $|P|^2$) ordered start, in red (with small $|P|^2$) random start.}
    \label{fig:ff1}
\end{figure}

Fig.~\ref{fig:ff1} shows MC histories at two points near the transition. At
$\left( N_t=6, h=0.0076 \right)$ the system flip-flops between two phases with
small and large $|P|^2$ (we show data for $N_s=80$ to have a reasonable tunnelling time). At
$\left( N_t=3, h=0.0028 \right)$ two apparently stable phases (with $N_s = 96$)
are observed depending on the initialization.

\begin{figure}[h!]
    \centering
        \centering
        \includegraphics[width=0.7\textwidth]{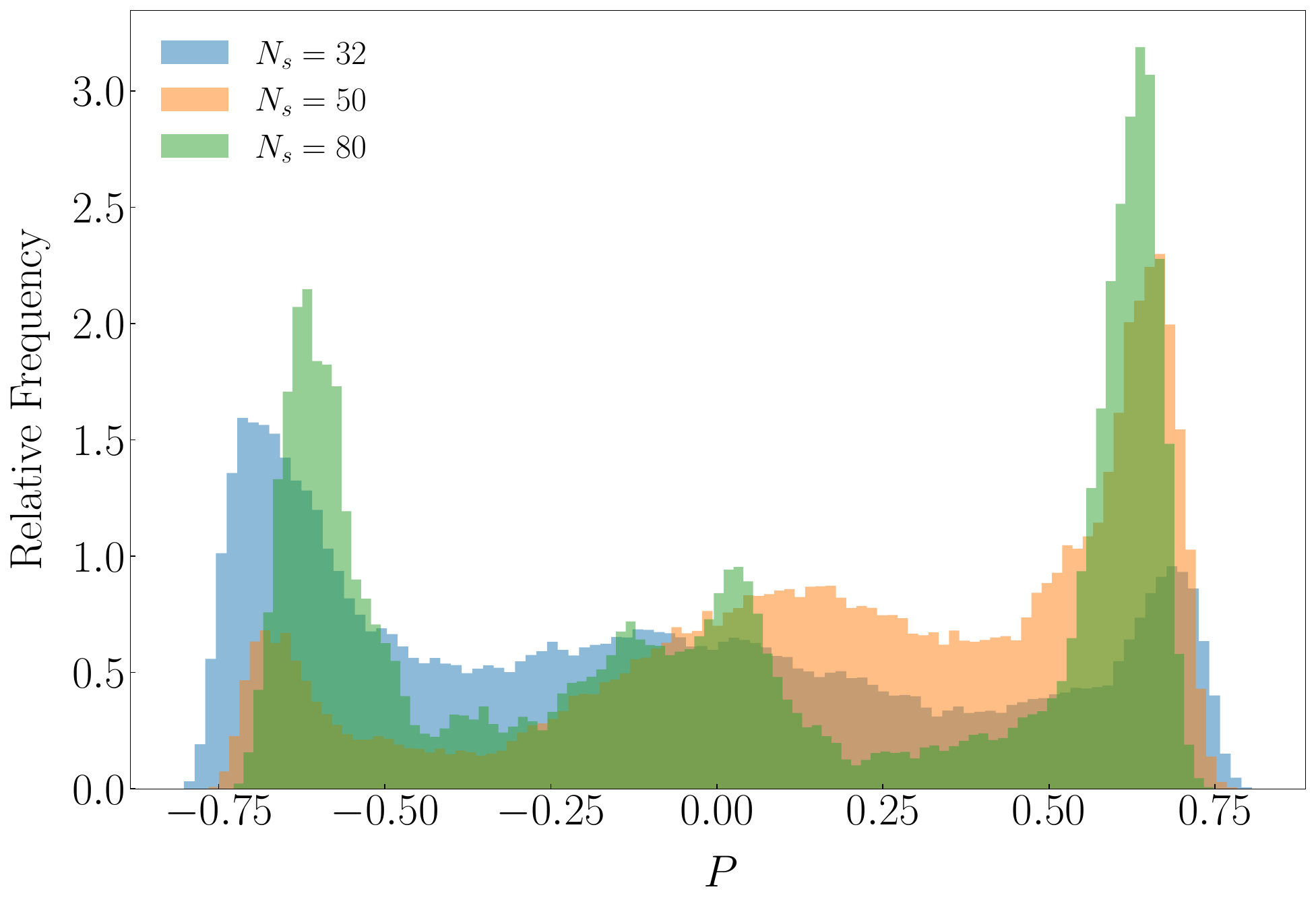}
    \caption{Polyakov-loop histogram at $\left( N_t\!=\!6,\,h\!=\!0.0075 \right)$.}
    \label{fig:P2histo}
\end{figure}

Moreover, for values of $h$ around the pseudocritical point, the Polyakov-loop
histogram develops three peaks (Fig.~\ref{fig:P2histo}), again pointing to
phase coexistence.

\begin{figure}[htb!]
    \begin{center}
        \includegraphics[width=0.7\textwidth]{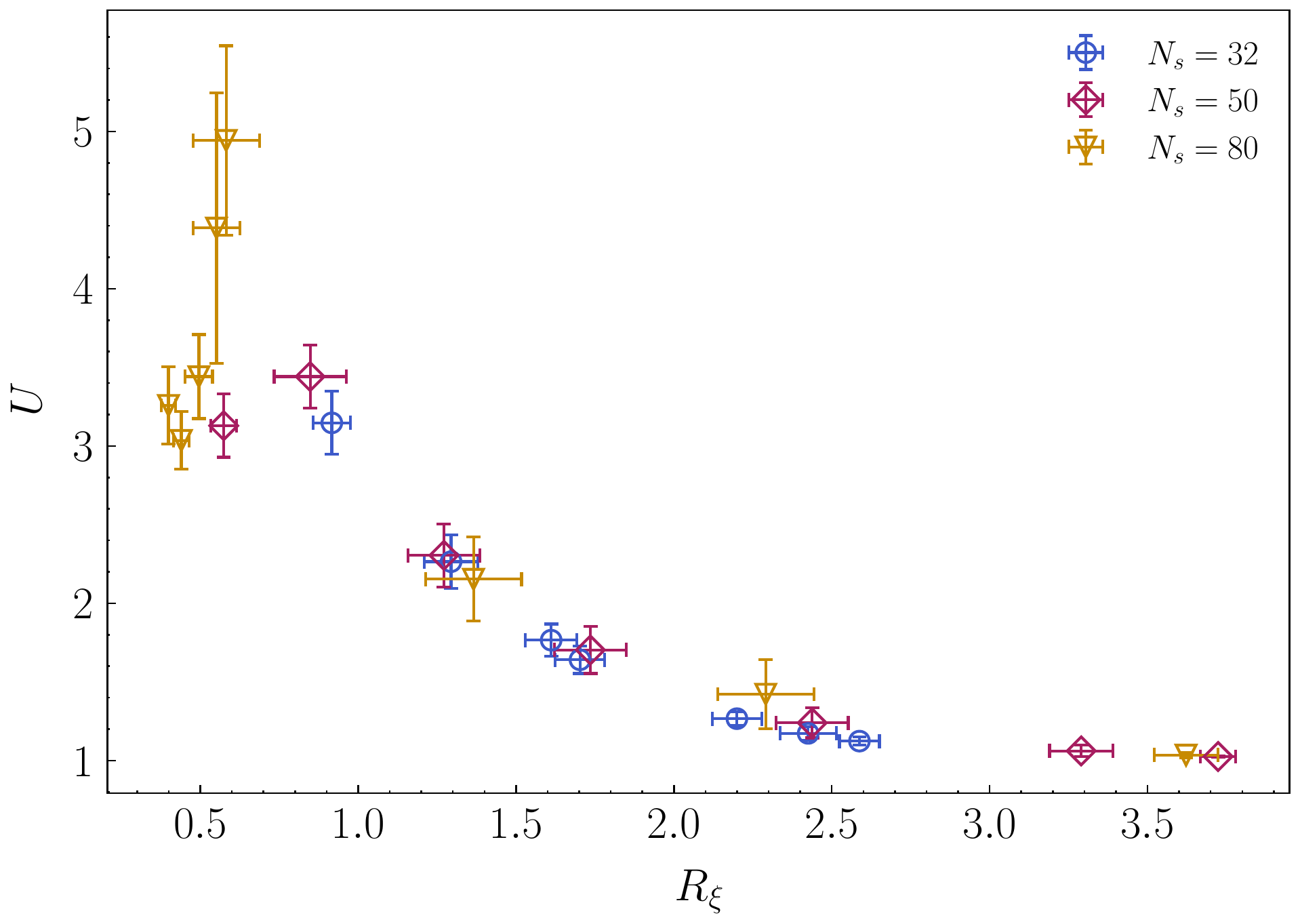}
    \end{center}
    \caption{Universal scaling curve $U(R_{\xi })$ for $N_t\!=\!6$, $\beta = 23.3805$.}
    \label{fig:U_xi_fo}
\end{figure}

Finally, Fig.~\ref{fig:U_xi_fo} shows the Binder cumulant as a function of the second-moment correlation length. Modelling the Polyakov-loop distribution as a central Gaussian of variance $\sigma^2/V$ and weight $w$ together with two ordered peaks of total weight $1-w$ at $x=\pm x_0$, a short calculation gives $U_{\max}\simeq (x_0^2/4\sigma^2)\,V\propto N_s^2$, attained at $1-w\sim 1/V$ (see e.g.~\cite{Challa:1986fo}). Since $\xi$ stays finite, $R_{\xi}=\xi/N_s\to 0$, so in the $U(R_\xi)$ plane the large-$U$ branch is pushed toward $R_\xi\simeq 0$ and grows unboundedly with $V$, in this sense $U(R_\xi)$ diverges at a first order transition.

\subsection{Conclusion}\label{sec:appB_concl}
In this appendix we carried out an exploratory study of the phase diagram of the trace deformed $\SU(2)$ theory.

At small $h$ and $N_{t}=10$, close to the pure, non-trace deformed Yang-Mills regime, where the deconfinement transition is known to be second order in the 2D Ising universality class, the crossing pattern and the approximate collapse of $U(R_\xi)$ are compatible with the reconfinement transition retaining the same continuous, Ising-like character. The finite-size-scaling ansatz yields $h_c=0.0035(5)$ and $\nu=0.9(1)$, consistent with the 2D Ising values within uncertainties. The sizeable scaling corrections prevent a sharp characterization.

At larger $h$ and higher temperature ($N_{t}=6$) the Ising exponents become incompatible with the data ($\chi^2/\text{dof}\sim 5$), while unconstrained fits drive $\nu$ towards the first order value $1/d=1/2$. In parallel dependence on hot/cold starts, long plateaus separated by rare tunnellings, multi-peaked Polyakov-loop distributions, and the volume-divergent Binder peak in the $U(R_\xi)$ plane consistently point to a discontinuous (first order) transition.

Taken together, these results support a phase-diagram scenario in which the critical line is continuous at small $h$ and becomes discontinuous beyond a tricritical point $h^\star$, in qualitative agreement with the change of effective-string description discussed in the main text.

%% file: phase_diagram_sketch.tex
\begin{tikzpicture}[scale=1.0, transform shape]

    \draw[thick,-{Latex[length=3mm]}] (0,0) -- (7.2,0)
        node[midway, below=6pt] {$h$};

    \draw[thick,-{Latex[length=3mm]}] (0,0) -- (0,5.2)
        node[midway, left=8pt] {$N_t$};

    \node[below left] at (0,0) {$0$};

    \coordinate (P2)  at (1.0,4.65);
    \coordinate (P3)  at (1.65,2.45);
    \coordinate (P4)  at (2.25,1.35);
    \coordinate (T)   at (2.70,1.05);
    \coordinate (P5)  at (2.85,0.95);
    \coordinate (P7)  at (4.05,0.60);
    \coordinate (P8)  at (4.65,0.50);
    \coordinate (P9)  at (5.25,0.43);
    \coordinate (P10) at (5.85,0.38);

    \draw[very thick]
        plot[smooth] coordinates {(P2) (P3) (P4) (T)};

    \draw[very thick, dashed]
        plot[smooth] coordinates {(T) (P5) (P7) (P8) (P9) (P10)};

    \coordinate (TEnd) at ($(T) + (50:3.4)$);
    \foreach \i in {0,1,...,13} {
        \pgfmathsetmacro\opa{1.0 - 0.9*max(\i-5,0)/8}
        \pgfmathsetmacro\fa{\i/14}
        \pgfmathsetmacro\fb{(\i+1)/14}
        \draw[black, line width=1.0pt, opacity=\opa, dotted]
            ($(T)!\fa!(TEnd)$) -- ($(T)!\fb!(TEnd)$);
    }

    \fill[white] (T) circle (8pt);
    \draw[magenta, thick, fill=white, pattern=north east lines, pattern color=magenta, draw=none]
    (T) circle (8pt);

    \node at (4.2,1.35) {PY};
    \node at (2.35,3.9) {BNG};

    \node at (2.5,3.5) {second order};
    \node at (4.75,0.95) {first order};

\end{tikzpicture}

%% file: appendixB.tex
\section{Fit results}\label{sec:appendixB}

Here we report the fit results from the first stage of our analysis, namely the
fits to the Polyakov-loop correlators $G(R)$ and to the flux-tube profiles, at
the different values of $N_t$, $\beta$ and $h$ considered in this work.

We devote particular attention to the first ten tables, Tabs.~\ref{tab:E0_beta23_h000}--\ref{tab:E0_beta27_h005}, which
contain the fits to the Polyakov-loop correlators.
For fixed bare parameters the correlators at different separations $R$ are
measured on the same gauge configurations, and are therefore strongly
cross-correlated. This effect is further enhanced by the trace deformation term
in the action, where the Polyakov loop appears explicitly. Fits to highly
correlated data are known to be delicate, see Ref.~\cite{Michael:1993yj} for a
standard discussion. More recently, Ref.~\cite{Bruno:2022mfy} addressed fits to
correlated and autocorrelated data, including the problem of assigning a
goodness of fit to modified or uncorrelated fitting prescriptions. We also refer
to Ref.~\cite{Yoon:2011wdw} for practical strategies in lattice-QCD covariance
fits, including covariance-matrix modifications, cutoffs, Bayesian constraints
and eigenmode-based prescriptions.

The difficulty can be traced to the correlation matrix. Its estimator is noisy,
and in highly correlated data most eigenvalues are very close to zero, with large
relative uncertainties. When the matrix is inverted, these poorly determined
directions are amplified and can dominate the correlated $\chi^2$, making the
minimization unstable and the resulting goodness-of-fit unreliable. We therefore
adopted the simplest robust strategy: we performed the central fits using a
diagonal covariance matrix, while accounting for the cross-correlations through
a blocked bootstrap analysis.

This procedure gives stable parameter estimates and reliable uncertainties, but
the diagonal-fit $\chi^2$ no longer has its usual statistical interpretation. In
our data these values were typically of order $10^{-3}$ to $10^{-4}$, with
bootstrap fluctuations of order $50\%$ to $100\%$ of the central value. We also
tested a fully correlated estimate of the $\chi^2$, building the covariance
matrix from blocked time series, in order to account for autocorrelations, and
using residuals from the uncorrelated fit. This did not lead to a stable
diagnostic either, as the resulting $\chi^2$ values fluctuated strongly with
the blocking size, namely with how much autocorrelation, in units of
$\tau_{\rm int}$, was included (a contributing factor is likely also related to the proximity to the critical
region and the very fine lattice spacings considered here). Also an analysis with the $\Gamma$ method led
to violent fluctuations of the $\chi^2$, depending on the ``windowing strategy''. In particular
different estimates of the ratio between the exponential and integrated autocorrelation time 
(whose determination from the data is very noisy)
 used in input to the standard determination of the integration window, as described in 
Ref.~\cite{Wolff:2003sm}, led to drastic differences in the $\chi^2$. For this reason we do not report $\chi^2$ values
in Tabs.~\ref{tab:E0_beta23_h000}--\ref{tab:E0_beta27_h005}, since they would not provide additional physical
information.

\begin{table}[h]
    \centering
    \begin{adjustbox}{max width=\textwidth}
    \begin{tabular}{|c|c|c|}
        \hline
        $(N_t\times N_s^2,h)$ & $A$ & $E_0$ \\ \hline
        $(16\times96^2,0.000)$ & 0.02684(62) & 0.00879(44) \\ \hline
        $(17\times96^2,0.000)$ & 0.02888(63) & 0.01492(58) \\ \hline
        $(18\times96^2,0.000)$ & 0.02950(61) & 0.02120(68) \\ \hline
        $(19\times96^2,0.000)$ & 0.03004(61) & 0.02827(78) \\ \hline
        $(20\times96^2,0.000)$ & 0.03063(64) & 0.03573(88) \\ \hline
        $(21\times96^2,0.000)$ & 0.03038(63) & 0.04188(92) \\ \hline
    \end{tabular}
    \end{adjustbox}
    \caption{Results from the fit of $G(R)$ with Eq.~\eqref{eq:fS} for $\beta = 23.3805$ and $h = 0$.}
    \label{tab:E0_beta23_h000}
\end{table}

\begin{table}[h]
    \centering
    \begin{adjustbox}{max width=\textwidth}
    \begin{tabular}{|c|c|c|}
        \hline
        $(N_t\times N_s^2,h)$ & $A$ & $E_0$ \\ \hline
        $(15\times96^2,0.001)$ & 0.02802(60) & 0.00923(43) \\ \hline
        $(16\times96^2,0.001)$ & 0.02974(56) & 0.01520(52) \\ \hline
        $(17\times96^2,0.001)$ & 0.03089(55) & 0.02186(60) \\ \hline
        $(18\times96^2,0.001)$ & 0.03121(62) & 0.02838(78) \\ \hline
        $(19\times96^2,0.001)$ & 0.03194(65) & 0.03558(86) \\ \hline
        $(20\times96^2,0.001)$ & 0.03122(65) & 0.04074(93) \\ \hline
        $(21\times96^2,0.001)$ & 0.03017(65) & 0.0459(10) \\ \hline
    \end{tabular}
    \end{adjustbox}
    \caption{Results from the fit of $G(R)$ with Eq.~\eqref{eq:fS} for $\beta = 23.3805$ and $h = 0.001$.}
    \label{tab:E0_beta23_h001}
\end{table}

\begin{table}[h]
    \centering
    \begin{adjustbox}{max width=\textwidth}
    \begin{tabular}{|c|c|c|}
        \hline
        $(N_t\times N_s^2,h)$ & $A$ & $E_0$ \\ \hline
        $(13\times96^2,0.002)$ & 0.02692(59) & 0.00516(30) \\ \hline
        $(14\times96^2,0.002)$ & 0.03049(62) & 0.01153(48) \\ \hline
        $(15\times96^2,0.002)$ & 0.03231(62) & 0.01834(61) \\ \hline
        $(16\times96^2,0.002)$ & 0.03260(66) & 0.02399(73) \\ \hline
        $(17\times96^2,0.002)$ & 0.03284(63) & 0.02997(77) \\ \hline
        $(18\times96^2,0.002)$ & 0.03189(62) & 0.03436(83) \\ \hline
        $(19\times96^2,0.002)$ & 0.03269(64) & 0.04180(89) \\ \hline
        $(20\times96^2,0.002)$ & 0.03162(65) & 0.04595(96) \\ \hline
        $(21\times96^2,0.002)$ & 0.03194(67) & 0.0524(10) \\ \hline
    \end{tabular}
    \end{adjustbox}
    \caption{Results from the fit of $G(R)$ with Eq.~\eqref{eq:fS} for $\beta = 23.3805$ and $h = 0.002$.}
    \label{tab:E0_beta23_h002}
\end{table}

\begin{table}[h]
    \centering
    \begin{adjustbox}{max width=\textwidth}
    \begin{tabular}{|c|c|c|}
        \hline
        $(N_t\times N_s^2,h)$ & $A$ & $E_0$ \\ \hline
        $(12\times96^2,0.003)$ & 0.03054(62) & 0.00825(40) \\ \hline
        $(13\times96^2,0.003)$ & 0.03253(60) & 0.01494(53) \\ \hline
        $(14\times96^2,0.003)$ & 0.03309(66) & 0.02070(68) \\ \hline
        $(15\times96^2,0.003)$ & 0.03400(64) & 0.02719(74) \\ \hline
        $(16\times96^2,0.003)$ & 0.03318(66) & 0.03141(83) \\ \hline
        $(17\times96^2,0.003)$ & 0.03374(66) & 0.03774(86) \\ \hline
        $(18\times96^2,0.003)$ & 0.03346(69) & 0.04232(95) \\ \hline
        $(19\times96^2,0.003)$ & 0.03277(70) & 0.0471(10) \\ \hline
        $(20\times96^2,0.003)$ & 0.03159(65) & 0.05031(98) \\ \hline
        $(21\times96^2,0.003)$ & 0.03140(67) & 0.0552(10) \\ \hline
    \end{tabular}
    \end{adjustbox}
    \caption{Results from the fit of $G(R)$ with Eq.~\eqref{eq:fS} for $\beta = 23.3805$ and $h = 0.003$.}
    \label{tab:E0_beta23_h003}
\end{table}

\begin{table}[h]
    \centering
    \begin{adjustbox}{max width=\textwidth}
    \begin{tabular}{|c|c|c|}
        \hline
        $(N_t\times N_s^2,h)$ & $A$ & $E_0$ \\ \hline
        $(10\times96^2,0.004)$ & 0.03104(67) & 0.00611(37) \\ \hline
        $(11\times96^2,0.004)$ & 0.03377(65) & 0.01442(57) \\ \hline
        $(12\times96^2,0.004)$ & 0.03464(61) & 0.02143(64) \\ \hline
        $(13\times96^2,0.004)$ & 0.03464(65) & 0.02635(75) \\ \hline
        $(14\times96^2,0.004)$ & 0.03491(65) & 0.03181(80) \\ \hline
        $(15\times96^2,0.004)$ & 0.03418(64) & 0.03535(82) \\ \hline
        $(16\times96^2,0.004)$ & 0.03509(69) & 0.04089(91) \\ \hline
        $(17\times96^2,0.004)$ & 0.03478(69) & 0.04589(95) \\ \hline
        $(18\times96^2,0.004)$ & 0.03389(70) & 0.04865(98) \\ \hline
        $(19\times96^2,0.004)$ & 0.03279(65) & 0.05157(96) \\ \hline
        $(20\times96^2,0.004)$ & 0.03214(71) & 0.0558(11) \\ \hline
        $(21\times96^2,0.004)$ & 0.03200(73) & 0.0604(11) \\ \hline
    \end{tabular}
    \end{adjustbox}
    \caption{Results from the fit of $G(R)$ with Eq.~\eqref{eq:fS} for $\beta = 23.3805$ and $h = 0.004$.}
    \label{tab:E0_beta23_h004}
\end{table}

\begin{table}[h]
    \centering
    \begin{adjustbox}{max width=\textwidth}
    \begin{tabular}{|c|c|c|}
        \hline
        $(N_t\times N_s^2,h)$ & $A$ & $E_0$ \\ \hline
        $(9\times96^2,0.005)$ & 0.03304(52) & 0.01251(55) \\ \hline
        $(10\times96^2,0.005)$ & 0.03488(45) & 0.02399(61) \\ \hline
        $(11\times96^2,0.005)$ & 0.03567(54) & 0.03059(76) \\ \hline
        $(12\times96^2,0.005)$ & 0.03658(56) & 0.03615(79) \\ \hline
        $(13\times96^2,0.005)$ & 0.03669(53) & 0.03997(75) \\ \hline
        $(14\times96^2,0.005)$ & 0.03619(57) & 0.04264(83) \\ \hline
        $(15\times96^2,0.005)$ & 0.0349(11) & 0.0445(12) \\ \hline
        $(16\times96^2,0.005)$ & 0.0340(14) & 0.0471(14) \\ \hline
        $(17\times96^2,0.005)$ & 0.0331(20) & 0.0499(18) \\ \hline
        $(18\times96^2,0.005)$ & 0.0317(19) & 0.0519(18) \\ \hline
        $(19\times96^2,0.005)$ & 0.0345(23) & 0.0588(21) \\ \hline
        $(20\times96^2,0.005)$ & 0.0323(21) & 0.0601(21) \\ \hline
        $(21\times96^2,0.005)$ & 0.0312(23) & 0.0641(24) \\ \hline

    \end{tabular}
    \end{adjustbox}
    \caption{Results from the fit of $G(R)$ with Eq.~\eqref{eq:fS} for $\beta = 23.3805$ and $h = 0.005$.}
    \label{tab:E0_beta23_h005}
\end{table}

\begin{table}[h]
    \centering
    \begin{adjustbox}{max width=\textwidth}
    \begin{tabular}{|c|c|c|}
        \hline
        $(N_t\times N_s^2,h)$ & $A$ & $E_0$ \\ \hline
        $(8\times96^2,0.006)$ & 0.03291(48) & 0.02199(73) \\ \hline
        $(9\times96^2,0.006)$ & 0.03491(45) & 0.03582(71) \\ \hline
        $(10\times96^2,0.006)$ & 0.03634(45) & 0.04221(68) \\ \hline
        $(11\times96^2,0.006)$ & 0.03675(45) & 0.04503(67) \\ \hline
        $(12\times96^2,0.006)$ & 0.03728(52) & 0.04802(76) \\ \hline
        $(13\times96^2,0.006)$ & 0.03722(52) & 0.05023(76) \\ \hline
        $(14\times96^2,0.006)$ & 0.03761(52) & 0.05323(74) \\ \hline
        $(15\times96^2,0.006)$ & 0.03660(74) & 0.0547(10) \\ \hline
        $(16\times96^2,0.006)$ & 0.03606(77) & 0.0562(11) \\ \hline
        $(17\times96^2,0.006)$ & 0.03569(77) & 0.0589(11) \\ \hline
        $(18\times96^2,0.006)$ & 0.03409(72) & 0.0598(11) \\ \hline
        $(19\times96^2,0.006)$ & 0.03330(72) & 0.0619(11) \\ \hline
        $(20\times96^2,0.006)$ & 0.03318(74) & 0.0660(11) \\ \hline
        $(21\times96^2,0.006)$ & 0.03078(68) & 0.0661(11) \\ \hline
    \end{tabular}
    \end{adjustbox}
    \caption{Results from the fit of $G(R)$ with Eq.~\eqref{eq:fS} for $\beta = 23.3805$ and $h = 0.006$.}
    \label{tab:E0_beta23_h006}
\end{table}

\begin{table}[h]
    \centering
    \begin{adjustbox}{max width=\textwidth}
    \begin{tabular}{|c|c|c|}
        \hline
        $(N_t\times N_s^2,h)$ & $A$ & $E_0$ \\ \hline
        $(7\times96^2,0.007)$ & 0.03142(58) & 0.0299(10) \\ \hline
        $(8\times96^2,0.007)$ & 0.03358(52) & 0.04460(89) \\ \hline
        $(9\times96^2,0.007)$ & 0.03559(54) & 0.05132(87) \\ \hline
        $(10\times96^2,0.007)$ & 0.03685(54) & 0.05579(82) \\ \hline
        $(11\times96^2,0.007)$ & 0.03729(60) & 0.05742(90) \\ \hline
        $(12\times96^2,0.007)$ & 0.03750(53) & 0.05892(79) \\ \hline
        $(13\times96^2,0.007)$ & 0.03780(58) & 0.06050(90) \\ \hline
        $(14\times96^2,0.007)$ & 0.03781(52) & 0.06188(74) \\ \hline
        $(15\times96^2,0.007)$ & 0.03696(75) & 0.0624(10) \\ \hline
        $(16\times96^2,0.007)$ & 0.03688(79) & 0.0643(11) \\ \hline
        $(17\times96^2,0.007)$ & 0.03593(76) & 0.0652(11) \\ \hline
        $(18\times96^2,0.007)$ & 0.03506(77) & 0.0662(11) \\ \hline
        $(19\times96^2,0.007)$ & 0.03398(73) & 0.0683(11) \\ \hline
        $(20\times96^2,0.007)$ & 0.03331(76) & 0.0705(12) \\ \hline
        $(21\times96^2,0.007)$ & 0.03043(69) & 0.0693(11) \\ \hline
    \end{tabular}
    \end{adjustbox}
    \caption{Results from the fit of $G(R)$ with Eq.~\eqref{eq:fS} for $\beta = 23.3805$ and $h = 0.007$.}
    \label{tab:E0_beta23_h007}
\end{table}

\begin{table}[h]
    \centering
    \begin{adjustbox}{max width=\textwidth}
    \begin{tabular}{|c|c|c|}
        \hline
        $(N_t\times N_s^2,h)$ & $A$ & $E_0$ \\ \hline
        $(11\times96^2,0.004)$ & 0.03349(54) & 0.01953(62) \\ \hline
        $(12\times96^2,0.004)$ & 0.0341(10) & 0.02558(96) \\ \hline
        $(13\times96^2,0.004)$ & 0.0326(24) & 0.0285(16) \\ \hline
        $(14\times96^2,0.004)$ & 0.03527(59) & 0.03223(78) \\ \hline
        $(15\times96^2,0.004)$ & 0.03563(59) & 0.03583(86) \\ \hline
        $(16\times96^2,0.004)$ & 0.03488(51) & 0.03664(80) \\ \hline
        $(17\times96^2,0.004)$ & 0.03464(53) & 0.03852(84) \\ \hline
        $(18\times96^2,0.004)$ & 0.03443(50) & 0.04062(81) \\ \hline
        $(19\times96^2,0.004)$ & 0.03390(56) & 0.04270(93) \\ \hline
    \end{tabular}
    \end{adjustbox}
    \caption{Results from the fit of $G(R)$, with Eq.~\eqref{eq:fS} for $\beta = 27.4745$ and $h = 0.004$}
    \label{tab:E0_beta27_h004}
\end{table}

\begin{table}[h]
    \centering
    \begin{adjustbox}{max width=\textwidth}
    \begin{tabular}{|c|c|c|}
        \hline
        $(N_t\times N_s^2,h)$ & $A$ & $E_0$ \\ \hline
        $(9\times96^2,0.005)$ & 0.03266(73) & 0.0333(37) \\ \hline
        $(10\times96^2,0.005)$ & 0.03359(47) & 0.03628(79) \\ \hline
        $(11\times96^2,0.005)$ & 0.03443(69) & 0.0404(11) \\ \hline
        $(12\times96^2,0.005)$ & 0.03549(82) & 0.0431(11) \\ \hline
        $(13\times96^2,0.005)$ & 0.03621(88) & 0.0453(11) \\ \hline
        $(14\times96^2,0.005)$ & 0.03539(97) & 0.0447(11) \\ \hline
        $(15\times96^2,0.005)$ & 0.0355(12) & 0.0473(12) \\ \hline
        $(16\times96^2,0.005)$ & 0.03468(70) & 0.04734(82) \\ \hline
        $(17\times96^2,0.005)$ & 0.03357(72) & 0.04765(82) \\ \hline
    \end{tabular}
    \end{adjustbox}
    \caption{Results from the fit of $G(R)$, with Eq.~\eqref{eq:fS} for $\beta = 27.4745$ and $h = 0.005$}
    \label{tab:E0_beta27_h005}
\end{table}

\begin{table}
    \centering
    \begin{tabular}{|c|c|c|c|}
        \hline
        $R / a$ & $\tilde A$ & $p$ & $\chi^2 / \mathrm{dof}$ \\ \hline
        7  & 0.000852(43) & 2.579(43) & 1.68 \\ \hline
        9  & 0.00109(11)  & 2.430(71) & 1.47 \\ \hline
        11 & 0.00186(36)  & 2.52(12) & 1.07  \\ \hline
        13 & 0.00193(58)  & 2.34(16) & 0.6   \\ \hline
        15 & 0.0034(17)   & 2.50(25) & 1.54  \\ \hline
        17 & 0.0032(22)   & 2.31(31) & 0.7   \\ \hline
        19 & 0.021(25)    & 3.05(53) & 0.76  \\ \hline
        21 & 0.009(13)    & 2.52(63) & 0.91  \\ \hline
        23 & 0.012(22)    & 2.50(76) & 1.34  \\ \hline
    \end{tabular}
    \caption{Fit results for the profile, according to Eq.~\eqref{eq:rho_fitmodel}, for $N_t = 23$ and $h = 0$.}
    \label{tab:fitres_profile_nt23}
\end{table}
        
\begin{table}
    \centering
    \begin{tabular}{|c|c|c|c|}
        \hline
        $R / a$ & $\tilde A$ & $p$ & $\chi^2 / \mathrm{dof}$ \\ \hline
        7  & 0.000954(47)     & 2.672(42) & 1.62 \\ \hline
        9  & 0.00135(14)  & 2.571(70) & 1.29 \\ \hline
        11 & 0.00223(42)  & 2.63(11) & 0.64  \\ \hline
        13 & 0.0039(12)   & 2.74(17) & 0.66  \\ \hline
        15 & 0.0060(29)   & 2.77(24) & 1.15  \\ \hline
        17 & 0.0089(66)   & 2.81(34) & 1.23  \\ \hline
        19 & 0.0066(70)   & 2.55(47) & 1.6   \\ \hline
        21 & 0.020(33)    & 2.92(68) & 1.26  \\ \hline
        23 & 0.10(23)     & 3.45(94) & 0.73  \\ \hline
    \end{tabular}
    \caption{Fit results for the profile, according to Eq.~\eqref{eq:rho_fitmodel}, for $N_t = 21$ and $h = 0.001$.}
    \label{tab:fitres_profile_nt21}
\end{table}
        
\begin{table}
    \centering
    \begin{tabular}{|c|c|c|c|}
        \hline
        $R / a$ & $\tilde A$ & $p$ & $\chi^2 / \mathrm{dof}$ \\ \hline
        7  & 0.001192(58)     & 2.817(42) & 1.48 \\ \hline
        9  & 0.00172(18)  & 2.718(72) & 1.2  \\ \hline
        11 & 0.00253(48)  & 2.70(11) & 1.44  \\ \hline
        13 & 0.0045(15)   & 2.80(18) & 0.88  \\ \hline
        15 & 0.0079(41)   & 2.91(26) & 1.82  \\ \hline
        17 & 0.0096(79)   & 2.87(38) & 1.71  \\ \hline
        19 & 0.00082(81)  & 1.65(44) & 0.35  \\ \hline
        21 & 0.0033(45)   & 2.12(56) & 0.93  \\ \hline
        23 & 0.006(13)    & 2.31(82) & 0.6   \\ \hline
    \end{tabular}
    \caption{Fit results for the profile, according to Eq.~\eqref{eq:rho_fitmodel}, for $N_t = 20$ and $h = 0.002$.}
    \label{tab:fitres_profile_nt20}
\end{table}
        
\begin{table}
    \centering
    \begin{tabular}{|c|c|c|c|}
        \hline
        $R / a$ & $\tilde A$ & $p$ & $\chi^2 / \mathrm{dof}$ \\ \hline
        7  & 0.001529(63)     & 3.039(35) & 1.87 \\ \hline
        9  & 0.00267(24)  & 3.023(62) & 2.14 \\ \hline
        11 & 0.00449(77)  & 3.06(10) & 1.29  \\ \hline
        13 & 0.0072(20)   & 3.07(15) & 1.48  \\ \hline
        15 & 0.0095(41)   & 3.01(21) & 1.13  \\ \hline
        17 & 0.020(13)    & 3.20(31) & 0.55  \\ \hline
        19 & 0.16(16)     & 3.95(45) & 1.04  \\ \hline
        21 & 0.053(65)    & 3.28(52) & 0.88  \\ \hline
        23 & 0.046(73)    & 3.08(65) & 1.03  \\ \hline
    \end{tabular}
    \caption{Fit results for the profile, according to Eq.~\eqref{eq:rho_fitmodel}, for $N_t = 19$ and $h = 0.00405$.}
    \label{tab:fitres_profile_nt19}
\end{table}
        
\begin{table}
    \centering
    \begin{tabular}{|c|c|c|c|}
        \hline
        $R / a$ & $\tilde A$ & $p$ & $\chi^2 / \mathrm{dof}$ \\ \hline
        7 & 0.001594(64)  & 3.051(35) & 1.92 \\ \hline
        9 & 0.00293(26)   & 3.069(62) & 0.46 \\ \hline
        11 & 0.00484(80)  & 3.075(99) & 1.01 \\ \hline
        13 & 0.0044(11)   & 2.79(14) & 1.05  \\ \hline
        15 & 0.0104(45)   & 3.05(21) & 1.17  \\ \hline
        17 & 0.0115(74)   & 2.94(30) & 1.47  \\ \hline
        19 & 0.033(33)    & 3.27(45) & 1.34  \\ \hline
        21 & 0.039(55)    & 3.21(60) & 1.19  \\ \hline
        23 & 0.024(47)    & 2.90(82) & 1.12  \\ \hline
    \end{tabular}
    \caption{Fit results for the profile, according to Eq.~\eqref{eq:rho_fitmodel}, for $N_t = 18$ and $h = 0.00466$.}
    \label{tab:fitres_profile_nt18}
\end{table}
        
\begin{table}
    \centering
    \begin{tabular}{|c|c|c|c|}
        \hline
        $R / a$ & $\tilde A$ & $p$ & $\chi^2 / \mathrm{dof}$ \\ \hline
        7 & 0.001822(74)  & 3.168(35) & 5.12 \\ \hline
        9 & 0.00333(31)   & 3.181(64) & 3.07 \\ \hline
        11 & 0.00485(83)  & 3.11(10) & 1.42  \\ \hline
        13 & 0.0097(29)   & 3.27(16) & 1.47  \\ \hline
        15 & 0.0092(40)   & 3.02(21) & 1.16  \\ \hline
        17 & 0.0134(86)   & 3.03(30) & 0.41  \\ \hline
        19 & 0.017(15)    & 2.98(40) & 1.1   \\ \hline
        21 & 0.0039(48)   & 2.26(51) & 0.69  \\ \hline
        23 & 0.0010(16)   & 1.62(66) & 0.84  \\ \hline
    \end{tabular}
    \caption{Fit results for the profile, according to Eq.~\eqref{eq:rho_fitmodel}, for $N_t = 17$ and $h = 0.00496$.}
    \label{tab:fitres_profile_nt17}
\end{table}
        
\begin{table}
    \centering
    \begin{tabular}{|c|c|c|c|}
        \hline
        $R / a$ & $\tilde A$ & $p$ & $\chi^2 / \mathrm{dof}$ \\ \hline
        7 & 0.002149(89)  & 3.287(36) & 3.69 \\ \hline
        9 & 0.00519(51)   & 3.462(68) & 1.0  \\ \hline
        11 & 0.0107(20)   & 3.56(11) & 1.61  \\ \hline
        13 & 0.0245(79)   & 3.73(17) & 2.23  \\ \hline
        15 & 0.084(46)    & 4.11(27) & 1.8   \\ \hline
        17 & 0.23(20)     & 4.37(41) & 0.96  \\ \hline
        19 & 0.36(45)     & 4.37(57) & 0.92  \\ \hline
        21 & 1.6(3.0)     & 4.82(81) & 1.67  \\ \hline
        23 & 0.09(20)     & 3.48(93) & 0.4   \\ \hline
    \end{tabular}
    \caption{Fit results for the profile, according to Eq.~\eqref{eq:rho_fitmodel}, for $N_t = 16$ and $h = 0.00548$.}
    \label{tab:fitres_profile_nt16}
\end{table}
        
\begin{table}
    \centering
    \begin{tabular}{|c|c|c|c|}
        \hline
        $R / a$ & $\tilde A$ & $p$ & $\chi^2 / \mathrm{dof}$ \\ \hline
        7 & 0.002735(97) & 3.468(31) & 8.91  \\ \hline
        9 & 0.00716(61)  & 3.673(59) & 2.86  \\ \hline
        11 & 0.0153(25)  & 3.768(98) & 1.32  \\ \hline
        13 & 0.0321(91)  & 3.89(15) & 1.7    \\ \hline
        15 & 0.092(43)   & 4.17(24) & 1.29   \\ \hline
        17 & 0.055(34)   & 3.69(30) & 1.11   \\ \hline
        19 & 0.088(79)   & 3.72(40) & 0.84   \\ \hline
        21 & 0.9(1.2)    & 4.54(57) & 1.29   \\ \hline
        23 & 1.4(2.7)    & 4.55(78) & 1.91   \\ \hline
    \end{tabular}
    \caption{Fit results for the profile, according to Eq.~\eqref{eq:rho_fitmodel}, for $N_t = 15$ and $h = 0.00572$.}
    \label{tab:fitres_profile_nt15}
\end{table}
        
\begin{table}
    \centering
    \begin{tabular}{|c|c|c|c|}
        \hline
        $R / a$ & $\tilde A$ & $p$ & $\chi^2 / \mathrm{dof}$ \\ \hline
        7 & 0.00299(10)  & 3.518(31) & 10.55 \\ \hline
        9 & 0.00755(62)  & 3.680(58) & 3.9   \\ \hline
        11 & 0.0237(40)  & 4.01(10) & 2.8    \\ \hline
        13 & 0.077(23)   & 4.34(16) & 1.55   \\ \hline
        15 & 0.129(61)   & 4.32(24) & 1.37   \\ \hline
        17 & 0.140(96)   & 4.13(32) & 1.42   \\ \hline
        19 & 0.24(24)    & 4.18(44) & 0.93   \\ \hline
        21 & 0.37(54)    & 4.20(62) & 0.76   \\ \hline
        23 & 0.25(50)    & 3.89(82) & 1.55   \\ \hline
    \end{tabular}
    \caption{Fit results for the profile, according to Eq.~\eqref{eq:rho_fitmodel}, for $N_t = 14$ and $h = 0.00596$.}
    \label{tab:fitres_profile_nt14}
\end{table}
        
\begin{table}
    \centering
    \begin{tabular}{|c|c|c|c|}
        \hline
        $R / a$ & $\tilde A$ & $p$ & $\chi^2 / \mathrm{dof}$ \\ \hline
        7 & 0.00342(12)  & 3.597(30) & 15.91 \\ \hline
        9 & 0.00984(82)  & 3.840(59) & 6.74  \\ \hline
        11 & 0.0335(57)  & 4.20(10) & 3.81   \\ \hline
        13 & 0.074(22)   & 4.31(16) & 2.1    \\ \hline
        15 & 0.190(95)   & 4.52(25) & 1.05   \\ \hline
        17 & 0.18(13)    & 4.25(34) & 1.42   \\ \hline
        19 & 0.23(24)    & 4.16(46) & 1.0    \\ \hline
        21 & 0.7(1.1)    & 4.48(67) & 1.11   \\ \hline
        23 & 0.22(45)    & 3.84(82) & 1.04   \\ \hline
    \end{tabular}
    \caption{Fit results for the profile, according to Eq.~\eqref{eq:rho_fitmodel}, for $N_t = 13$ and $h = 0.00617$.}
    \label{tab:fitres_profile_nt13}
\end{table}
        
\begin{table}
    \centering
    \begin{tabular}{|c|c|c|c|}
        \hline
        $R / a$ & $\tilde A$ & $p$ & $\chi^2 / \mathrm{dof}$ \\ \hline
        7 & 0.00421(14)  & 3.717(30) & 21.41 \\ \hline
        9 & 0.0142(12)   & 4.044(60) & 8.59  \\ \hline
        11 & 0.0509(89)  & 4.41(11) & 4.12   \\ \hline
        13 & 0.180(58)   & 4.76(18) & 2.71   \\ \hline
        15 & 0.35(18)    & 4.78(26) & 2.17   \\ \hline
        17 & 1.7(1.4)    & 5.27(40) & 2.21   \\ \hline
        19 & 32(44)      & 6.33(63) & 1.36   \\ \hline
        21 & 11(19)      & 5.59(76) & 1.7    \\ \hline
        23 & 63(171)     & 6.1(1.1) & 1.69   \\ \hline
    \end{tabular}
    \caption{Fit results for the profile, according to Eq.~\eqref{eq:rho_fitmodel}, for $N_t = 12$ and $h = 0.00637$.}
    \label{tab:fitres_profile_nt12}
\end{table}

\clearpage

%% file: reconfinement.bbl
\begin{thebibliography}{10}

\bibitem{Eguchi:1982nm}
T.~Eguchi and H.~Kawai, \emph{{Reduction of Dynamical Degrees of Freedom in the
  Large N Gauge Theory}},
  \href{https://doi.org/10.1103/PhysRevLett.48.1063}{\emph{Phys. Rev. Lett.}
  {\bfseries 48} (1982) 1063}.

\bibitem{Bhanot:1982sh}
G.~Bhanot, U.~M. Heller and H.~Neuberger, \emph{{The Quenched Eguchi-Kawai
  Model}}, \href{https://doi.org/10.1016/0370-2693(82)90106-X}{\emph{Phys.
  Lett.} {\bfseries B113} (1982) 47}.

\bibitem{Gross:1982at}
D.~J. Gross and Y.~Kitazawa, \emph{{A Quenched Momentum Prescription for Large
  N Theories}}, \href{https://doi.org/10.1016/0550-3213(82)90278-4}{\emph{Nucl.
  Phys.} {\bfseries B206} (1982) 440}.

\bibitem{Gonzalez-Arroyo:1982hwr}
A.~Gonzalez-Arroyo and M.~Okawa, \emph{{A Twisted Model for Large $N$ Lattice
  Gauge Theory}},
  \href{https://doi.org/10.1016/0370-2693(83)90647-0}{\emph{Phys. Lett. B}
  {\bfseries 120} (1983) 174}.

\bibitem{Gonzalez-Arroyo:1982hyq}
A.~Gonzalez-Arroyo and M.~Okawa, \emph{{The Twisted Eguchi-Kawai Model: A
  Reduced Model for Large N Lattice Gauge Theory}},
  \href{https://doi.org/10.1103/PhysRevD.27.2397}{\emph{Phys. Rev. D}
  {\bfseries 27} (1983) 2397}.

\bibitem{Narayanan:2003fc}
R.~Narayanan and H.~Neuberger, \emph{{Large N reduction in continuum}},
  \href{https://doi.org/10.1103/PhysRevLett.91.081601}{\emph{Phys. Rev. Lett.}
  {\bfseries 91} (2003) 081601}
  [\href{https://arxiv.org/abs/hep-lat/0303023}{{\ttfamily hep-lat/0303023}}].

\bibitem{Gonzalez-Arroyo:2010omx}
A.~Gonzalez-Arroyo and M.~Okawa, \emph{{Large $N$ reduction with the Twisted
  Eguchi-Kawai model}},
  \href{https://doi.org/10.1007/JHEP07(2010)043}{\emph{JHEP} {\bfseries 07}
  (2010) 043} [\href{https://arxiv.org/abs/1005.1981}{{\ttfamily 1005.1981}}].

\bibitem{Unsal:2008ch}
M.~Unsal and L.~G. Yaffe, \emph{{Center-stabilized Yang-Mills theory:
  Confinement and large N volume independence}},
  \href{https://doi.org/10.1103/PhysRevD.78.065035}{\emph{Phys. Rev. D}
  {\bfseries 78} (2008) 065035}
  [\href{https://arxiv.org/abs/0803.0344}{{\ttfamily 0803.0344}}].

\bibitem{Myers:2007vc}
J.~C. Myers and M.~C. Ogilvie, \emph{{New phases of SU(3) and SU(4) at finite
  temperature}}, \href{https://doi.org/10.1103/PhysRevD.77.125030}{\emph{Phys.
  Rev.} {\bfseries D77} (2008) 125030}
  [\href{https://arxiv.org/abs/0707.1869}{{\ttfamily 0707.1869}}].

\bibitem{Dobozy:2026edc}
B.~Dobozy and E.~Poppitz, \emph{{Metamorphosis of fractional instantons on a
  twisted $T^4$ with a double-trace deformation: a numerical study}},
  \href{https://arxiv.org/abs/2606.22078}{{\ttfamily 2606.22078}}.

\bibitem{Poppitz:2021cxe}
E.~Poppitz, \emph{{Notes on Confinement on R3 \texttimes{} S1: From
  Yang\textendash{}Mills, Super-Yang\textendash{}Mills, and QCD (adj) to
  QCD(F)}}, \href{https://doi.org/10.3390/sym14010180}{\emph{Symmetry}
  {\bfseries 14} (2022) 180}
  [\href{https://arxiv.org/abs/2111.10423}{{\ttfamily 2111.10423}}].

\bibitem{Kovtun:2007py}
P.~Kovtun, M.~{\"U}nsal and L.~G. Yaffe, \emph{{Volume independence in large
  N(c) QCD-like gauge theories}},
  \href{https://doi.org/10.1088/1126-6708/2007/06/019}{\emph{JHEP} {\bfseries
  0706} (2007) 019} [\href{https://arxiv.org/abs/hep-th/0702021}{{\ttfamily
  hep-th/0702021}}].

\bibitem{Myers:2009df}
J.~C. Myers and M.~C. Ogilvie, \emph{{Phase diagrams of SU(N) gauge theories
  with fermions in various representations}},
  \href{https://doi.org/10.1088/1126-6708/2009/07/095}{\emph{JHEP} {\bfseries
  0907} (2009) 095} [\href{https://arxiv.org/abs/0903.4638}{{\ttfamily
  0903.4638}}].

\bibitem{Cossu:2009sq}
G.~Cossu and M.~D'Elia, \emph{{Finite size phase transitions in QCD with
  adjoint fermions}},
  \href{https://doi.org/10.1088/1126-6708/2009/07/048}{\emph{JHEP} {\bfseries
  0907} (2009) 048} [\href{https://arxiv.org/abs/0904.1353}{{\ttfamily
  0904.1353}}].

\bibitem{Athenodorou:2020clr}
A.~Athenodorou, M.~Cardinali and M.~D'Elia, \emph{{Spectrum of trace deformed
  Yang-Mills theories}},
  \href{https://doi.org/10.1103/PhysRevD.104.074510}{\emph{Phys. Rev. D}
  {\bfseries 104} (2021) 074510}
  [\href{https://arxiv.org/abs/2010.03618}{{\ttfamily 2010.03618}}].

\bibitem{Bonati:2020lal}
C.~Bonati, M.~Cardinali, M.~D'Elia, M.~Giordano and F.~Mazziotti,
  \emph{{Reconfinement, localization and thermal monopoles in $SU(3)$
  trace-deformed Yang-Mills theory}},
  \href{https://doi.org/10.1103/PhysRevD.103.034506}{\emph{Phys. Rev. D}
  {\bfseries 103} (2021) 034506}
  [\href{https://arxiv.org/abs/2012.13246}{{\ttfamily 2012.13246}}].

\bibitem{Bonati:2018rfg}
C.~Bonati, M.~Cardinali and M.~D'Elia, \emph{{$\theta$ dependence in trace
  deformed $SU(3)$ Yang-Mills theory: a lattice study}},
  \href{https://doi.org/10.1103/PhysRevD.98.054508}{\emph{Phys. Rev. D}
  {\bfseries 98} (2018) 054508}
  [\href{https://arxiv.org/abs/1807.06558}{{\ttfamily 1807.06558}}].

\bibitem{Bonati:2019kmf}
C.~Bonati, M.~Cardinali, M.~D'Elia and F.~Mazziotti, \emph{{$\theta$-dependence
  and center symmetry in Yang-Mills theories}},
  \href{https://doi.org/10.1103/PhysRevD.101.034508}{\emph{Phys. Rev. D}
  {\bfseries 101} (2020) 034508}
  [\href{https://arxiv.org/abs/1912.02662}{{\ttfamily 1912.02662}}].

\bibitem{Ambjorn:1984me}
J.~Ambj{\o}rn, P.~Olesen and C.~Peterson, \emph{{Observation of a string in
  three-dimensional SU(2) lattice gauge theory}},
  \href{https://doi.org/10.1016/0370-2693(84)91352-2}{\emph{Phys. Lett.}
  {\bfseries B142} (1984) 410}.

\bibitem{Teper:1998te}
M.~J. Teper, \emph{{SU(N) gauge theories in (2+1)-dimensions}},
  \href{https://doi.org/10.1103/PhysRevD.59.014512}{\emph{Phys. Rev.}
  {\bfseries D59} (1999) 014512}
  [\href{https://arxiv.org/abs/hep-lat/9804008}{{\ttfamily hep-lat/9804008}}].

\bibitem{Caselle:2004er}
M.~Caselle, M.~Pepe and A.~Rago, \emph{{Static quark potential and effective
  string corrections in the (2+1)-d SU(2) Yang-Mills theory}},
  \href{https://doi.org/10.1088/1126-6708/2004/10/005}{\emph{JHEP} {\bfseries
  0410} (2004) 005} [\href{https://arxiv.org/abs/hep-lat/0406008}{{\ttfamily
  hep-lat/0406008}}].

\bibitem{Caselle:2011vk}
M.~Caselle, A.~Feo, M.~Panero and R.~Pellegrini, \emph{{Universal signatures of
  the effective string in finite temperature lattice gauge theories}},
  \href{https://doi.org/10.1007/JHEP04(2011)020}{\emph{JHEP} {\bfseries 1104}
  (2011) 020} [\href{https://arxiv.org/abs/1102.0723}{{\ttfamily 1102.0723}}].

\bibitem{Bringoltz:2006zg}
B.~Bringoltz and M.~Teper, \emph{{A Precise calculation of the fundamental
  string tension in SU(N) gauge theories in 2+1 dimensions}},
  \href{https://doi.org/10.1016/j.physletb.2006.12.056}{\emph{Phys. Lett.}
  {\bfseries B645} (2007) 383}
  [\href{https://arxiv.org/abs/hep-th/0611286}{{\ttfamily hep-th/0611286}}].

\bibitem{Brandt:2010bw}
B.~B. Brandt, \emph{{Probing boundary-corrections to Nambu-Goto open string
  energy levels in 3d SU(2) gauge theory}},
  \href{https://doi.org/10.1007/JHEP02(2011)040}{\emph{JHEP} {\bfseries 1102}
  (2011) 040} [\href{https://arxiv.org/abs/1010.3625}{{\ttfamily 1010.3625}}].

\bibitem{Athenodorou:2016kpd}
A.~Athenodorou and M.~Teper, \emph{{Closed flux tubes in D = 2 + 1 SU(N) gauge
  theories: dynamics and effective string description}},
  \href{https://doi.org/10.1007/JHEP10(2016)093}{\emph{JHEP} {\bfseries 10}
  (2016) 093} [\href{https://arxiv.org/abs/1602.07634}{{\ttfamily
  1602.07634}}].

\bibitem{Brandt:2017yzw}
B.~B. Brandt, \emph{{Spectrum of the open QCD flux tube and its effective
  string description I: 3d static potential in SU(N = 2, 3)}},
  \href{https://doi.org/10.1007/JHEP07(2017)008}{\emph{JHEP} {\bfseries 07}
  (2017) 008} [\href{https://arxiv.org/abs/1705.03828}{{\ttfamily
  1705.03828}}].

\bibitem{Brandt:2018fft}
B.~B. Brandt, \emph{{Spectrum of the open QCD flux tube and its effective
  string description}}, \href{https://doi.org/10.22323/1.336.0039}{\emph{PoS}
  {\bfseries Confinement2018} (2018) 039}
  [\href{https://arxiv.org/abs/1811.11779}{{\ttfamily 1811.11779}}].

\bibitem{Brandt:2021kvt}
B.~B. Brandt, \emph{{Revisiting the flux tube spectrum of 3d SU(2) lattice
  gauge theory}},
  \href{https://doi.org/10.1007/s12648-021-02127-9}{\emph{Indian Journal of
  Physics} {\bfseries 95} (2021) 1613}
  [\href{https://arxiv.org/abs/2102.06413}{{\ttfamily 2102.06413}}].

\bibitem{Bonati:2021vbc}
C.~Bonati, M.~Caselle and S.~Morlacchi, \emph{{The Unreasonable effectiveness
  of effective string theory: The case of the 3D SU(2) Higgs model}},
  \href{https://doi.org/10.1103/PhysRevD.104.054501}{\emph{Phys. Rev. D}
  {\bfseries 104} (2021) 054501}
  [\href{https://arxiv.org/abs/2106.08784}{{\ttfamily 2106.08784}}].

\bibitem{Caristo:2021tbk}
F.~Caristo, M.~Caselle, N.~Magnoli, A.~Nada, M.~Panero and A.~Smecca,
  \emph{{Fine corrections in the effective string describing SU(2) Yang-Mills
  theory in three dimensions}},
  \href{https://arxiv.org/abs/2109.06212}{{\ttfamily 2109.06212}}.

\bibitem{Caselle:2024zoh}
M.~Caselle, N.~Magnoli, A.~Nada, M.~Panero, D.~Panfalone and L.~Verzichelli,
  \emph{{Confining strings in three-dimensional gauge theories beyond the
  Nambu-Got\={o} approximation}},
  \href{https://doi.org/10.1007/JHEP08(2024)198}{\emph{JHEP} {\bfseries 08}
  (2024) 198} [\href{https://arxiv.org/abs/2407.10678}{{\ttfamily
  2407.10678}}].

\bibitem{Polchinski:1992ty}
J.~Polchinski and Z.~Yang, \emph{{High temperature partition function of the
  rigid string}}, \href{https://doi.org/10.1103/PhysRevD.46.3667}{\emph{Phys.
  Rev. D} {\bfseries 46} (1992) 3667}
  [\href{https://arxiv.org/abs/hep-th/9205043}{{\ttfamily hep-th/9205043}}].

\bibitem{Bonati:2025hik}
C.~Bonati, M.~Caselle, A.~Negro, D.~Panfalone and L.~Verzichelli,
  \emph{{Effective string description of the reconfined phase in the trace
  deformed SU(2) Yang-Mills theory in (2+1) dimensions}},
  \href{https://doi.org/10.22323/1.466.0394}{\emph{PoS} {\bfseries LATTICE2024}
  (2025) 394} [\href{https://arxiv.org/abs/2501.13684}{{\ttfamily
  2501.13684}}].

\bibitem{Kennedy:1985nu}
A.~Kennedy and B.~Pendleton, \emph{{Improved Heat Bath Method for Monte Carlo
  Calculations in Lattice Gauge Theories}},
  \href{https://doi.org/10.1016/0370-2693(85)91632-6}{\emph{Phys. Lett.}
  {\bfseries B156} (1985) 393}.

\bibitem{Creutz:1987xi}
M.~Creutz, \emph{{Overrelaxation and Monte Carlo Simulation}},
  \href{https://doi.org/10.1103/PhysRevD.36.515}{\emph{Phys. Rev. D} {\bfseries
  36} (1987) 515}.

\bibitem{Metropolis:1953am}
N.~Metropolis, A.~Rosenbluth, M.~Rosenbluth, A.~Teller and E.~Teller,
  \emph{{Equation of state calculations by fast computing machines}},
  \href{https://doi.org/10.1063/1.1699114}{\emph{J. Chem. Phys.} {\bfseries 21}
  (1953) 1087}.

\bibitem{Edwards:2009qw}
S.~Edwards and L.~von Smekal, \emph{{SU(2) lattice gauge theory in 2+1
  dimensions: Critical couplings from twisted boundary conditions and
  universality}},
  \href{https://doi.org/10.1016/j.physletb.2009.10.063}{\emph{Phys. Lett. B}
  {\bfseries 681} (2009) 484}
  [\href{https://arxiv.org/abs/0908.4030}{{\ttfamily 0908.4030}}].

\bibitem{Polyakov:1978vu}
A.~M. Polyakov, \emph{{Thermal Properties of Gauge Fields and Quark
  Liberation}}, \href{https://doi.org/10.1016/0370-2693(78)90737-2}{\emph{Phys.
  Lett.} {\bfseries B72} (1978) 477}.

\bibitem{McLerran:1981pb}
L.~D. McLerran and B.~Svetitsky, \emph{{Quark Liberation at High Temperature: A
  Monte Carlo Study of SU(2) Gauge Theory}},
  \href{https://doi.org/10.1103/PhysRevD.24.450}{\emph{Phys. Rev.} {\bfseries
  D24} (1981) 450}.

\bibitem{Aharony:2013ipa}
O.~Aharony and Z.~Komargodski, \emph{{The Effective Theory of Long Strings}},
  \href{https://doi.org/10.1007/JHEP05(2013)118}{\emph{JHEP} {\bfseries 1305}
  (2013) 118} [\href{https://arxiv.org/abs/1302.6257}{{\ttfamily 1302.6257}}].

\bibitem{Brandt:2016xsp}
B.~B. Brandt and M.~Meineri, \emph{{Effective string description of confining
  flux tubes}}, {\emph{Int. J. Mod. Phys.} {\bfseries A31} (2016) 1643001}
  [\href{https://arxiv.org/abs/1603.06969}{{\ttfamily 1603.06969}}].

\bibitem{Caselle:2021eir}
M.~Caselle, \emph{{Effective String Description of the Confining Flux Tube at
  Finite Temperature}},
  \href{https://doi.org/10.3390/universe7060170}{\emph{Universe} {\bfseries 7}
  (2021) 170} [\href{https://arxiv.org/abs/2104.10486}{{\ttfamily
  2104.10486}}].

\bibitem{Luscher:2004ib}
M.~L{\"u}scher and P.~Weisz, \emph{{String excitation energies in SU(N) gauge
  theories beyond the free-string approximation}},
  \href{https://doi.org/10.1088/1126-6708/2004/07/014}{\emph{JHEP} {\bfseries
  0407} (2004) 014} [\href{https://arxiv.org/abs/hep-th/0406205}{{\ttfamily
  hep-th/0406205}}].

\bibitem{Nambu:1974zg}
Y.~Nambu, \emph{{Strings, Monopoles and Gauge Fields}},
  \href{https://doi.org/10.1103/PhysRevD.10.4262}{\emph{Phys. Rev.} {\bfseries
  D10} (1974) 4262}.

\bibitem{Goto:1971ce}
T.~Got{\={o}}, \emph{{Relativistic quantum mechanics of one-dimensional
  mechanical continuum and subsidiary condition of dual resonance model}},
  \href{https://doi.org/10.1143/PTP.46.1560}{\emph{Prog. Theor. Phys.}
  {\bfseries 46} (1971) 1560}.

\bibitem{Arvis:1983fp}
J.~F. Arvis, \emph{{The exact q anti-q potential in Nambu string theory}},
  \href{https://doi.org/10.1016/0370-2693(83)91640-4}{\emph{Phys. Lett.}
  {\bfseries B127} (1983) 106}.

\bibitem{Dubovsky:2012sh}
S.~Dubovsky, R.~Flauger and V.~Gorbenko, \emph{{Effective String Theory
  Revisited}}, \href{https://doi.org/10.1007/JHEP09(2012)044}{\emph{JHEP}
  {\bfseries 1209} (2012) 044}
  [\href{https://arxiv.org/abs/1203.1054}{{\ttfamily 1203.1054}}].

\bibitem{EliasMiro:2019kyf}
J.~Elias~Mir{\'o}, A.~L. Guerrieri, A.~Hebbar, J.~Penedones and P.~Vieira,
  \emph{{Flux Tube S-matrix Bootstrap}},
  \href{https://doi.org/10.1103/PhysRevLett.123.221602}{\emph{Phys. Rev. Lett.}
  {\bfseries 123} (2019) 221602}
  [\href{https://arxiv.org/abs/1906.08098}{{\ttfamily 1906.08098}}].

\bibitem{EliasMiro:2021nul}
J.~Elias~Mir{\'o} and A.~Guerrieri, \emph{{Dual EFT bootstrap: QCD flux
  tubes}}, \href{https://doi.org/10.1007/JHEP10(2021)126}{\emph{JHEP}
  {\bfseries 10} (2021) 126}
  [\href{https://arxiv.org/abs/2106.07957}{{\ttfamily 2106.07957}}].

\bibitem{Caselle:2014eka}
M.~Caselle, M.~Panero, R.~Pellegrini and D.~Vadacchino, \emph{{A different kind
  of string}}, \href{https://doi.org/10.1007/JHEP01(2015)105}{\emph{JHEP}
  {\bfseries 1501} (2015) 105}
  [\href{https://arxiv.org/abs/1406.5127}{{\ttfamily 1406.5127}}].

\bibitem{Aharony:2024ctf}
O.~Aharony, N.~Barel and T.~Sheaffer, \emph{{Effective strings in QED$_{3}$}},
  \href{https://doi.org/10.1007/JHEP03(2025)143}{\emph{JHEP} {\bfseries 03}
  (2025) 143} [\href{https://arxiv.org/abs/2412.01313}{{\ttfamily
  2412.01313}}].

\bibitem{Teper:2009uf}
M.~Teper, \emph{{Large N and confining flux tubes as strings - a view from the
  lattice}}, {\emph{Acta Phys. Polon.} {\bfseries B40} (2009) 3249}
  [\href{https://arxiv.org/abs/0912.3339}{{\ttfamily 0912.3339}}].

\bibitem{Aharony:2009gg}
O.~Aharony and E.~Karzbrun, \emph{{On the effective action of confining
  strings}}, \href{https://doi.org/10.1088/1126-6708/2009/06/012}{\emph{JHEP}
  {\bfseries 0906} (2009) 012}
  [\href{https://arxiv.org/abs/0903.1927}{{\ttfamily 0903.1927}}].

\bibitem{Polyakov:1976fu}
A.~M. Polyakov, \emph{{Quark Confinement and Topology of Gauge Groups}},
  \href{https://doi.org/10.1016/0550-3213(77)90086-4}{\emph{Nucl. Phys.}
  {\bfseries B120} (1977) 429}.

\bibitem{Gopfert:1981er}
M.~G{\"o}pfert and G.~Mack, \emph{{Proof of Confinement of Static Quarks in
  Three-Dimensional U(1) Lattice Gauge Theory for All Values of the Coupling
  Constant}}, \href{https://doi.org/10.1007/BF01961240}{\emph{Commun. Math.
  Phys.} {\bfseries 82} (1981) 545}.

\bibitem{Caselle:2016mqu}
M.~Caselle, M.~Panero and D.~Vadacchino, \emph{{Width of the flux tube in
  compact U(1) gauge theory in three dimensions}},
  \href{https://doi.org/10.1007/JHEP02(2016)180}{\emph{JHEP} {\bfseries 1602}
  (2016) 180} [\href{https://arxiv.org/abs/1601.07455}{{\ttfamily
  1601.07455}}].

\bibitem{Peliti:1985eo}
L.~Peliti and S.~Leibler, \emph{{Effects of Thermal Fluctuations on Systems
  with Small Surface Tension}},
  \href{https://doi.org/10.1103/PhysRevLett.54.1690}{\emph{Phys. Rev. Lett.}
  {\bfseries 54} (1985) 1690}.

\bibitem{Helfrich:1985eo}
W.~Helfrich, \emph{{Effect of thermal undulations on the rigidity of fluid
  membranes and interfaces}},
  \href{https://doi.org/10.1051/jphys:019850046070126300}{\emph{J. Phys.
  France} {\bfseries 46} (1985) 1263}.

\bibitem{Forster:1986ot}
D.~F{\"o}rster, \emph{{On the scale dependence, due to thermal fluctuations, of
  the elastic properties of membranes}},
  \href{https://doi.org/http://dx.doi.org/10.1016/0375-9601(86)90536-0}{\emph{Phys.
  Lett.} {\bfseries A114} (1986) 115}.

\bibitem{Polyakov:1986cs}
A.~M. Polyakov, \emph{{Fine Structure of Strings}},
  \href{https://doi.org/10.1016/0550-3213(86)90162-8}{\emph{Nucl. Phys.}
  {\bfseries B268} (1986) 406}.

\bibitem{Kleinert:1986bk}
H.~Kleinert, \emph{{The Membrane Properties of Condensing Strings}},
  \href{https://doi.org/10.1016/0370-2693(86)91111-1}{\emph{Phys. Lett.}
  {\bfseries B174} (1986) 335}.

\bibitem{Braaten:1986bz}
E.~Braaten, R.~D. Pisarski and S.-M. Tse, \emph{{The Static Potential for
  Smooth Strings}},
  \href{https://doi.org/10.1103/PhysRevLett.58.93}{\emph{Phys. Rev. Lett.}
  {\bfseries 58} (1987) 93}.

\bibitem{German:1989vk}
G.~Germ{\'a}n and H.~Kleinert, \emph{{Perturbative Two Loop Quark Potential of
  Stiff Strings in Any Dimension}},
  \href{https://doi.org/10.1103/PhysRevD.40.1108}{\emph{Phys. Rev.} {\bfseries
  D40} (1989) 1108}.

\bibitem{Ambjorn:2014rwa}
J.~Ambj\o{}rn, Y.~Makeenko and A.~Sedrakyan, \emph{{Effective QCD string beyond
  the Nambu-Goto action}},
  \href{https://doi.org/10.1103/PhysRevD.89.106010}{\emph{Phys. Rev. D}
  {\bfseries 89} (2014) 106010}
  [\href{https://arxiv.org/abs/1403.0893}{{\ttfamily 1403.0893}}].

\bibitem{Aharony:2010db}
O.~Aharony and N.~Klinghoffer, \emph{{Corrections to Nambu-Goto energy levels
  from the effective string action}},
  \href{https://doi.org/10.1007/JHEP12(2010)058}{\emph{JHEP} {\bfseries 1012}
  (2010) 058} [\href{https://arxiv.org/abs/1008.2648}{{\ttfamily 1008.2648}}].

\bibitem{Billo:2012da}
M.~Bill{\'o}, M.~Caselle, F.~Gliozzi, M.~Meineri and R.~Pellegrini, \emph{{The
  Lorentz-invariant boundary action of the confining string and its universal
  contribution to the inter-quark potential}},
  \href{https://doi.org/10.1007/JHEP05(2012)130}{\emph{JHEP} {\bfseries 1205}
  (2012) 130} [\href{https://arxiv.org/abs/1202.1984}{{\ttfamily 1202.1984}}].

\bibitem{Caselle:2026coc}
M.~Caselle, E.~Cellini, A.~Nada, D.~Panfalone and L.~Verzichelli,
  \emph{{Intrinsic Width of the Flux Tube as a tool to explore confining
  mechanisms in Lattice Gauge Theories}},
  \href{https://arxiv.org/abs/2601.19520}{{\ttfamily 2601.19520}}.

\bibitem{DiGiacomo:1990hc}
A.~Di~Giacomo, M.~Maggiore and {\v{S}}.~Olejn{\'{\i}}k, \emph{{Confinement and
  Chromoelectric Flux Tubes in Lattice {QCD}}},
  \href{https://doi.org/10.1016/0550-3213(90)90567-W}{\emph{Nucl. Phys.}
  {\bfseries B347} (1990) 441}.

\bibitem{Svetitsky:1982gs}
B.~Svetitsky and L.~G. Yaffe, \emph{{Critical Behavior at Finite Temperature
  Confinement Transitions}},
  \href{https://doi.org/10.1016/0550-3213(82)90172-9}{\emph{Nucl. Phys.}
  {\bfseries B210} (1982) 423}.

\bibitem{Zierenberg:2016zoh}
J.~Zierenberg, N.~G. Fytas, M.~Weigel, W.~Janke and A.~Malakis, \emph{{Scaling
  and universality in the phase diagram of the 2D Blume-Capel model}},
  \href{https://doi.org/10.1140/epjst/e2016-60337-x}{\emph{Eur. Phys. J.
  Special Topics} {\bfseries 226} (2017) 789}
  [\href{https://arxiv.org/abs/1612.02138}{{\ttfamily 1612.02138}}].

\bibitem{Mozolenko:2024blc}
V.~Mozolenko and L.~N. Shchur, \emph{{Blume-Capel model analysis with
  microcanonical population annealing method}},
  \href{https://doi.org/10.1103/PhysRevE.109.045306}{\emph{Phys. Rev. E}
  {\bfseries 109} (2024) 045306}
  [\href{https://arxiv.org/abs/2402.18985}{{\ttfamily 2402.18985}}].

\bibitem{condmat0012164}
A.~Pelissetto and E.~Vicari, \emph{{Critical Phenomena and
  Renormalization-Group Theory}},
  \href{https://doi.org/10.1016/S0370-1573(02)00219-3}{\emph{Phys. Rept.}
  {\bfseries 368} (2002) 549}
  [\href{https://arxiv.org/abs/cond-mat/0012164}{{\ttfamily
  cond-mat/0012164}}].

\bibitem{Bonati:2024gh}
C.~Bonati, A.~Pelissetto and E.~Vicari, \emph{{Three-dimensional Abelian and
  non-Abelian gauge Higgs theories}},
  \href{https://arxiv.org/abs/2410.05823}{{\ttfamily 2410.05823}}.

\bibitem{Ferrenberg:1988yz}
A.~M. Ferrenberg and R.~H. Swendsen, \emph{{New Monte Carlo Technique for
  Studying Phase Transitions}},
  \href{https://doi.org/10.1103/PhysRevLett.61.2635}{\emph{Phys. Rev. Lett.}
  {\bfseries 61} (1988) 2635}.

\bibitem{Challa:1986fo}
M.~S.~S. Challa, D.~P. Landau and K.~Binder, \emph{{Finite-size effects at
  temperature-driven first-order transitions}},
  \href{https://doi.org/10.1103/PhysRevB.34.1841}{\emph{Phys. Rev. B}
  {\bfseries 34} (1986) 1841}.

\bibitem{Michael:1993yj}
C.~Michael, \emph{{Fitting correlated data}},
  \href{https://doi.org/10.1103/PhysRevD.49.2616}{\emph{Phys. Rev. D}
  {\bfseries 49} (1994) 2616}
  [\href{https://arxiv.org/abs/hep-lat/9310026}{{\ttfamily hep-lat/9310026}}].

\bibitem{Bruno:2022mfy}
M.~Bruno and R.~Sommer, \emph{{On fits to correlated and auto-correlated
  data}}, \href{https://doi.org/10.1016/j.cpc.2022.108643}{\emph{Comput. Phys.
  Commun.} {\bfseries 285} (2023) 108643}
  [\href{https://arxiv.org/abs/2209.14188}{{\ttfamily 2209.14188}}].

\bibitem{Yoon:2011wdw}
B.~Yoon, Y.-C. Jang, C.~Jung and W.~Lee, \emph{{Covariance fitting of highly
  correlated data in lattice QCD}},
  \href{https://doi.org/10.3938/jkps.63.145}{\emph{J. Korean Phys. Soc.}
  {\bfseries 63} (2013) 145} [\href{https://arxiv.org/abs/1101.2248}{{\ttfamily
  1101.2248}}].

\bibitem{Wolff:2003sm}
{\scshape ALPHA} collaboration, U.~Wolff, \emph{{Monte Carlo errors with less
  errors}}, \href{https://doi.org/10.1016/S0010-4655(03)00467-3,
  10.1016/j.cpc.2006.12.001}{\emph{Comput. Phys. Commun.} {\bfseries 156}
  (2004) 143} [\href{https://arxiv.org/abs/hep-lat/0306017}{{\ttfamily
  hep-lat/0306017}}].

\end{thebibliography}
